\newif\iffigs\figstrue
\documentclass[10pt,a4paper]{article}
\usepackage{latexsym,amssymb,lscape,graphics,setspace}
\usepackage{amsmath}
\usepackage{graphicx}        
\usepackage{longtable}
\usepackage{multirow}
\usepackage{booktabs}
\usepackage{multirow}
\usepackage{color}
\usepackage{slashed,epsfig}
\usepackage{amsfonts}
\usepackage{cite}
\usepackage{verbatim}
\usepackage{colortbl}
\usepackage[table]{xcolor}
\usepackage{hyperref}       
\usepackage{array}
\usepackage{fancyhdr}
\usepackage{tikz}
\usepackage{multicol}
\usepackage{lscape}
\graphicspath{{images/}}

\newcommand{\mathsym}[1]{{}}

\newtheorem{definizione}{Definition}[section]
\newtheorem{teorema}{Theorem}[section]

\newtheorem{dimostrazione}{Proof}[teorema]

\newcommand{\bd}{\begin{definizione}}
\newcommand{\ed}{\end{definizione}}

\def\IC{\relax\,\hbox{$\inbar\kern-.3em{\rm C}$}}
\def\IG{\relax\,\hbox{$\inbar\kern-.3em{\rm G}$}}
\def\IB{\relax{\rm I\kern-.18em B}}
\def\ID{\relax{\rm I\kern-.18em D}}
\def\IL{\relax{\rm I\kern-.18em L}}
\def\IF{\relax{\rm I\kern-.18em F}}
\def\IH{\relax{\rm I\kern-.18em H}}
\def\II{\relax{\rm I\kern-.17em I}}
\def\IN{\relax{\rm I\kern-.18em N}}
\def\IP{\relax{\rm I\kern-.18em P}}
\def\IQ{\relax\,\hbox{$\inbar\kern-.3em{\rm Q}$}}
\def\bfzero{\relax\,\hbox{$\inbar\kern-.3em{\rm 0}$}}
\def\IK{\relax{\rm I\kern-.18em K}}
\def\IG{\relax\,\hbox{$\inbar\kern-.3em{\rm G}$}}
 \font\cmss=cmss10 \font\cmsss=cmss10 at 7pt
\def\IR{\relax{\rm I\kern-.18em R}}
\def\ZZ{\relax\ifmmode\mathchoice
{\hbox{\cmss Z\kern-.4em Z}}{\hbox{\cmss Z\kern-.4em Z}}
{\lower.9pt\hbox{\cmsss Z\kern-.4em Z}} {\lower1.2pt\hbox{\cmsss
Z\kern-.4em Z}}\else{\cmss Z\kern-.4em Z}\fi}
\def\bfone{\relax{\rm 1\kern-.35em 1}}

\def\inbar{\vrule height1.5ex width.4pt depth0pt}
\def\bfzero{\relax{\rm I\kern-.18em 0}}
\def\bfone{\relax{\rm 1\kern-.35em 1}}

\DeclareFontFamily{U}{rsf}{} \DeclareFontShape{U}{rsf}{m}{n}{
  <5> <6> rsfs5 <7> <8> <9> rsfs7 <10-> rsfs10}{}
\DeclareMathAlphabet\Scr{U}{rsf}{m}{n}

\newcommand{\ft}[2]{{\textstyle\frac{#1}{#2}}}

\def\1bar{1\hskip -.275cm -}
\def\2bar{2\hskip -.275cm -}
\def\3bar{3\hskip -.275cm -}

\newsavebox{\uuunit}
\sbox{\uuunit}
{\setlength{\unitlength}{0.825em}
\begin{picture}(0.6,0.7)
\thinlines
\put(0,0){\line(1,0){0.5}}
\put(0.15,0){\line(0,1){0.7}}
\put(0.35,0){\line(0,1){0.8}}
\multiput(0.3,0.8)(-0.04,-0.02){10}{\rule{0.5pt}{0.5pt}}
\end {picture}}

\makeatletter \@addtoreset{equation}{section} \makeatother

\def\bfone{\relax{\rm 1\kern-.35em 1}}

\def\bfone{\relax{\rm 1\kern-.35em 1}}
\font\cmss=cmss10 \font\cmsss=cmss10 at 7pt

\newcommand{\so}{\mathfrak{so}}
\newcommand{\su}{\mathfrak{su}}

\newcommand{\uu}{\mathfrak{u}}
\newcommand{\sym}{\mathfrak{sp}}

\begin{document}
\begin{titlepage}
\begin{center}
\vskip 0.2cm
{{\large {\sc The macroscopic K\"ahler metric of Geometric Thermodynamics\\
versus the microscopic one  on the  Event Manifold:\\}}
\vskip 0.1cm 
{\large{\it Exact Partition Functions  on {\sc Calabi Vesentini} microscopic manifolds:\\
 extended Souriau temperatures and {\sc spontaneous magnetizations} ${}^\dagger$}} }\\
 \vskip 1cm {\sc Pietro Fr\'e\,$^{a,b}$,  \\ Alexander S. Sorin\,$^{c,b}$
 and Mario Trigiante\,$^{d, e}$} \vskip 0.5cm
\smallskip
{\sl \small \frenchspacing ${}^a\,$ {\tt Emeritus Professor of}  Dipartimento di Fisica, 
Universit\`a di Torino, Via P. Giuria 1, I-10125 Torino, Italy \\[2pt]
${}^{b}\,${\tt Senior Consultant of } Additati\&Partners Consulting s.r.l, 
Via Filippo Pacini 36, I-51100 Pistoia, Italy \\[2pt]
${}^c\,$ Center for Quantum Science and Technology, Tel-Aviv University, Tel-Aviv 69978, Israel \\[2pt]
${}^d\,$Dipartimento DISAT, Politecnico di Torino,
C.so Duca degli Abruzzi 24, I-10129 Torino, Italy\\[2pt]
${}^e\,$INFN, Sezione di Torino\\[2pt]
E-mail:  {\tt pietro.fre@unito.it, asorin@tauex.tau.ac.il,\\
 mario.trigiante@polito.it, } }
\begin{abstract}
In this paper we clarify the relation between Geometric Thermodynamics and Information Geometry based on the Fisher matrix. On the macroscopic  odd-dimensional contact manifold  of thermodynamic variables, we introduce for the first time a metric, whose pull-back on the isoentropic symplectic submanifolds transverse to the Reeb field is K\"ahlerian. The pull-back of such metric on equilibrium states, that are lagrangian submanifolds, is the Fisher Hessian. Then we consider the Souriau-like Thermodynamics that uses Calabi-Vesentini (CV) manifolds as  K\"ahlerian microscopic event manifolds and the Killing moment maps as observable functions. A systematic use of the theory of compact abelian structures and the setup of Special K\"ahler Geometry in which CV manifolds are encoded allows us to perform the explicit integration defining the partition function for any entry in the CV Tits Satake universality class. The additional actions
completing the abelian structure are non linear Casimir functions of the Killing moment-maps and suggest a generalization of Souriau  thermodynamics that partially breaks the isometry group symmetry by means of the non vanishing mean values of the Casimir functions in a manner similar to the spontaneous magnetization in ferromagnetism. Our new exact Gibbs distributions provide the analogue for Cartan Neural Networks of the Gaussian probability distributions in flat space used in conventional Machine Learning.
\end{abstract}
\vfill
\end{center}
\noindent \parbox{175mm}{\hrulefill}
\par
${}^\dagger$ P.G. Fr\'e acknowledges support by the Company \textit{Additati\&Partners 
Consulting s.r.l} during the development of 
the present research.\\
The work of A.S. Sorin is supported in part by the Center for Integration in Science of the Israel Ministry of Aliyah
and Integration.
\\[5pt]
\end{titlepage}
{\small \tableofcontents} \noindent {}
\newpage
\section{Introduction} \label{introibo}
The conceptual environment, in which the  issues addressed in the present paper are located, was extensively discussed in a previous paper recently published  by the same authors \cite{geotermico}. The target that we pursue in this article can be described as a conceptual systematic reorganization of the logic behind so named Information Geometry\cite{raone,cenzone,amarone}, namely its identification with Geometrical Thermodynamics\cite{gianno1,gianno2,lychaginlecture,Kushner_2020,Lychagin_2020,ludaed,ludaed2,
Ruppeiner_2012b,Ruppeiner_2010,Ruppeiner_2012,Ruppeiner_2013,ruppoRdiag} and, then, the development of mathematical techniques for the explicit calculation of partition functions in those cases where the microscopic event manifold is a
K\"ahlerian non-compact symmetric space $\mathrm{U/H}$. 
\par 
Such a choice is motivated by the research project on Cartan Neural Networks\cite{pgtstheory,naviga,TSnaviga,tassellandum} that aims at the construction of Neural Networks that are required to be both covariant and interpretable, thanks to the removal of point-wise activation functions, the necessary non linearity device being provided, instead, by the exponential map from a solvable Lie algebra to its corresponding solvable Lie group. In that context a rather tight logic based on classical theorems and on  mathematical structures that were developed in the context of Supergravity Theory (see in  particular \cite{pgtstheory} and \cite{TSnaviga}) lead to the conclusion that the most appropriate mathematical modelling of Neural Network layers is indeed provided by non-compact symmetric spaces $\mathrm{U/H}$ (where $\mathrm{U}$ is a simple non-compact Lie group and $\mathrm{H}\subset \mathrm{U}$ its maximal compact subgroup\footnote{More generally, as it starts emerging from a new ongoing research project\cite{toinepietromario}, the mathematical modeling of Cartan Neural Network layers can be provided by all Special K\"ahler Homogeneous manifolds, the
non-symmetric among which might possibly be interpreted as Special K\"ahler Symmetric Spaces with a deformed metric partially breaking the full Isometry Group U.}). 
\par
Probability distributions on this type of layers cannot be classical Gaussians (in all their multivariate versions) that, instead, are appropriate to Euclidian flat spaces $\mathbb{E}^n \, =\, \mathbb{R}^n$, rather they must be  \textbf{Gibbs probability measures} on $\mathrm{U/H}$, consistent with the isometry group $\mathrm{U}$ (or a subgroup thereof) of such spaces. As we stress in the sequel the construction of such probability measures is what goes under the name of \textbf{Souriau thermodynamics}\cite{souriaub1,souriaub2,barbarpapad4,
marlentropia,caldobarbaresco,barbaresco2,barbaresco3,marlegibbs,barbarpapad1,barbarpapad2,barbarpapad3,nebbo} that in this paper \textbf{we generalize} while developing suitable mathematical methods \textbf{to calculate explicitly and exactly} its underlying \textbf{partition functions}. 
\par
Souriau thermodynamics and its generalization, however, have to be seen in the more general context of \textbf{geometric thermodynamics} that we summarize and conceptually reorganize in the next section \ref{infogeomacro}.
\par 
In Machine Learning literature authors frequently utilize the concept of \textit{Information Geometry} based on the notion of Fisher Information matrix whose
identification with a metric tensor in a space of parameters is usually credited to
Rao, Chentsov and Amari \cite{raone,cenzone,amarone}.
Let $\Omega$ be a space of events (either discrete or continuous or mixed), whose points we denote by $\mathbf{q}\in \Omega$ (typically an array of coordinates) and let $\mho$ be a space of parameters whose points we denote by $\boldsymbol{\lambda}\in \mho$ (typically another array of coordinates). Let moreover $p(\mathbf{x},\boldsymbol{\lambda})$ be a probability distribution on $\Omega$, depending on  $n$ parameters $\lambda^i$ and  properly normalized:
\begin{equation}\label{probanorma}
  \int_\Omega \, p(\mathbf{q},\boldsymbol{\lambda}) \mathrm{d}[\mathbf{q}] \, = \, 1
\end{equation}
where $\mathrm{d}[\mathbf{q}]$ denotes the integration measure. The Fisher Information matrix is defined as the following expectation value:
\begin{equation}\label{coriandolo}
  M_{ij}^{Fish}(\boldsymbol{\lambda}) \, = \, \mathbb{E}\left[\frac{\partial^2}{\partial\lambda^i\partial\lambda^j}\,
  \log\left[p(\mathbf{q},
  \boldsymbol{\lambda})\right]\right] \, \equiv  \, \int_\Omega \frac{\partial^2}{\partial\lambda^i\partial\lambda^j}\,
  \log\left[p(\mathbf{q},
  \boldsymbol{\lambda})\right] \times p(\mathbf{q},
  \boldsymbol{\lambda}) \, \mathrm{d}[\mathbf{q}]
\end{equation}
and it is connected to the Kullbach-Leibler divergence and other statistical tokens
(see for instance the comprehensive review \cite{fioresi2023deep}). According to a general consensus in the Statistical/Machine Learning scientific community, if the underlying stochastic processes at the basis of the considered parameterized probability distribution  $p(\mathbf{q},\boldsymbol{\lambda})$ are of \textbf{Markovian type} then the latter is of Gibbs type namely:
\begin{equation}\label{ciacolone}
  p(\mathbf{q},\boldsymbol{\lambda}) \, = \, \frac{\exp\left[-\boldsymbol{\lambda}\cdot \mathbf{X}(\mathbf{q})\right]}{Z(\boldsymbol{\lambda})} \quad ; \quad 
  Z(\boldsymbol{\lambda}) \equiv \int_\Omega \,\exp\left[-\boldsymbol{\lambda}\cdot \mathbf{X}(\mathbf{q})\right] \, \mathrm{d}[\mathbf{q}]
\end{equation}
where $\mathbf{X}(\mathbf{q})\, = \, X^{i,\dots,n}(\mathbf{q})$ is a suitable collection, depending on the addressed problem, of $n$ observable functions defined over the manifold of events $\Omega$ 
and the normalization denominator $Z(\boldsymbol{\lambda})$ is named \textbf{the partition function}. It is a matter of calculation "on the back of an envelope" to verify that, for Gibbs type probability distributions, Fisher Information Matrix reduces to the Hessian of what is named the stochastic hamiltonian:
\begin{equation}\label{calottapolare}
  M_{ij}^{Fish}(\boldsymbol{\lambda}) \, = \,\frac{\partial^2 \mathcal{H}^{sto}(\boldsymbol{\lambda})}{\partial\lambda^i\partial\lambda^j}
  \quad ; \quad \mathcal{H}^{sto}(\boldsymbol{\lambda})\, = \, - \, \log\left[Z(\boldsymbol{\lambda}) \right]
\end{equation}
But why should this matrix be interpreted as a Riemannian metric tensor? The deep reason for this will be explained in section
\ref{infogeomacro} where we establish a logical development and a robust conceptual framework that puts into a unique comprehensive overview Information Geometry and Geometrical Thermodynamics, reconstructing also the historical path that leads to this unified overview. Contact Geometry and its reduction to symplectic geometry on the isoentropic leaves transverse to the Reeb field is the conceptual frame in which the Fisher matrix acquires the status of a metric on lagrangian submanifolds describing equilibrium states. Indeed it is the pull-back of the K\"ahler metric existing on all isoentropic leaves for whatever type of equilibrium thermodynamics derived from whatever Gibbs-type probability distribution. 
\par
It is within this general conceptual framework that we consider 
the Gibbs type of probability distributions defined on non-compact, symmetric spaces fulfilling covariance with respect to the isometry group $\mathrm{U}$. Naming, as it is appropriate to do, \textbf{generalized temperatures} the parameters $\boldsymbol{\lambda}\,=\,\boldsymbol{\beta}$ that, in this case, form a vector in the coadjoint representation of the isometry Lie algebra $\mathbb{U}$ and are associated with functions $\mathbf{X(q)}$ on $\Omega=\mathrm{U/H}$ that are now the moment-maps of  Killing vector fields $\mathfrak{P}_{\mathfrak{k}}(\mathbf{q})$,  one obtains what goes under the name of Souriau thermodynamics\cite{souriaub1,souriaub2,barbarpapad4,
marlentropia,caldobarbaresco,barbaresco2,barbaresco3,marlegibbs,barbarpapad1,barbarpapad2,barbarpapad3,nebbo} and provides the appropriate generalization to non-compact symmetric spaces of the classical gaussian distributions pertaining to flat Euclidian space. 
\par
As we emphasize in the course of the present article, the ability to calculate explicitly the partition function integral $Z(\boldsymbol{\beta})$ and to determine in this way the subspace of the Lie Algebra
$\mathbb{T} \subset \mathbb{U}$ whose dual $\mathbb{T}^\star$ contains the available temperature vectors $\boldsymbol{\beta} \in \mathbb{T}^\star$ for which the corresponding integral converges, strictly depends on the theory of \textbf{Abelian Structures}, namely on the classification of canonical Darboux coordinate bases on the  microscopic K\"ahler event manifold $\Omega$, where the microscopic K\"ahler $2$-form takes the same appearance as in classical Hamiltonian Mechanics \textit{i.e.}:
\begin{eqnarray}\label{cosimoprimo}
  \boldsymbol{\mathcal{K}} &\equiv& \mathit{i}\, \frac{\partial^{2} }{\partial z^i\,\partial \bar{z}^{j^\star}}\, \mathcal{K}(z,\bar{z}) \, \mathrm{d}z^i \, \wedge \, \mathrm{d}\bar{z}^{j^\star} \nonumber \\
 \null &=& \sum_{a=1}^{n} \mathrm{d}\mathfrak{p}_a(z,\bar{z}) \,\wedge \, \mathrm{d}\mathfrak{q}^a(z,\bar{z})
\end{eqnarray}
where $z^i$ ($i=1,\dots,n$) is some set of complex coordinates for $\Omega$, $n$ being defined as the complex dimension of the K\"ahler manifold 
\begin{equation}\label{dimensionotto}
  n \, \equiv \, \mathrm{dim}_{\mathbb{C}} \, \Omega \quad ; \quad \mathrm{dim}_{\mathbb{R}} \, \Omega \, = \, 2\, n
\end{equation}
and $\mathcal{K}(z,\bar{z})$ is the K\"ahler potential generating the K\"ahler metric, while $\mathfrak{p}_a(z,\bar{z})$ and $\mathfrak{q}^a(z,\bar{z})$ are \textbf{two sets of real functions} of the complex coordinates $z,\bar{z}$ (or of any other set of $2n$ real coordinates covering the manifold $\Omega$) that are in involution with respect to the Poisson brackets induced by  the K\"ahler metric and its inverse, namely:
\begin{equation}\label{coricidinoA}
  \left\{\mathfrak{p}_a(z,\bar{z})\, , \,\mathfrak{p}_b(z,\bar{z})\right\} \, = \, 0 \quad; \quad
  \left\{\mathfrak{q}^a(z,\bar{z})\, , \,\mathfrak{q}^b(z,\bar{z})\right\} \, = \, 0 \quad; \quad
  \left\{\mathfrak{p}_a(z,\bar{z})\, , \,\mathfrak{q}^b(z,\bar{z})\right\} \, = \, \delta^{b}_a
\end{equation}
where, by definition:
\begin{equation}\label{PB1}
  \left\{\boldsymbol{\mathit{f}}(z,\bar{z})\, , \,\boldsymbol{\mathit{g}}(z,\bar{z})\right\} \, \equiv\, 
  \mathit{i}\, g^{ij^\star} \, \left(\frac{\partial \boldsymbol{\mathit{f}} }{\partial z^i}
  \,\frac{\partial \boldsymbol{\mathit{g}}}{\partial \bar{z}^{j^\star}} \, - \,
  \frac{\partial \boldsymbol{\mathit{g}} }{\partial z^i}
  \,\frac{\partial \boldsymbol{\mathit{f}} }{\partial \bar{z}^{j^\star}}\right) 
  \end{equation}
and 
\begin{equation}\label{cartolina}
  g_{ij^\star}\, \equiv  \, \frac{\partial^{2} }{\partial z^i\,\partial \bar{z}^{j^\star}}\, \mathcal{K}(z,\bar{z}) \quad ; \quad  g^{ij^\star} \, = \, 
   \left(g^{-1}\right)^{ij^\star}
\end{equation}
are the K\"ahler metric and its inverse.
\par
From the Hamiltonian viewpoint, a system is said to be {\bf Liouville integrable}, or {\bf completely canonically integrable},  if one can find $n$ functions $\mathfrak{p}_a(z,\bar{z})$ that satisfy the first of  (\ref{PB1}) and which include the Hamiltonian of the system itself. In this case, $\mathfrak{p}_a(z,\bar{z})$, named \textbf{actions}, represent a complete set of Hamiltonians in involution and, as such, are constants of motion. The definition of a completely canonically integrable system also requires the level manifold of first integrals $\boldsymbol{\mathcal{Q}}_n\subset \Omega$, spanned by the canonically dual coordinates $\mathfrak{q}^a(z,\bar{z})$, to be compact.  By Arnol'd Theorem, $\mathcal{Q}_n$ is then diffeomorphic to an n-torus: $\boldsymbol{\mathcal{Q}}_n\sim \boldsymbol{\mathcal{T}}^n$:
\begin{equation}\label{ntorus}
  \boldsymbol{\mathcal{T}}^{n} \equiv \underbrace{\mathbb{S}^1\times ...\times \mathbb{S}^1}_{n-times}
\end{equation}
Howwever, the existence of n integrals of motion in involution does not require $\boldsymbol{\mathcal{Q}}_n$, spanned by the $\mathfrak{q}^a(z,\bar{z})$ to be compact. With a slight misuse of terminology, we shall extend the definition of Liouville integrable systems to these situations. A particularly relevant completely opposite case occurs when $\boldsymbol{\mathcal{Q}}^{n}\simeq \mathbb{R}^n$ and the K\"ahler potential does not depend on the variables  variables $\mathfrak{q}^a(z,\bar{z})$ that can be identified with the real parts of the complex coordinates in a suitable complex basis: 
\begin{equation}\label{fragerolamo}
  \mathfrak{q}^a(z,\bar{z}) \, = \, \ft 12 \, (z^a+\bar{z}^{a^\star}) \equiv u^a
\end{equation}
It follows that $\mathcal{K}(z,\bar{z})$ depends only on the imaginary parts:
\begin{equation}\label{cortinadilegno}
  \mathcal{K}(z,\bar{z}) \, = \, \mathcal{K}(\boldsymbol{v}) \quad ; \quad v^a \, - \, \ft 12 \, \mathit{i} \, (z^a -\bar{z}^{a^\star}) \quad a=1\,\dots,n  
\end{equation}
and the manifold $\Omega$ admits $n$ translational abelian isometries. 
\par
As we are going to see this is the case of the K\"ahler metric of the most general macroscopic thermodynamic space $\mho$, the abelian non-compact isometries corresponding to generic shifts in the average values of the observable functions defined over the microscopic event manifold $\Omega$, irrespectively whether the latter is also K\"alerian or not and irrespectively of what such observable functions are.
\par
Noticeably the same is also true for all \textbf{Homogeneous Special K\"ahler Manifolds} among which we find the \textbf{Non-Compact Symmetric Spaces of K\"ahler type} that we utilize as mathematical models of the layers in Cartan Neural Networks.
This happens because on the solvable groups $\mathcal{S}_{\mathfrak{hsk}}$ that, by the metric equivalence which is essential for the Cartan Neural Network architectures (see \cite{pgtstheory}), one constructs the relevant K\"ahler metric  introducing a K\"ahler potential of the type described in eq.(\ref{cortinadilegno}) with the specific form:
\begin{equation}\label{cubicone}
  \mathcal{K}(\boldsymbol{v}) \, = \, - \, \log \left(d_{abc} \,v^a\,v^b \,v^c\right)
\end{equation}
where $d_{abc}$ is a constant symmetric tensor with three indices. Indeed the classification of homogeneous special K\"ahler geometries amounts to a classification of the cubic forms defined by the $3$-tensor $d_{abc}$ and endowed with specific necessary properties singled out in the 1980s and early 1990s \cite{deWit:1995tf,toineugenio,SKGaggio3,SKGaggio2,SKGaggio1,specHomgeoA2,specHomgeoA1}.
The translational isometries encoded in this general structure of special K\"ahler manifolds correspond, at the level of their metric equivalent solvable group $\mathcal{S}_{\mathfrak{hsk}}$ to a \textbf{maximal abelian ideal} $\mathcal{AI}_{max}\subset\mathcal{S}_{\mathfrak{hsk}}$ of dimension equal to one half the dimension of $\mathcal{S}_{\mathfrak{hsk}}$ which is possessed by all such solvable Lie groups. 
\par
The abelian structure where the  $n$-dimensional submanifold $\boldsymbol{\mathcal{Q}}^{n} \simeq  \mathcal{AI}_{max}$ is isomorphic not to a $n$-torus, rather to the maximal abelian ideal of the solvable group $\mathcal{S}_{\mathfrak{hsk}}$ exists on all homogeneous  special K\"ahler manifolds and might have different interesting applications (see section \ref{mariosecta}) yet it is not the one appropriate to derive  Gibbs distributions that provide the generalization of Gaussian distributions on Cartan Neural Network layers. As we showed in \cite{geotermico}, in order to obtain Gaussian-like Gibbs distributions centered at the origin of the manifold and then transported by the action of the solvable group to be centered around any other point of the manifold, the temperature vector $\boldsymbol{\beta} \in \mathbb{T}^\star \subset \mathbb{U}^\star$ must belong to the co-adjoint orbit of the compact subalgebra $\mathbb{H}\subset \mathbb{U}$ so that it can be further reduced to the Cartan subalgebra of the latter. So we rather need an abelian structure where the n-dimensional submanifold $\boldsymbol{\mathcal{Q}}^{n}\, = \,\boldsymbol{\mathcal{T}}^{n} $ is compact and therefore a true $n$-torus. The difficult point and the main obstacle on this road is that the  K\"ahler non-compact symmetric spaces $\mathrm{U/H}$ of real dimension $2n$ that we utilize as network layers do not possess $n$ abelian compact isometries, while they always have non compact isometries forming an $n$-dimensional abelian group. Indeed the number of compact abelian isometries of $\mathrm{U/H}$ is only $\mathit{k} < n $  where
\begin{equation}\label{integromezzo}
  \mathit{k} \, \equiv \, \text{Integer Part of $\frac{n}{2}$} 
\end{equation}
This does not mean that a compact abelian structure $\mathfrak{p}_a(z,\bar{z}), \mathfrak{q}^a(z,\bar{z})$ cannot be constructed and that the compact $n$-torus $\boldsymbol{\mathcal{T}}^{n}$ cannot be found, yet it has the very important consequence that only half of the angles $\mathfrak{q}^a(z,\bar{z})$ in $\boldsymbol{\mathcal{T}}^{n}$ are orbits of 
compact isometries and, symmetrically, that only half of the conjugate actions $\mathfrak{p}_a(z,\bar{z})$ are moment-maps of Killing vector fields, which can be 
associated to the compact Cartan generators. 
\par
We focus on the Calabi-Vesentini manifolds:
\begin{equation}\label{Cvmanigoldi}
  \mathcal{M}_{CV}^{[2,q]}\,  \equiv \, \frac{\mathrm{SO(2,2+q)}}{\mathrm{SO(2)\times SO(2+q)}} 
\end{equation}
each of which, at all time,  must be regarded as one of the two cofactors constituting an item in following infinite series of Special K\"ahler manifolds:
\begin{equation}\label{speckal}
 \mathcal{SK}_{3+q} \, \equiv \, \frac{\mathrm{SL(2,\mathbb{R})}}{\mathrm{SO(2)}}\times
 \frac{\mathrm{SO(2,2+q)}}{\mathrm{SO(2) \times SO(2+q)}}
\end{equation} 
The second part of the present article deals with the task of constructing the compact abelian structure mentioned above, deriving the missing $n-\mathit{k}$ actions $\mathfrak{p}_{\mathit{k}+i}$ and their conjugate angles $\mathfrak{q}^{\mathit{k}+i}$, inquiring on their nature and properties. Because of the general properties of  symplectic/Poissonian geometry established by the K\"ahler $2$-form, the missing $n-\mathit{k}$ actions are the moment-maps of as  many hamiltonian vector fields that are not Killing vectors of the K\"ahler metric, yet they generate transformations that respect a smaller subalgebra of the full isometry algebra $\mathbb{U}$. As the reader will discover  reading section \ref{sashasecta},
there is a precise algorithm that allows the identification and the construction of the extra actions that turn out to be associated with a principal  sequence of subalgebras of the compact subalgebra $\mathbb{H}\subset \mathbb{U}$ and formed as the square roots of invariant quadratic polynomials in the moment maps of their generators. Their dual angles arise with some complicated labor from the same subalgebra sequence. 
\par
The important point is that in the Darboux coordinate basis, 
following from the construction of the abelian structure, the integral defining the partition function becomes explicitly computable in simple terms. 
\par
This important result is not limited to the explicit calculation of the partition function with Souriau temperatures in the co-adjoint representation of $\mathbb{U}$: the very architecture of the abelian structure suggests that one can include new generalized temperatures  associated with each one of the extra actions. This breaks the full symmetry under $\mathbb{U}$ leading to a sort of phase transition similar to that occurring in ferromagnetism. The derivative of the stochastic Hamiltonian with respect to each of the extra temperatures yields a non-vanishing expectation value to the corresponding action which is a new observable function on the manifold of events. Such non-vanishing expectation value breaks the $\mathrm{U}$-symmetry to a smaller subgroup and it is the statistical analogue of spontaneous magnetization in ferromagnets.
\section{From Information Entropy to Geometrical Thermodynamics}
\label{infogeomacro}
As we emphasized in our recent paper \cite{geotermico}, the notion of \textbf{Information Geometry}, quite popular and  frequently used in the Machine Learning
literature and, there, typically credited to Fisher, Rao, Chentsov and Amari\cite{raone,cenzone,amarone}, should be placed in the context of a much broader and deeper historical conceptual development, that starts in the 1940s with Shannon \cite{shannone} and von Neumann\footnote{In his own memories Shannon states that it was John von Neumann the one who encouraged him to name his functional \textit{entropy} since - he said - it exactly coincides with the physical concept of entropy. In his fundamental papers of 1957, Jaynes explicitly refers to both Shannon and von Neumann for the conception of entropy from which he elaborates his own results.}, continues quite noticeably in the 1950s with Jaynes \cite{gianno1,gianno2} and  arrives to the 1990s and early 2000s with Lychagin and Roop \cite{lychaginlecture,ludaed,ludaed2,Lychagin_2020,Kushner_2020}, also sided by the independent quite inspiring contributions of Ruppeiner and collaborators \cite{Ruppeiner_2010,Ruppeiner_2012,Ruppeiner_2012b,Ruppeiner_2013,ruppoRdiag,Ruppeiner_2020}. Such historical recollection is not only motivated by a most rightful recognition of intellectual priorities, rather it is also very important in order to put the geometrical structures associated with 
both thermodynamics and its source, namely statistical mechanics, into an appropriate logical and geometrical order.   
\par
Indeed there are two geometries involved in this problem:
\begin{description}
  \item[A)] A \textbf{macroscopic geometry} whose carrier space $\mho$ is spanned by the generalized \textit{thermodynamic variables}, that are \textit{average values} of microscopic functions plus their duals, generically named \textit{temperatures}. So named \textit{information geometry} is a derivation of this macroscopic geometry, as we presently will show.
  \item[B)] A \textbf{microscopic geometry} whose carrier space $\Omega$ is instead spanned, in physical statistical mechanics by the \textit{variables/coordinates} of the \textit{underlying mechanical/dynamical system},  or, \textit{in more general statistical models}, by the \textbf{manifold} to which  one maps the \textbf{data},  whatever they are.  
\end{description}
\par
In the macroscopic case A) the \textbf{thermodynamic manifold} $\mho$  is always an odd-dimensional \textbf{contact manifold} and its \textit{iso-entropic leaves}, transverse to the Reeb field, are always symplectic manifolds endowed, as we are going to see, with a K\"ahler metric. 
\par
In the microscopic case B) there are many different options for $\Omega$, yet the specific form of the \textit{macroscopic K\"ahler metric}  is determined by the \textit{choices of the microscopic geometry}, in this way generalizing to  diverse situations the  corner stone of statistical physics, namely the principle that microscopic dynamics is what determines the macroscopic thermodynamic behavior of matter. Hence we must always write $\mho =\mho\left[\Omega,\mathbf{X}\right]$ to
put into evidence that the geometric structure of the macroscopic contact manifold
is determined by the geometry (encrypted in the chosen observables $\mathbf{X}$) of the microscopic manifold $\Omega$.
\par Before proceeding further in the present discussion we remind the reader that all the necessary mathematical concepts and definitions (contact and symplectic geometry, Reeb field, etc) have been summarized in detailed appendices of our previous paper \cite{geotermico} and we do not repeat them here. Furthermore we anticipate that, what we preliminary discussed in \cite{geotermico} as Souriau thermodynamics, corresponds to the case where also the B) microscopic manifold is K\"ahlerian, just as, by the way, happens with the phase--space of any dynamical system, yet it is  also a symmetric space $\mathrm{U/H}$ which brings \textbf{group theory} into the thermodynamic setup.  
\subsection{Jaynes' fundamental idea}
In his two crystal-clear, mathematically rigorous and philosophically oriented papers
\cite{gianno1,gianno2}, Jaynes took upon himself the task of justifying the building blocks of statistical mechanics for equilibrium states from a general principle: that
of the maximization of Shannon's entropy  at fixed mean values of a collection of observable functions defined over the underlying microscopic manifold. 
Jaynes aimed at the mathematical/philosophical foundations of physical statistical mechanics, yet he was perfectly aware of the possible generalization of his construction to much more general systems of data and mentioned such a perspective more than once in his two articles. Rephrased in the modern geometrical set up of probability theory, Jaynes' setup goes as follows (see appendix B of \cite{geotermico} for a summary of the fundamental principles and concepts of probability theory as exposed in standard textbooks like \cite{sinaikarolo}).
\par
Let $\rho$ be a probability density defined on some measurable space $\Omega$, named the space of events. More specifically let $\mathbf{q}\in \Omega$  be an event, namely a point in the stochastic space we consider, let $\mathrm{d}\mu(\mathbf{q})$ be the  integration measure on $\Omega$ and let $\rho(\mathbf{q}) \in [0,1]$  be the value  in $\mathbf{q}$ of the probability density  that is obviously normalized
as follows:
\begin{equation}\label{normaliz}
 \mathrm{N}[\rho] \,\equiv \,   \int_\Omega \, \rho(\mathbf{q}) \,d\mu(\mathbf{q}) \, = \, 1
\end{equation}
The measure of indeterminacy (or missing information) contained in the probability distribution $\rho$ was  defined by Shannon by means of the following {\bf Entropy Functional}
\begin{equation}\label{shaninfo}
    \mathcal{I}\left[\rho\right] \, \equiv \, - \, \int_{\Omega}\,
    \rho(\mathbf{q}) \, \log \, \left[ \rho(\mathbf{q})\right] \,d\mu(\mathbf{q})
\end{equation}
\subsubsection{Conditional minimalization of information and the partition function}\label{parcondicio}
Jaynes formulated the following problem: determine the probability distribution that extremizes the functional
$\mathcal{I}\left[\rho\right]$ under the following two conditions:
\begin{description}
  \item[A)] The correct normalization (\ref{normaliz}) should hold true.
  \item[B)] The average value of a certain  vector $\mathbf{X}(q)$
  of $n$ measurable functions
  \begin{equation}\label{stocavetto}
   \mathbf{X} \, = \, \left\{ X_1(q), X_2(q),\dots, X_i(q), \dots ,X_n(q)\right\}
  \end{equation}
  defined over the space of events $\Omega$,
  should be fixed to a certain precise  $n$-dimensional vector $\mathbf{x}\in
  \mathbb{R}^n$:
  \begin{equation}\label{muccacarolina}
    \langle \mathbf{X}\rangle \, \equiv \, \int_{\Omega}
    \,\mathbf{X}(\mathbf{q})\, \rho({\mathbf{q}})\,d\mu(\mathbf{q}) \, =
    \,\mathbf{x}\in\mathbb{R}^n
  \end{equation}
\end{description}
The classical way to solve this problem is to use variational calculus in the presence of Lagrange multipliers. One
introduces $n+1$ multipliers: $\lambda_0$ associated with the
normalization constraint (\ref{normaliz}) and $n$
multipliers $\lambda_i$,  which we can regard as the components
of a vector  $\pmb{\lambda}\in \mathbb{R}^n$, that are associated with
the constraints (\ref{muccacarolina}). Thus the new functional to be extremized is 
\begin{equation}\label{functlag}
    \mathcal{F}[\rho]\, = \, -\,\mathcal{I}\left[\rho\right] \, - \,
    \lambda_0 \, \left(\mathrm{N}[\rho] \, -\, 1 \right) +
    \pmb{\lambda}\cdot\left(\langle \mathbf{X}\rangle -\mathbf{x}
    \right)
\end{equation}
The variation of the functional   in $\delta\rho$  yields:
\begin{equation}\label{ralloppo}
    \frac{\delta\mathcal{F}[\rho]}{\delta\rho} \, = \,\log[\rho] +1
    -\lambda_0 +\pmb{\lambda}\cdot\mathbf{X} \, = \,0
\end{equation}
which implies:
\begin{equation}\label{calindro}
    \rho(\mathbf{q}) \, = \,\exp\left[\lambda_0 - 1-
    \pmb{\lambda}\cdot \mathbf{X}(\mathbf{q})\right]
\end{equation}
Imposing the normalization constraint(\ref{normaliz}) fixes the value of
$\lambda_0$  so that the  final expression of the extremal probability distribution is the following :
\begin{equation}\label{distribuzionepesci}
    \rho_{ex}(\mathbf{q}) \, = \,
    \frac{\exp\left[-\pmb{\lambda}\cdot\mathbf{X}\left(\mathbf{q}
    \right)\right]}{Z\left(\pmb{\lambda}\right)}
\end{equation}
where:
\begin{equation}\label{partifungo}
    Z\left(\pmb{\lambda}\right)\, \equiv \, \int_{\Omega}
    \,\exp\left[-\pmb{\lambda}\cdot\mathbf{X}\left(\mathbf{q}\right)
    \right]
    \,d\mu(\mathbf{q})
\end{equation}
is the \textbf{Partition Function} and, for reasons that will become immediately clear,
the following object:
\begin{equation}\label{tremoamillo}
    \mathcal{H}^{stoch}\left(\pmb{\lambda}\right) \, = \, - \log
    \left[Z\left(\pmb{\lambda}\right)\right]
\end{equation}
is named the \textbf{stochastic Hamiltonian}.
As a consequence of the
definition (\ref{tremoamillo}) the vector $\mathbf{x}$ of mean values  imposed on
the vectors of microscopic functions $\mathbf{X}(q)$  is obtained from the  hamiltonian by means of a derivative:
\begin{equation}\label{variabiliestensive}
    \mathbf{x}\,= \, \mathrm{d}_{\pmb{\lambda}}\mathcal{H}^{stoch}\left(\pmb{\lambda}
    \right) \,
    \Rightarrow\,
    \text{short-hand for }\null \, x_i \, = \, \frac{\partial}{\partial
    \lambda^i} \,\mathcal{H}^{stoch}\left(\lambda_1,\,\dots \, , \, \lambda_n\right)
\end{equation}
Calculating Shannon Entropy Functional (\ref{shaninfo}) on the extremal
probability distribution
(\ref{distribuzionepesci}) with elementary algebra we obtain:
\begin{equation}
   - \mathcal{I}\left[\rho_{ex}\right]\, = \,
    \mathcal{H}^{stoch}\left(\pmb{\lambda}\right) - \pmb{\lambda}\cdot
    \mathbf{x} \, = \, \mathcal{H}^{stoch}\left(\pmb{\lambda}\right)\, -\,
    \lambda^i \, \frac{\partial}{\partial
    \lambda^i} \,\mathcal{H}^{stoch}\left(\pmb{\lambda}\right)
\label{leggendoleggo}
\end{equation}
that  has the  form of a  Legendre transform. Hence the Shannon functional plays the same role as that of a Lagrangian, and the the stochastic Hamiltonian is indeed a Hamiltonian, since it is the Legendre transform of the Lagrangian. The intensive variables of Thermodynamics (\textit{i.e.} the Lagrange multipliers $\pmb{\lambda}$ ) are the momenta and the average values $x^i$ are the coordinates.  
\subsection{Lychagin's and Roop's fundamental idea: Contact Geometry encodes the Principles of Classical Thermodynamics} 
The brilliant and simple idea presented in various papers and author combinations \cite{Kushner_2020,Lychagin_2020,ludaed,ludaed2}  by Lychagin and Roop and systematically reviewed by Lychagin in his lectures \cite{lychaginlecture} is the following. We are all familiar, since high school and early university education, with the first and second principles of classical thermodynamics that are always expressed in differential form:
\begin{alignat}{5}
\label{primosecondoprincipio}
&1)& \quad\mathrm{d}U + P \,\mathrm{d}V &\,  = \,& \mathrm{d}Q & \quad \text{First Principle} \nonumber\\
&2)&  \quad \mathrm{d}S &\,= \, &\frac{1}{T} \,\mathrm{d} Q &\quad \text{Second Principle}
\end{alignat}
where $U$ is the internal energy, $P$ the pressure, $V$ the volume, $T$ the temperature, $S$ the entropy and $Q$ the heat. As we learnt at school, the $4$ differentials $\mathrm{d}U$, $\mathrm{d}S$, $\mathrm{d}T$   are exact differentials, namely the differentials of $4$ variables, or, in thermodynamic parlance, \textit{functions of state}, characterizing an equilibrium state. The same is true of the pressure $P$, which together with the other four, makes a quintuplet of coordinates of the thermodynamical space. Instead, as it is well known, infinitesimal heat $dQ$ is not an exact differential, namely there exists \textit{no heat function of state}. Note also that the internal energy $U$ and the volume $V$ are extensive variables, while the temperature $T$ and the pressure $P$ are intensive ones. The brilliant idea by Lychagin and Roop, which geometrizes classical thermodynamics and makes it coordinate independent, is expressed as follows. On the five dimensional space spanned by the coordinates $S,T,U,P,V$ (keep in mind they being an odd number) let us introduce the following differential one form:
\begin{eqnarray}
\label{contactfurma1}
 \boldsymbol{\alpha} \, = - \mathrm{d}S\, + \,   \,T^{-1}\, \mathrm{d}U\,
 + \, T^{-1}\,P\, \mathrm{d}V
\end{eqnarray}
The first and second principle of thermodynamics - this was the crucial observation by Lychagin and Roop - can be summarized by saying that in the $5$-dimensional thermodynamic space, the equilibrium states can be identified with the loci where the $1$-form $\alpha$ vanishes.
\subsubsection{Formalizing the idea in terms of contact geometry}
Once the above observation is made, the road to a full-fledged geometrization of thermodynamics is opened up. The key notion is that of \textbf{contact manifold} and \textbf{contact geometry} (see Appendix A of \cite{geotermico} for a short but essential summary). In simple words a contact manifold $\mathcal{M}_{2n+1}$ is a $(2n+1)$-dimensional manifold ($n\in \mathbb{N})$,  endowed with a $1$-form $\alpha$ (\textit{the contact form}) that is of maximal rank, namely such that:
\begin{equation}\label{cortallo}
  \alpha \wedge \underbrace{\mathrm{d}\alpha \wedge \mathrm{d}\alpha\wedge\dots \wedge \mathrm{d}\alpha}_{\text{$n$-times}} \, \neq \, 0
\end{equation}
The contact form introduces what is named a \textit{contact structure} $\xi_{\alpha}$, namely a sub-bundle of the tangent bundle $\mathcal{TM}_{2n+1} \longrightarrow \mathcal{M}_{2n+1}$ spanned by all those vector fields $\boldsymbol{\frak{t}}$ that are in the kernel of $\alpha$:
\begin{equation}\label{lofoten}
  \text{ker}(\alpha) \, = \, \left\{ \boldsymbol{\frak{t}}\in \Gamma\left[\mathcal{TM}_{2n+1},\mathcal{M}_{2n+1}\right]
   \, \mid \, \alpha\left(\boldsymbol{\frak{t}}\right) \, = \, 0  \right\}
\end{equation}
A classical theorem by Darboux states that in every open neighborhood of a contact manifold one can find a coordinate patch $x_0,x_i,\lambda^i$ ($i=1,\dots,n$) in which the contact $1$-form $\alpha$ takes the following local form:
\begin{eqnarray}
  \alpha &=& \mathrm{d}x_0 \, + \, \sum_{i=1}^n \, \lambda^i \, \mathrm{d}x_i \label{darbouxform1}  \\
  \null& \Downarrow & \null \nonumber\\
  \mathrm{d}\alpha & = &  \sum_{i=1}^n \,  \mathrm{d}\lambda^i \wedge \mathrm{d}x_i\label{darbouxform2}
\end{eqnarray}
The $1$-form in eq.(\ref{contactfurma1}) is precisely a contact $1$-form, satisfying the defining constraint (\ref{cortallo}) and it can be easily put into Darboux form (\ref{darbouxform1}) upon the simple coordinate redefinition:
\begin{equation}\label{senigallia}
  x_0 \, = \, - S \quad ; \quad x_1 \, = \, U \quad ; \quad  x_2 \, = \, V \quad ; \quad 
  \lambda_1 \, = \, \frac{1}{T} \quad ; \quad \lambda_2 \, = \, \frac{P}{T} 
\end{equation}
Note that the subdivision of the thermodynamic coordinates into the two Darboux subsets $x_i$ and $\lambda_i$ has a clearcut statistical mechanical interpretation.
The $x$.s are the extensive variables corresponding to the  \textit{macroscopic mean values} of \textit{microscopic functions}, respectively, $x_1 = U$ the total energy of the mechanical system composed by an Avogadro number of molecules, $x_2 \, = \,V$ the mean value of the volume of the portion of $\mathbb{R}^3$ space occupied by the
same number of molecules.  The $\lambda_i$ are instead the intensive variables that characterize an equilibrium state of the system and have the same value in each of its parts, the inverse temperature $\lambda_1 \, = \, \frac{1}{T}$ and the pressure per temperature $\lambda_2 \, = \, \frac{P}{T}$, respectively.
\par 
It is also clear that once such identifications are made, the contact manifold framework can be applied to any statistical system described by probability distributions $\rho_{ex}(\mathbf{q})$ as those is eq.s (\ref{distribuzionepesci}-\ref{partifungo}) that maximize the Shannon entropy 
(\ref{shaninfo}) on whatever \textit{microscopic manifold of events $\Omega$} and that, from now on, we name \textbf{Gibbs distributions on $\Omega$}. Therefore the \textit{macroscopic thermodynamic manifold $\mho\left[\Omega,\mathbf{X}\right]$}, singled out by the choice of a microscopic manifold $\Omega=\mathcal{M}^{micro}_{m}$ of dimension $m$ and by a choice of $n$ observable functions $\mathbf{X}(q)$ defined over $\mathcal{M}^{micro}_{m}$ is always a $(2n+1)$-dimensional contact manifold $\mho \, = \,\mathcal{M}^{macro}_{2n+1}$  with contact form:
\begin{equation}\label{alphamacro}
  \underbrace{\alpha}_{\text{defined on $\mho\left[\Omega,\mathbf{X}\right]$}} \, = \, \mathrm{d}\mathcal{I} \, + \, \sum_{i=1}^n \lambda_i \, \mathrm{d}x^i
\end{equation}
\subsubsection{Equilibrium states as Legendrian submanifolds} 
The definition (A.7) of Legendrian submanifolds given in appendix A of 
\cite{geotermico} states that they are
isotropic submanifolds of maximal dimension $n$ of a contact manifold $\mathcal{M}_{2n+1}$.  On the other hand we recall that
an isotropic submanifold is a submanifold such that its tangent bundle is in the kernel of the contact form, namely the $1$-form $\alpha$ vanishes on each Legendrian submanifold.  The great intuition of the authors
of \cite{ludaed,ludaed2} has been that of identifying, regardless of the utilized coordinates and hence in an intrinsic way,
 the \textbf{thermodynamic equilibrium states} with the points
of \textbf{Legendrian submanifolds} of the ambient space.
 In terms of the theory  of conditional minimization
discussed in section \ref{parcondicio} it is very simple
to define  the Legendrian submanifolds that represents the termodynamic equilibrium states.
\begin{definizione}
\label{defingolegend} Any admissible thermodynamic state can be  identified with a point in the following  $n$-dimensional submanifold  of the contact manifold:
\begin{equation}
\label{crocolo}
    \mathcal{L}_n \, = \, \left\{ \mathcal{I} =
    \mathcal{I}\left(\pmb{\lambda},\mathbf{x}\right)\, , \, x_i \, = \,
    \frac{\partial}{\partial \lambda^i}
    \mathcal{H}\left(\pmb{\lambda}\right)\right\} \, \subset \,\mathcal{M}_{2n+1}
\end{equation}
\end{definizione}
\par
\begin{teorema}
\label{gargantua} The submanifold $\mathcal{L}_n$ defined by means of equation
(\ref{crocolo}) is isotropic and hence legendrian.
\end{teorema}
\begin{dimostrazione}
The proof is extremely simple. It suffices to recall equation
(\ref{leggendoleggo}). Using that relation we can evaluate the
total differential $\mathrm{d}\mathcal{I}$, as it follows:
\begin{equation}
\mathrm{d}\mathcal{I} \, = \,
\mathrm{d}\mathcal{I}\left(\pmb{\lambda},\mathbf{x}\right)\, =
\,\mathrm{d}\left(\mathcal{H}(\pmb{\lambda})-\pmb{\lambda}\cdot\mathbf{x}\right)\,
=\, \left(\frac{\partial}{\partial
\lambda^i}\mathcal{H}(\pmb{\lambda})-x^i\right)\mathrm{d}\lambda^i \, - \, \lambda^i
\,dx_i \, = \, - \, \lambda^i \,dx_i
\end{equation}
Hence on the submanifold (\ref{crocolo}) we have
$d\mathcal{I}+\lambda^i \,dx_i=0$
\end{dimostrazione}
\subsubsection{From Legendrian to Lagrangian submanifolds transverse to the Reeb field}
\label{logranofino}
Given the original contact variety $\mathcal{M}^{2n+1}$
with the contact $1$-form given in eq.(\ref{alphamacro})
we see at once that the Reeb vector field (see appendix A of \cite{geotermico} for its definition) is
\begin{equation}\label{Reebtermo}
    \mathbf{R}\, = \,\frac{\partial}{\partial \mathcal{I}}
\end{equation}
namely the entropy gradient. In fact, it satisfies the two conditions:
\begin{equation}\label{cominternus}
    \alpha\left(\mathbf{R}\right) \, = \, 1 \quad ; \quad
    d\alpha\left(\mathbf{R}, \mathbf{X}\right)\, = \, 0 \quad
    \forall \mathbf{X} \in \Gamma[\mathcal{TM}^{2n+1},\mathcal{M}^{2n+1}]
\end{equation}
On the other hand, from the general discussion in appendix A of \cite{geotermico}
we know that every $2n$-dimensional submanifold
of a contact manifold $\mathcal{M}^{2n+1}$  that is
transverse to the Reeb vector field  is a symplectic
manifold $\mathcal{S}^{2n}$ whose symplectic  $2$-form is the restriction
to $\mathcal{S}^{2n}$ of the exterior differential of the contact 1-form:
\begin{equation}\label{rovereto}
    \omega \, = \,
   \mathrm{d}\alpha \mid_{\mathcal{S}^{2n}}
\end{equation}
Hence applying these general notions to the case at hand, we see that the  symplectic variety transverse to Reeb vector
(\ref{Reebtermo}) is given by the following projection map:
\begin{equation}\label{trasversus}
    \pi_{TR} \, : \, \mho \, = \,\mathcal{M}^{2n+1} \, \rightarrow \,
    \mathcal{S}^{2n} \subset \mho \quad ; \quad
    \pi_{TR}(\mathcal{I},\pmb{\lambda},\mathbf{x})\, = \, (\pmb{\lambda},\mathbf{x})
\end{equation}
and the  symplectic $2$-form  is as follows
\begin{equation}\label{adige}
    \omega \, = \, \sum_{i=1}^{n} \mathrm{d}\lambda^i \wedge dx_i \, \quad
    ;\quad \mathrm{d}\alpha \, = \, \pi^\star_{TR}(\omega)
\end{equation}
Let us now consider the Legendrian submanifold $\mathcal{L}_n
\subset \mathcal{M}^{2n+1}$ which contains the thermodynamic equilibrium states introduced in Definition \ref{defingolegend}.
It is obvious that we can consider its image through the projection map (\ref{trasversus}):
\begin{equation}\label{corneliatutta}
    \mathcal{S}^{2n} \supset \mathfrak{L}_n \, \equiv \, \pi_{TR}\left(\mathcal{L}_n\right)
\end{equation}
The important result is that the submanifold $\mathfrak{L}_n$
thus defined is a Lagrangian submanifold, namely one on which the symplectic $2$-form completely vanishes. The demonstration of this fact follows immediately from the definition. In fact, we can translate equation
(\ref{corneliatutta}) into the following constructive definition:
\begin{equation}\label{tarapizo}
    \mathfrak{L}_n \, = \,\left\{ x^i \, = \, \frac{\partial}{\partial \lambda^i}
    \mathcal{H}\left(\pmb{\lambda}\right)\right\}
\end{equation}
Using the latter we find:
\begin{equation}\label{isonzo}
    \left.\omega\right|_{\mathfrak{L}_n} \, = \,
    \,\sum_{i,j=1}^{n} d\lambda^i \wedge d\lambda^j \, \partial_i\partial_j
    \mathcal{H}\left(\pmb{\lambda}\right) \, = \,0
\end{equation}
which follows because of the commutativity of the partial derivatives. 
\subsection{The metric on the contact manifold from its reduction to the symplectic hypersurfaces transverse to the Reeb field}
\label{metraturacontatta}
So far, on the macroscopic contact manifold $\mho[\Omega,\mathbf{X}]$, we have introduced just the contact $1$-form $\alpha$ but not yet any Riemannian metric. This is an issue that neither Lychagin and Roop touched upon in their papers nor we did consider in our previous paper \cite{geotermico}. We come to it here for the first time and, as the reader will see, this is the fundamental issue in relation with the so named Information Geometry. Indeed the natural  problem which arises when considering the above geometric reformulation of thermodynamics is the following:
\begin{description}
  \item[a)] Since the submanifolds $\mathcal{S}^{2n}$ that are transverse to the Reeb field are symplectic with a symplectic $2$-form already in Darboux coordinates provided by eq.(\ref{adige}), (compare with eq.(\ref{darbouxform2})),
   what is the most general  metric on $\mho[\Omega,\mathbf{X}]$ such that
  its restriction to $\mathcal{S}^{2n}$ admits the 2-form of eq.(\ref{adige}) as the K\"ahler $2$-form of a K\"ahler metric on the same submanifold? 
\end{description}
Due to the very simple structure of the projection map $\pi_{TR}$, the general answer to  question a) is as follows:
\begin{equation}\label{contattametrica}
  ds^2_{\mho}\, = \, \mathrm{d}{\mathcal{I}}^2 \, + \, ds^2_{\mathcal{S}^{2n}}
\end{equation}
where $ds^2_{\mathcal{S}^{2n}}$ is a K\"ahler metric on $\mathcal{S}^{2n}$ and its K\"ahler form $\boldsymbol{\mathcal{K}} $  is equal to $\omega$ as given in eq.(\ref{adige}).
\par
That above is a classical problem in K\"ahler geometry and it goes under the name of 
\textbf{action/angle} formalism when it is applied to the phase-space of a dynamical system, or under the name of \textbf{AMSY symplectic formalism} \cite{abreu,abreu2009toric,Martelli:2005tp,Bykov:2017mgc} when it is applied to  K\"ahler geometry on toric varieties. Its AMSY incarnation is reviewed in the lectures \cite{fre2023lectures} by one of us and it is extensively applied to orbifold resolutions in the context of $D3$-brane effective supergravity theories in  \cite{Bianchi_2021,bruzzo2023d3brane}. For the action/angle incarnation the reader  is instead referred both to the discussion already anticipated in the introduction and to section \ref{sashasecta} of the present paper where the action/angle coordinate basis is extensively used  in a strategic way, but not on the macroscopic manifold $\mho[\Omega,\mathbf{X}]$, rather on the microscopic one $\Omega$, while solving the problem of computing the partition function as defined in eq.(\ref{partifungo}). 
\par
In all cases the symplectic formalism for K\"ahler metrics comes into play when the K\"ahler metric has a non-trivial dependence only on one half of the $2n$ coordinates, because there are $n$ abelian isometries that can be compact $\mathrm{U(1)}$.s as it is the case of toric varieties where the angles $\Theta^i$ are the phases of the $n$ complex coordinates or, as in  the case at hand, where the $n$ abelian isometries form some other non-compact abelian group.  We see later on that, for Calabi-Vesentini manifolds, as for all the homogeneous Special K\"ahler manifolds introduced and classified in Supergravity literature (see \cite{pgtstheory} for a review finalized to Machine Learning applications and specially compiled for Data Scientists and the already quoted original literature 
\cite{deWit:1995tf,toineugenio,SKGaggio3,SKGaggio2,SKGaggio1,specHomgeoA2,specHomgeoA1} ), the non-compact invariance group is made of translations of the real parts of the complex coordinates that physically are connected with the axion fields.  
\par 
In all cases the K\"ahler potential $\mathcal{K}$ is supposed to depend only on one half of the coordinates. 
In our case let us say that it only depends on the mean imaginary part $v_i$ of complex variables whose real parts are the mean values $x_i$, and introducing the momenta variables as:
\begin{equation}\label{coreuta}
  \lambda^i \, \equiv \, \frac{\partial \mathcal{K}(\mathbf{v})}{\partial v_i}
\end{equation}
one defines the symplectic real potential by means of the Legendre transform:
\begin{equation}\label{legendroG}
  G(\boldsymbol{\lambda}) \, = \sum_{i=1}^n\, v_i \, \lambda^i \, - \, \mathcal{K}(\mathbf{v})
\end{equation}
which makes sense as long as one is able to invert the relation (\ref{coreuta}) and write the $v_i$ as  functions of the $\lambda^j$, which is typically a hard analytic task, when one starts from the K\"ahler potential. Viceversa if, for some reason, one has the symplectic potential $G(\boldsymbol{\lambda})$, one can retrieve the K\"ahler potential by inverse Legendre transform:
\begin{equation}\label{legendroK}
  \mathcal{K}(\mathbf{v}) \, = \, \sum_{i=1}^n\, v_i \, \lambda^i \, - \, G(\boldsymbol{\lambda})
\end{equation}
where now we have:
\begin{equation}\label{cetra}
  v_i \, \equiv \, \frac{\partial {G}(\boldsymbol{\lambda})}{\partial \lambda^i}
\end{equation}
The very important thing is that utilizing the real symplectic potential $G(\boldsymbol{\lambda})$ we can always write the full K\"ahler metric in terms of the Hessian:
\begin{equation}\label{tantopercantar}
  \mathrm{H}_{ij}\left(\boldsymbol{\lambda}\right) \, = \, \frac{\partial^2 G(\boldsymbol{\lambda})}{\partial \lambda^i \partial\lambda^j}
\end{equation}
We have:
\begin{eqnarray}
\label{isoentropic}
ds^2_{\mathcal{S}^{2n}} \, = \, \ft 12 \left(\sum_{i,j} \, \mathrm{H}_{ij}\left(\boldsymbol{\lambda}\right) \,\mathrm{d}\lambda^i\,\mathrm{d}\lambda^j\, + \,
\sum_{i,j} \, \mathrm{H}^{-1|ij}\left(\boldsymbol{\lambda}\right) \,\mathrm{d}x_i\,\mathrm{d}x_j\,\right)
\end{eqnarray}
In matrix notation putting all the $2n$ coordinates in a single array $(\lambda^i,x_j)$ enumerated by indices $A,B,\dots$ we have that the metric and the complex structure tensor are respectively given by: 
\begin{eqnarray}
\label{metricandCC}
g_{AB} & = & \ft 12 \, \left( \begin{array}{c|c}
\mathrm{H}_{ij} & 0 \\
\hline
0 & \mathrm{H}^{-1}_{ij}\\
\end{array}\right)\quad ; \quad J^A_{\phantom{A}B} \, = \, \left( \begin{array}{c|c}
0&\mathrm{H}^{-1|ij}  \\
\hline
- \,\mathrm{H}_{ij} & 0\\
\end{array}\right)
\end{eqnarray}
satisfying the obligatory relations:
\begin{equation}\label{tafferuglio}
  J^2 \, = \, - \, \mathrm{Id} \quad ; \quad J^T\cdot g \cdot J \, = \, g
\end{equation}
the first telling us that $J$ is indeed a complex structure, the second telling us that the metric is hermitian with respect to it. The components of the K\"ahler 2-form are given, according with their definition by\footnote{In the following syntetic formula we utilize the convention that the capital indices $A,B$ run on the values $\null^i$ and $\null_i$ for the coordinates $x_i$ and the momenta $\lambda^i$, respectively.}
\begin{eqnarray}
\label{calloforma}
K_{AB} & = &  \left(g \cdot J\right)_{AB} \, = \, \ft 12 \, \left( \begin{array}{c|c}
0&\mathbf{1}_{ij}  \\
\hline
- \,\mathbf{1}_{ij} & 0\\
\end{array}\right)
\end{eqnarray}
which implies that the K\"ahler $2$-form is indeed the symplectic form $\omega$ of eq. (\ref{adige}):
\begin{equation}\label{callo2forma}
  \boldsymbol{\mathcal{K}} \, = \,   \sum_{i} \mathrm{d}\lambda^i \wedge \mathrm{d}x_i  
\end{equation}
As we stressed, the above result of eq.s (\ref{isoentropic}-\ref{metricandCC}) is completely general for all K\"ahler manifolds for which the K\"ahler 2-form, by using an appropriate coordinate system $(\lambda^i,x_i)$, can be put into the Darboux form (\ref{callo2forma}). A particular case is the one where the K\"ahler potential for a complex manifold spanned by $n$ complex coordinates:
\begin{equation}\label{complessini}
  z^i \, = \, u^i \, + \, \mathit{i} \, v^i
\end{equation}
depends only on the imaginary parts $v^i$ and not on the real ones:
\begin{equation}\label{carnevalino}
  \mathcal{K}(\mathbf{z},\bar{\mathbf{z}}) \, = \,  \mathcal{F}(\mathbf{v})
\end{equation}
In this case, from the very definition of the K\"ahler potential, one obtains:
\begin{equation}\label{kallometrica}
  ds^2 \, =  \, \sum_{i,j} \mathrm{d}{z}^i \times \mathrm{d}\bar{z}^j\, \frac{\partial^2\mathcal{K}}{\partial z^i \partial \bar{z}^j} \, = \,\ft 12 \, \sum_{i,j}\left( \mathrm{d}u^i \mathrm{d}u^j \, \mathcal{F}_{ij}(\mathbf{v}) \, + \, 
  \mathrm{d}v^i \mathrm{d}v^j \, \mathcal{F}_{ij}(\mathbf{v})\right)
\end{equation}
where:
\begin{equation}\label{hesso1}
 \mathcal{F}_{ij}(\mathbf{v})\, = \, \frac{\partial^2\mathcal{K}}{\partial v^i \partial v^j} 
\end{equation}
is the hessian of $\mathcal{F}(\mathbf{v})$.
Next, if we introduce the \textit{momenta}:
\begin{equation}\label{gongolo}
  \mathfrak{m}_i(\mathbf{v})\, \equiv \, \frac{\partial\mathcal{F}(\mathbf{v}) }{\partial v^i}\quad \Rightarrow \quad \mathrm{d}v^i \, = \, \mathcal{F}^{-1|ij} \, \mathrm{d}\mathfrak{m}_j
\end{equation}
the K\"ahler metric becomes
\begin{equation}\label{tantaroba}
  ds^2 \, = \,\ft 12 \left(\sum_{i,j} \, \mathcal{F}^{-1|ij} \,\mathrm{d}\mathfrak{m}_i\,\mathrm{d}\mathfrak{m}_j\, + \,
\sum_{i,j} \, \mathcal{F}_{ij} \,\mathrm{d}u^i\,\mathrm{d}u^j\,\right)\, = \, \ft 12 \left(\sum_{i,j} \, \mathcal{G}^{ij}(\boldsymbol{\mathfrak{m}}) \,\mathrm{d}\mathfrak{m}_i\,\mathrm{d}\mathfrak{m}_j\, + \,
\sum_{i,j} \, \mathcal{G}^{-1}_{ij}(\boldsymbol{\mathfrak{m}}) \,\mathrm{d}u^i\,\mathrm{d}u^j\,\right)
\end{equation}
where:
\begin{equation}\label{hesso2}
  \mathcal{G}^{ij}(\boldsymbol{\mathfrak{m}}) \, = \,\frac{\partial^2\mathcal{G}(\boldsymbol{\mathfrak{m}})}
  {\partial \mathfrak{m}_i \partial \mathfrak{m}_j} \, = \,\mathcal{F}^{-1|ij}(\boldsymbol{{v}})  
\end{equation}
is the hessian of the symplectic potential defined as the Legendre transform of the K\"ahler potential. The new coordinates $(\mathfrak{m}_i,u^i)$ are Darboux coordinates since the K\"ahler $2$-form, in completely analogous way to eq.(\ref{callo2forma}) and in agreement with the general scheme of the action/angle setup becomes:
\begin{equation}\label{fradiavolo}
  \boldsymbol{\mathcal{K}} \, = \, \sum_{i} \mathrm{d}\mathfrak{m}_i \wedge \mathrm{d}u^i 
\end{equation}
yet the coordinates $u^i$ are not angles and they take all real
values, exactly as the mean values $x_i$. This as we anticipated in the introduction
is true for all Homogeneous Special K\"ahler manifolds, and will be illustrated in detail in section \ref{baracco} in the case of Calabi-Vesentini manifolds, that are a possible and significant instance of \textbf{microscopic manifolds of events $\Omega$}. The \textbf{isometry Lie algebra $\mathbb{U}$} of these symmetric spaces $\mathrm{U/H}$ has a maximal abelian subalgebra of translations, with dimension equal to half the dimension of the manifold. As a consequence  of that, in an appropriate coordinate basis, the  K\"ahler potential of these symmetric space depends only on the other half of the coordinates and one obtains a Darboux coordinate basis where the \textit{"would be angles"} are \textit{"non-compact variables"} (see sect.s \ref{baracco}  and \ref{mariosecta}). 
\par
It is important to stress once again that this does not prevent the existence of \textbf{other Darboux coordinate bases} where the angles are true angles although half of them do not correspond to any isometry. This topic is addressed in \ref{sashasecta}. 
\par
Coming back to the \textbf{macroscopic thermodynamic contact manifold $\mho\left[\Omega,\mathbf{X}\right]$}, which is our concern in this section, the structure of the metric for the iso-entropic surfaces $\mathcal{S}^{2n}$, transverse to the Reeb field is always the one mentioned above in eq.(\ref{isoentropic}) and this happens, whatever are the underlying microscopic manifold $\Omega$ and the vector of functions $\mathbf{X}(q)$ defined over it. Indeed
we can summarize the discussion of the present subsection saying that the appropriate
Riemannian metric on $\mho\left[\Omega,\mathbf{X}\right]$, compatible with the contact structure defined by the contact form in eq.(\ref{alphamacro}) is the following one:
\begin{equation}\label{contactmetric}
  ds^2_{\mho} \, = \, \mathrm{d}{\mathcal{I}}^2 \, + \, \ft 12 \left(\sum_{i,j} \, \mathcal{H}_{ij}\left(\boldsymbol{\lambda}\right) \,\mathrm{d}\lambda^i\,\mathrm{d}\lambda^j\, + \,
\sum_{i,j} \, \mathcal{H}^{-1|ij}\left(\boldsymbol{\lambda}\right) \,\mathrm{d}x_i\,\mathrm{d}x_j\,\right)
\end{equation}
where:
\begin{equation}\label{pordenone}
  \mathcal{H}_{ij}(\boldsymbol{\lambda}) \, = \, \frac{\partial^2 \boldsymbol{\mathcal{H}}^{stoch}(\boldsymbol{\lambda})}{\partial \lambda^i \,\partial \lambda^j}
\end{equation}
is the Hessian of the stochastic hamiltonian defined in eq.s(\ref{partifungo}-\ref{tremoamillo}) that from a geometrical point of view plays the role of symplectic potential, the K\"ahler potential being its Legendre transform. 
\subsection{Pull-back of the contact metric on the equilibrium states
and Information Geometry} 
Following the original argument by Lychagin and Roop the thermodynamical equilibrium states are the Legendrian submanifolds $\mathcal{L}_{n} \subset \mho$ of the contact manifold defined by the embedding functions (\ref{crocolo}) or better the Lagrangian submanifolds $\mathfrak{L}_n \subset \mathcal{S}^{2n}$ obtained as the $\pi_{TR}$ projection of the former on the submanifold transverse to Reeb field. Let us then name:
\begin{equation}\label{carnenonvale}
  \iota_\mathcal{L} \, : \, \mathcal{L}_{n} \, \longrightarrow \, \mho
\end{equation}
the injection map of the Legendrian submanifolds into the contact thermodynamic space. The injection map of the Lagrangian submanifolds is consequently described as
follows:
\begin{eqnarray}\label{verduravale}
  \iota_\mathfrak{L} & : & \mathfrak{L}_{n} \, \longrightarrow \, \mho \nonumber\\
  \iota_\mathfrak{L} & = & \iota_\mathcal{L}\circ\pi_{TR}^{-1}
\end{eqnarray}
We are interested in the pull-back of the thermodynamical contact metric 
(\ref{contactmetric}) on the Lagrangian equilibrium manifolds. Using the embedding functions (\ref{crocolo}) we immediately find:
\begin{equation}\label{infogeodeduc}
 ds^2_{info}\, =\,  \iota_\mathfrak{L}^\star\left(ds^2_\mho\right) \, = \, \mathcal{H}_{ij}\left(\boldsymbol{\lambda}\right) \, \mathrm{d}\lambda^i \times \mathrm{d}\lambda^j
\end{equation}
which is the sort of metric used in Machine Learning under the name of Information Geometry metric\cite{raone,cenzone,amarone}. The logic in Machine Learning literature starts from the consideration of probability distributions of the Gibbs type as defined in eq.(\ref{distribuzionepesci}) that are labeled by a set of parameters $\boldsymbol{\lambda}$.  The main concept in ML is all the time the same, namely that of  \textbf{distance}. One would like to have a metric and be able to measure the distance between two probability distributions in order to know how similar or dissimilar they are. The Information Metric (\ref{infogeodeduc}) provides such a tool. What was missing both in Machine Learning literature and also in the literature on mathematical formalizations of thermodynamics was a general conceptual framework that leads from the basic principle of the minimization of the Shannon functional at fixed mean values, as introduced in the seminal and already much advanced papers by Jaynes   \cite{gianno1,gianno2}, to a uniquely well defined metric on the Lagrangian submanifolds describing macroscopic equilibrium states, that is what Information Geometry really is concerned with, although those who use it do not know its deep meaning and origin. On the other hand it must be noted that, apart from Information Geometry, the Thermodynamic Metric and its Curvature that were ingeniously introduced by Ruppeiner and his collaborators in the study of phase mixtures and Chemical Equations of State \cite{Ruppeiner_2010,Ruppeiner_2012,Ruppeiner_2012b,Ruppeiner_2013,ruppoRdiag,
Ruppeiner_2020} is exactly the same as that in eq.(\ref{infogeodeduc}). Ruppeiner et al were studying equations of state, both those deduced from a partition function and those phaenomenologically modeled  as the Van der Waals one: equations of state is just another name for the embedding functions of the Lagrangian submanifolds corresponding to equilibrium states. By means of intuitive arguments they guessed a metric defined in the space spanned by such coordinates. The so obtained  metric is just the one we deduced from first geometrical principles in this section. The important lesson to learn from our genial Chemical Physicist colleagues is that the singularities in the curvature are just the hallmark of phase transitions.  
\section{K\"ahler non-compact symmetric spaces as microscopic manifolds $\Omega$}
\label{Ktopolino}
As we stressed in our previous paper \cite{geotermico}, what is named Souriau's group thermodynamics\cite{souriaub1,souriaub2,barbarpapad4,
marlentropia,caldobarbaresco,barbaresco2,barbaresco3,marlegibbs,barbarpapad1,barbarpapad2,barbarpapad3,nebbo}, is an instance of the previously described general construction that we can formulate as that based on the following microscopic data:
\begin{equation}\label{crodinotrotato}
  \mho\left[\Omega, \boldsymbol{X} \right] \, = \, \mho\left[\mathrm{U/H}, \boldsymbol{\mathfrak{P}}_{Killing} \right]
\end{equation}
where $\mathrm{U/H}$ is a non-compact symmetric space, $\mathrm{U}$ being a simple non-compact Lie group and $\mathrm{H}\subset \mathrm{U}$ its maximally compact subgroup, equipped with its unique $\mathrm{U}$-invariant Riemannian metric, while $\boldsymbol{\mathfrak{P}}_{Killing}$ denotes the 
$\text{dim}\mathbb{U}$-dimensional vector of moment-maps associated with the $\text{dim}\mathbb{U}$-dimensional vector of 
\textbf{Killing vector fields}. The above phrasing implies that the considered $\mathrm{U/H}$ symmetric space cannot be generic, rather it should be such that the Killing vector fields are hamiltonian with respect to a symplectic structure compatible with the given metric. 
\par
Summarizing the results of our previous paper \cite{geotermico} we can state that there is a unique way of meeting such requirements: \textit{the symmetric space $\mathrm{U/H}$ equipped with its $\mathrm{U}$-invariant metric must be a K\"ahler manifold} and the symplectic $2$-form compatible with the metric is the K\"ahler $2$-form:
\begin{equation}\label{carlone}
  \omega \, = \, \boldsymbol{\mathcal{K}}_{\mathrm{U/H}}
\end{equation}
The necessary and sufficient condition for a symmetric space $\mathrm{U/H}$ of the considered type to be a K\"ahler manifold is formulated in terms of the maximal compact subalgebra $\mathbb{H}\subset \mathbb{U}$ as follows:
\begin{equation}\label{kronto}
  \mathbb{H} \, = \, \so_c(2) \oplus \mathbb{H}^\prime \quad ; \quad \left[\so_c(2)\, , \, \mathbb{H}^\prime\right] \, = \, 0
\end{equation}
namely the compact subalgebra should contain a one-dimensional central subalgebra $\so_c(2) \sim \uu(1)$. This provides the bridge with  original Souriau's conception, adopted also in more recent studies by Barbaresco, Marle and others \cite{marlentropia,caldobarbaresco,barbaresco2,barbaresco3,marlegibbs} and based on the notion of coadjoint orbits. Indeed, as we remarked in our previous paper \cite{geotermico}, every coadjoint orbit ${\text{CoAdj}}_{\mathrm{U}}[\mathbf{X}]$ with $\mathbf{X}\in \mathbb{U}$ is a quotient manifold:
\begin{equation}\label{kalino}
  {\text{CoAdj}}_{\mathrm{U}}[\mathbf{X}] \, \simeq \, \frac{\mathrm{U}}{\mathrm{H}}
\end{equation}
where $\mathrm{H} \subset \mathrm{U}$ is the subgroup generated by the Lie subalgebra $\mathbb{H}\subset \mathbb{U}$ defined as the stabilizer of the Lie algebra element $\mathbf{X} \in \mathbb{U}$, namely: 
\begin{equation}\label{crinolina}
  \mathbb{H} \, = \, \text{stab}[\mathbf{X}] \, = \,\left \{\mathbf{h}\in \mathbb{U} \, \mid \, \left[\mathbf{h}\, , \, \mathbf{X}\right]\, = \, 0 \right\}
\end{equation}
For K\"ahler symmetric spaces the compact subalgebra in eq.(\ref{kronto}) is the stabilizer of $X_c$ where with such a symbol we denote the generator of $\so_c(2)$:
\begin{equation}\label{artusi}
 \mathbb{H} \, = \, \so_c(2) \oplus \mathbb{H}^\prime \, = \, \text{stab}[X_c]
\end{equation}
\subsection{The general expression of the K\"ahler $2$-form}
Considering the adjoint representation of the central generator $X_c$
\begin{equation}\label{cartonista}
  {X}_{AB} \, = \, \text{adj}_{\mathbb{K}} \left(X_c\right)
\end{equation}
on the vector subspace $\mathbb{K}$ of coset generators, appearing in the orthogonal decomposition:
\begin{equation}\label{frissonate}
  \mathbb{U}\, = \, \mathbb{H} \oplus \mathbb{K}
\end{equation}
the K\"ahler 2-form of the symmetric space $\mathrm{U/H}$ has the following general expression in terms of the vielbein $V^A$
\begin{eqnarray}\label{cantico}
  \boldsymbol{\mathcal{K}}\, & = & \, {X}_{AB} \, V^A \wedge V^B \nonumber\\
  V^A \, & = & \, \text{Tr}\left( \Theta \, K^A\right) \quad ; \quad \Theta \, \equiv \, \mathbb{L}^{-1} \, \mathrm{d} \mathbb{L}
\end{eqnarray}
having denoted by $\mathbb{L}$ the coset representative, by $\Theta$ the left-invariant $1$-form, by $K^A$ an orthonormal basis of $\mathbb{K}$ and by $V^A$ the vielbein such that the symmetric space metric is:
\begin{equation}\label{krabbio}
  ds^2_{\mathrm{U/H}} \, = \, \sum_{A,B} \, V^A \times V^B \, \delta_{AB}
\end{equation}
For more details on the geometry of non-comapct symmetric spaces and their metric equivalence with solvable Lie groups, we refer the reader both to our previous paper on group thermodynamics \cite{geotermico} and to the foundational papers on  Cartan Neural Networks \cite{pgtstheory} and \cite{TSnaviga}.
\subsection{The general form of the moment-maps of Killing vector fields}
\label{deagostini} 
As explained in section 6 of \cite{geotermico}, given any basis $\boldsymbol{\mathfrak{k}}_\Lambda$ of the Killing vector fields generating the isometries of $\mathrm{U/H}$, each Killing vector being uniquely associated with a generator $J_A$ of the Lie algebra $\mathbb{U}$,  one obtains the corresponding moment map via a fully general formula which is the extremely simple following one\footnote{For convenience we always utilize a matrix representation of the generators $J_A$ such that
$Tr[J_A.J_B]= \pm \, \delta_{A B}$ where the plus sign occurs for non-compact generators and the minus sign for compact ones.}:
\begin{eqnarray}\label{gelindoelapecora}
  \mathfrak{P} & : & \mathbb{U} \, \longrightarrow \, \mathbb{C}^{\infty}\left(\frac{\mathrm{U}}{\mathrm{H}}\right)\nonumber\\
  \mathfrak{P}_A\left(\boldsymbol{\Upsilon}\right) &=& \ft 12 \, \text{Tr} \left[ X_c\cdot \mathbb{L}^{-1}
  \left(\boldsymbol{\Upsilon}\right)\cdot J_A \cdot \mathbb{L}
  \left(\boldsymbol{\Upsilon}\right)\right]
\end{eqnarray}
The functions $\mathfrak{P}_\Lambda\left(\boldsymbol{\Upsilon}\right)$  satisfy the necessary condition with respect to the Killing vector fields $\boldsymbol{\mathfrak{k}}_\Lambda$, namely:
\begin{equation}\label{lescontamines}
  i_{\boldsymbol{\mathfrak{k}}_\Lambda}\boldsymbol{\mathcal{K}}  \, = \, \mathrm{d}\mathfrak{P}_\Lambda\left(\boldsymbol{\Upsilon}\right)
\end{equation}
and their definition (\ref{gelindoelapecora}) has a very simple interpretation; they are the projection onto the $\so(2)_c$ central subalgebra defining the K\"ahler structure of the adjoint transformation of the generator $J_\Lambda$. An important comment is the following. Since all the non-compact symmetric spaces $\mathrm{U/H}$ are Hadamard-Cartan manifolds diffeomorphic to $\mathbb{R}^n$ (see \cite{TSnaviga}), they can be covered by just one open chart and the solvable coordinates $\boldsymbol{\Upsilon}$ provide such chart. For this reason the moment maps in solvable coordinates $\mathfrak{P}_\Lambda\left(\boldsymbol{\Upsilon}\right)$ are globally defined functions over the whole manifold and eq. (\ref{lescontamines}) holds true globally.
\section{Poissonian structure and the Souriau temperature problem}
From equation (\ref{lescontamines}) we immediately obtain:
\begin{equation}\label{pescioparenzo}
  i_{\boldsymbol{\mathfrak{k}}_B}\,i_{\boldsymbol{\mathfrak{k}}_A}
  \boldsymbol{\mathcal{K}} \, = \, i_{\boldsymbol{\mathfrak{k}}_B}\,
  \mathrm{d}\mathfrak{P}_A\left(\boldsymbol{\Upsilon}\right) \, = \, - \, 
  i_{\boldsymbol{\mathfrak{k}}_A}\,
  \mathrm{d}\mathfrak{P}_B\left(\boldsymbol{\Upsilon}\right)
\end{equation}
which shows that on the space $\mathbb{C}^{\infty}\left(\frac{\mathrm{U}}{\mathrm{H}}\right)$ of smooth functions over the symmetric space, namely over the solvable Lie Group manifold $\mathcal{S}_{\mathrm{U/H}}$, we can introduce a Poisson bracket:
\begin{equation}\label{corallino}
  \forall \mathit{f}(\boldsymbol{\Upsilon}),\mathit{g}(\boldsymbol{\Upsilon})\in \mathbb{C}^{\infty}\left(\frac{\mathrm{U}}{\mathrm{H}}\right) \, \simeq \, 
  \mathbb{C}^{\infty}\left(\mathcal{S}_{\mathrm{U}/\mathrm{H}}\right) \quad : \quad 
  \left\{f\, , \,  g\right\} \, \equiv  \, \boldsymbol{\Pi}^{\alpha\beta} \, \frac{\partial f}{\partial \Upsilon^\alpha}\,\frac{\partial g}{\partial \Upsilon^\beta}
\end{equation}
where the Poissonian bivector $\boldsymbol{\Pi}^{\alpha\beta}$ is just the inverse of the K\"ahler $2$-form $\boldsymbol{\mathcal{K}}_{\gamma\delta}$ regarded as a matrix. Explicitly one sets:
\begin{eqnarray}\label{lucullus}
  \boldsymbol{\mathcal{K}} & = & X_{AB} \, V^A \wedge V^B \, = \, X_{AB} \, V^A_\gamma \,  V^B_\delta \, \mathrm{d}\Upsilon^\gamma \wedge \mathrm{d}\Upsilon^\delta \, \equiv \, \boldsymbol{\mathcal{K}}_{\gamma\delta} \, \mathrm{d}\Upsilon^\gamma \wedge \mathrm{d}\Upsilon^\delta \nonumber\\
  \boldsymbol{\Pi}^{\alpha\beta} & \equiv & \left(\boldsymbol{\mathcal{K}}^{-1}\right)^{\alpha\beta} \nonumber\\
  & \Downarrow & \nonumber\\
  \boldsymbol{\Pi}^{\alpha\beta} & = & g^{\alpha\gamma} \, g^{\beta\delta} \, \boldsymbol{\mathcal{K}}_{\gamma\delta}
\end{eqnarray}
where $g^{\mu\nu}$  denotes the inverse metric tensor of the $\mathrm{U/H}$ symmetric space. 
\subsection{"Action/Angle" variables and Darboux bases}
In any solvable coordinate basis generically denoted $\boldsymbol{\Upsilon}$
\footnote{Solvable coordinates are those that explicitly realize the diffeomorphism between the $2n$-dimensional symmetric space $\mathrm{U/H}$, that is a Cartan Hadamard manifold, and $\mathbb{R}^{2n}$, as established by Cartan Hadamard theorem. They are real numbers $\Upsilon^A$ in one-to-one correspondence with a generator basis $T_A$ of the solvable Lie algebra $Solv_{\mathrm{U/H}}$. As parameters of the corresponding solvable group and hence as coordinates of the manifold $\mathrm{U/H}$, there are several solvable coordinate bases, depending on the chosen order of exponentiation from the Lie algebra to the group. In this paper we utilize two different solvable bases that we illustrate later one and that are useful for different purposes}  the manifold metric is very simple and the integration measure:
\begin{equation}\label{kukummo}
  \boldsymbol{\mathrm{d}\mu}\left(\boldsymbol{\Upsilon}\right) \, \equiv \,
  \underbrace{\boldsymbol{\mathcal{K}}\wedge\boldsymbol{\mathcal{K}}\wedge \dots \wedge\boldsymbol{\mathcal{K}}}_{n \text{-times}}   
\end{equation}
where $n\in \mathbb{N}$ is one-half of the dimensions of the K\"ahler symmetric space 
$\mathrm{U/H}$ is just the $\mathbb{R}^{2n}$ euclidian measure:
\begin{equation}\label{geremia}
  \boldsymbol{\mathrm{d}\mu}\left(\boldsymbol{\Upsilon}\right)\, = \, \text{const} \, \times \, \mathrm{d}\Upsilon^{1}
  \wedge \mathrm{d}\Upsilon^{2}\wedge \dots \wedge \mathrm{d}\Upsilon^{2n-1}
  \wedge \mathrm{d}\Upsilon^{2n}
\end{equation}
yet the K\"ahler $2$-form is not in  Darboux form:
\begin{equation}\label{cromatosko}
  \boldsymbol{\mathcal{K}} \, = \, \sum_{a=1}^{n} \, \mathrm{d}\mathfrak{p}_a \wedge \mathrm{d}\mathfrak{q}^a
\end{equation}
and a quite legitimate and interesting question is whether one can find a coordinate transformation:
\begin{alignat}{8}\label{cavalcione}
  \mathfrak{p}_a & = & \mathfrak{p}_a(\boldsymbol{\Upsilon}) &\quad ; \quad & 
  \mathfrak{q}^a & = & \mathfrak{q}^a(\boldsymbol{\Upsilon}) &\quad ; \quad& a=1,\dots, n
\end{alignat}
such that:
\begin{equation}\label{coricidino2}
  \left\{\mathfrak{p}_a(\boldsymbol{\Upsilon})\, , \,\mathfrak{p}_b(\boldsymbol{\Upsilon})\right\} \, = \, 0 \quad; \quad
  \left\{\mathfrak{q}^a(\boldsymbol{\Upsilon})\, , \,\mathfrak{q}^b(\boldsymbol{\Upsilon})\right\} \, = \, 0 \quad; \quad
  \left\{\mathfrak{p}_a(\boldsymbol{\Upsilon})\, , \,\mathfrak{q}^b(\boldsymbol{\Upsilon})\right\} \, = \, \delta^{a}_b
\end{equation}
Indeed, as we anticipated in the introduction (sect. \ref{introibo}), such a question is strongly related with the issue of Souriau thermodynamics in the way we have defined it above and it will be addressed for Calabi-Vesentini manifolds in sect.\ref{CVmanifesto}, sect.\ref{sashasecta} and sect.\ref{mariosecta}. Here we discuss the general problem.
\subsection{The Souriau temperature problem}
\label{topocaldo}
Coming back to eq.(\ref{crodinotrotato}), in view of the complete discussion of section \ref{infogeomacro}, Souriau's Gibbs distributions have the following structure:
\begin{alignat}{4}\label{cerusico}
  & G(\boldsymbol{\beta},\boldsymbol{\Upsilon}) &\quad = \quad &\quad  \frac{\exp\left[-\boldsymbol{\beta}\cdot\boldsymbol{\mathfrak{P}}
  \left(\boldsymbol{\Upsilon} \right)\right]}{Z(\boldsymbol{\beta})}\nonumber\\
  & Z(\boldsymbol{\beta}) &\quad = \quad &\quad \int_{\mathrm{U/H}}
  \, \exp\left[-\boldsymbol{\beta}\cdot\boldsymbol{\mathfrak{P}}
  \left(\boldsymbol{\Upsilon} \right)\right] \, \boldsymbol{\mathrm{d}\mu}\left(\boldsymbol{\Upsilon}\right)
\end{alignat}
where $\boldsymbol{\mathfrak{P}}_A\left(\boldsymbol{\Upsilon} \right)$ are the
Killing vector moment maps defined by eq.(\ref{gelindoelapecora}) that provide 
a Poissonian realization of the full $\mathbb{U}$-Lie algebra:
\begin{equation}\label{quirito}
  \left \{\boldsymbol{\mathfrak{P}}_A\left(\boldsymbol{\Upsilon} \right) \, , \,
  \boldsymbol{\mathfrak{P}}_B\left(\boldsymbol{\Upsilon} \right) \right\} \, = \, 
  \mathit{f}_{AB}^{\phantom{AB}C} \, \boldsymbol{\mathfrak{P}}_C\left(\boldsymbol{\Upsilon} \right)
\end{equation}
the structure constants $\mathit{f}_{\Lambda\Sigma}^{\phantom{\Lambda\Sigma}\Gamma}$
being the same that appear in the commutation relations of the corresponding generators in the chosen basis:
\begin{equation}\label{uguaglianza}
  \left[J_A \, ,\,J_B \right]\, = \, \mathit{f}_{AB}^{\phantom{AB}C} \, J_C
\end{equation}
Furthermore the scalar product on $\mathbb{U}$, denoted by $\cdot$ that appears in eq.(\ref{cerusico}), is provided by the Killing metric of the simple Lie algebra that we can always normalize in such a way that it is diagonal and has eigenvalue $1$ for all non-compact generators and $-1$ for all compact ones.
\par
This being clarified, the focal point is to determine the conditions on the generalized temperature vector $\boldsymbol{\beta} \in \mathbb{U}$ that guarantee the convergence of the integral in eq.(\ref{cerusico}) which defines the partition function.
As it was already noted in our previous paper \cite{geotermico}, thanks to the $\mathrm{U}$-invariance of the integration measure we can always perform an adjoint transformation that brings whatever $\boldsymbol{\beta}$ to have non vanishing components only along generators of the compact subalgebra:
\begin{equation}\label{cinciunciun}
 \forall \boldsymbol{\beta}\in \mathbb{U} \quad , \quad \exists g \in \mathrm{U}  \quad \text{such that} \quad \boldsymbol{\beta} \,  \stackrel{\text{Adj}[\mathfrak{g}]}{\longrightarrow} \, \hat{\boldsymbol{\beta}}\in \mathbb{H} \subset \mathbb{U}
\end{equation}
For more details see section 7 and subsection 7.1 of \cite{geotermico}. Next  by means of additional adjoint transformations with respect only to the compact subgroup $\mathrm{H}\subset \mathrm{U}$ we can further transform whatever $\hat{\boldsymbol{\beta}}\in \mathbb{H}$ to a $\boldsymbol{\beta}_c \in \boldsymbol{\mathcal{C}}$ the symbol $\boldsymbol{\mathcal{C}}$  denoting the Cartan subalgebra of $\mathbb{H}$. Hence relying on the metric equivalence of the symmetric space with the solvable Lie group $\mathcal{S}_{\mathrm{U/H}}$, that has a transitive free action on $\mathrm{U/H}$, we can write the canonical form of Souriau Gibbs distributions as in eq.s (7.16) of \cite{geotermico}, namely:
\begin{eqnarray}\label{carbonero}
  \mathrm{G}\left(\boldsymbol{\beta}_c\, ; \,  g\in \mathcal{S} \, \mid \, \boldsymbol{\Upsilon} \right) & \equiv & \frac{\exp\left[ - \sum_{a=0}^{\nu+1=\text{rank }\mathbb{H}^\prime} \,\boldsymbol{\beta}_c^{a} \,\boldsymbol{\mathfrak{P}}_a \left(g[\boldsymbol{\Upsilon}]\right) \right]} {Z(\boldsymbol{\beta}_c)} \nonumber\\
  \boldsymbol{\beta}_c^{a} & = & \text{temperatures associated with the compact Cartan generators $H^c_{0,1,\dots, \nu+1}$ of $\mathbb{H}$} \nonumber\\
  \boldsymbol{\mathfrak{P}}_a \left(\boldsymbol{\Upsilon}\right) &=& \text{moment maps of the compact Cartan generators $H^c_{0,1,\dots, \nu+1}$ of $\mathbb{H}$} \nonumber\\
  g & = & \text{any group element in $\mathcal{S}$} \nonumber\\
  {g}[\boldsymbol{\Upsilon}] & = & \text{new solvable parameters after a $g$-transformation}
  \end{eqnarray}
\subsubsection{Completion of the abelian structure including the compact Cartan subalgebra moment maps}
\label{abelprogram}  
Once the integrand in eq.(\ref{cerusico}) is reduced to have in the argument of its exponential function just a linear combination of moment maps belonging to the compact Cartan subalgebra $\boldsymbol{\mathcal{C}}$, it is clear that such moment maps commute among themselves in the Poissonian sense, namely;
\begin{equation}\label{cartilagine}
  \left\{\boldsymbol{\mathfrak{P}}_i(\boldsymbol{\Upsilon})\, , \, \boldsymbol{\mathfrak{P}}_j(\boldsymbol{\Upsilon})\right\} \, = \, 0 \quad \quad i,j \, = \, 0,1, \dots, \nu +1\, = \, \text{rank}[\mathbb{H}^\prime]
\end{equation}
and one is tempted to interpret them as Darboux coordinates, according with the definition of the latter in eq.s (\ref{cavalcione}-\ref{coricidino2}). Such an interpretation holds true only if we are able:
\begin{description}
  \item[A)] \textbf{To complete the list} of functions $\mathfrak{p}_i(\boldsymbol{\Upsilon})$ by adjoining to the first $\nu+2$, corresponding to the moment maps of the compact Cartan generators $\mathfrak{p}_a(\boldsymbol{\Upsilon})\, = \, \boldsymbol{\mathfrak{P}}_a(\boldsymbol{\Upsilon})$ ($a=0,1,\dots,\nu+1$) a convenient set of $n-\nu-2=\nu+1$ additional independent functions
      in involution among themselves and with the first ones. 
  \item[B)] \textbf{To find the complementary} $n$ functions $\mathfrak{q}^b(\boldsymbol{\Upsilon})$ satisfying with the $\mathfrak{p}_a(\boldsymbol{\Upsilon})$ the commutation relations (\ref{coricidino2})
\end{description}
Here comes a very delicate issue to be clarified at once in order to avoid misconceptions and logical errors. As we have already pointed out in previous subsections, the isometry structure of most $2n$-dimensional K\"ahlerian non-compact symmetric spaces $\mathrm{U/H}$
and, in particular, of all Calabi-Vesentini manifolds, displays the existence of \textbf{a $n$ dimensional maximal abelian ideal $\mathcal{AI}_{max}\subset Solv_{\mathrm{U/H}}$ of the solvable Lie algebra}, realized as  \textbf{$n$ non-compact translational isometries}. Hence, in an appropriately chosen coordinate basis the K\"ahler potential depends only on the imaginary parts of the complex coordinates and all what we described above from eq.(\ref{complessini}) to eq.(\ref{fradiavolo} ) applies, namely we have the Darboux coordinate basis $(\mathfrak{m}_{a},u^b)$ where the $\mathfrak{m}_{a}$ are the derivatives of the K\"ahler potential with respect to the imaginary parts $v^i$ and the cyclic variables $u^b$ are non-compact. This however has two problems:
\begin{enumerate}
  \item $\mathfrak{P}_i(\boldsymbol{\Upsilon}) \notin \bigcup_{a=1}^{n}\left\{ \mathfrak{m}_a\right\}$, for ($i=0,\dots, \nu+1$), namely $\bigcup_{a=1}^{n}\left\{ \mathfrak{m}_a\right\}$  is not a list of $n$-objects including any of the $\nu+2$ moment maps of the compact Cartan subalgebra.
  \item The complementary commuting variables $u^a$ are non-compact. 
\end{enumerate}
Hence the completion mentioned in the above points A) e B) has to be done with the different methods of \textit{Integrable System Theory} (see section \ref{sashasecta}) in such a way as to find:
\begin{enumerate}
  \item A list $\mathfrak{p}_a(\boldsymbol{\Upsilon})$ of $n=2\nu +2$ commuting functions that includes the $\nu+1 < n$ compact Cartan subalgebra moment maps $\mathfrak{P}_i(\boldsymbol{\Upsilon})$, ($i=1,\dots,\nu+1$) 
  \item The complementary commuting variables $\mathfrak{q}^b(\boldsymbol{\Upsilon})$ satisfying the Darboux relations (\ref{coricidino2}) and being all valued on a compact interval, namely being true angles. 
\end{enumerate}
When one succeeds in the above programme 1)-2) one says that 
\textbf{a compact abelian structure} has been found. 
\par
The main observation in this respect is that, once the Poissonian structure has been established by means of the 
K\"ahler $2$-form of the manifold, the Poissonian Lie bracket is extended from the space of moment maps of the Killing vector fields to much more general functions obtained from the enveloping algebra of the latter that are no longer the moment maps of any Killing vector field. It is in this more general enviroment that one looks for the abelian structure and for true "action/angles" variables.
\par
Typically the missing functions $\mathfrak{p}_a(\boldsymbol{\Upsilon})$ ($a=2+\nu+1,\dots,3+2\nu$), at least in the Calabi-Vesentini case, are square roots of quadratic polynomials in the compact subalgebra moment maps. 
\par
Once the abelian structure is found the change of coordinates from the original solvable ones to the abelian structure Darboux coordinates is also found and the integral defining the partition function  enormously simplifies:
\begin{equation}\label{cardioloso}
  Z(\boldsymbol{\beta}_c) \, = \, \int_{\mathcal{P}_n} \exp\left[-\, \sum_{a=0}^{\nu+1}
  \boldsymbol{\beta}_c^a \, \mathfrak{p}_a \right] \mathrm{d}^{n}\mathfrak{p} \,\times \,\int_{\boldsymbol{\mathcal{T}}^{n}} \mathrm{d}^{n}\mathfrak{q}
\end{equation}
In the above schematic formula $\boldsymbol{\mathcal{T}}^{n}$ is the $n$-torus predicted by Arnol'd theorem and mentioned in the introduction while 
$\mathcal{P}_n$ denotes the convex Polytope on which the action Darboux coordinates $\mathfrak{p}-a$ take values after the coordinate transformation (\ref{cavalcione}-\ref{coricidino2}). The actual topology 
of the image of the symmetric space in Darboux action/angle coordinates, \textit{i.e.} the polytope $\mathcal{P}_n$ has to be studied and determined in each abelian structure one is able to find. 
\par
Actually as we show in section \ref{sashasecta} and in its subsections the number of inequivalent integration architectures (namely polytope structures) can be very large as the dimension of the manifold grows, yet as far as
the partition function integral (\ref{cardioloso}) is concerned they all yield the same result as it should be. 
\par
Here we anticipate that, once one arrives at the abelian structure and at the expression (\ref{cardioloso}) for the original partition function, one realizes 
that the temperature vector $\boldsymbol{\beta}_c$ can be generalized with the inclusion of additional temperatures corresponding to the remaining action coordinates $\mathfrak{p}_a$($a=\nu+2,\dots, n$), writing:
\begin{equation}\label{cerebroleso}
  Z({\boldsymbol{\lambda}}) \, = \, \int_{\mathcal{P}_n} \exp\left[-\, \sum_{a=1}^{n}
  \boldsymbol{\lambda}^a \, \mathfrak{p}_a \right] \mathrm{d}^{n}\mathfrak{p} \, 
   \times \, \int_{\boldsymbol{\mathcal{T}}^{n}} \mathrm{d}^{n}\mathfrak{q}
\end{equation}
We refer the reader to section (\ref{sashasecta}) and in particular to eq.(\ref{maggiorana}-\ref{eq:extended_Hamiltonian}) for details.
\par
When the extra temperatures are switched on, the multitude of abelian structures introduce several \textbf{different generalizations of Souriau thermodynamics}.
The question to be answered is how do the extra components of the temperature vector $\boldsymbol{\lambda}$ transform under the action of the full group $\mathrm{U}$? This issue has to be studied case by case in order to evaluate how such extensions can be utilized in those contexts where Souriau thermodynamics provides convenient probability distributions. In any case, switching on the additional generalized temperatures breaks the $\mathrm{U}$-symmetry of the partition functions and introduces non vanishing average values for new functions defined over the event manifold (the extra acions), typically introducing in this way a phase transition
to a symmetry broken phase. 
\subsection{The general result for Calabi Vesentini manifolds}
\label{lodicoprima}
In the next section \ref{CVmanifesto} we analyze the structure and the properties of the Calabi-Vesentini symmetric spaces $\mathcal{M}_{CV}^{[2,q]}$, while in section \ref{sashasecta} we address the general construction of the corresponding partition functions on the basis  of abelian-structure techniques including also the analysis of the above mentioned generalization of Souriau thermodynamics. Restricting one's attention to the case of temperatures in the adjoint orbit of the compact Cartan subalgebra, one of the main achievement of the present paper is the explicit analytic expression of the partition function for all such cases which, setting
\begin{equation}\label{cavolonero}
  q \, = \, \left\{ \begin{array}{lcl}
                      2\, \nu \, + \, 1 & \nu \in \mathbb{N} & \text{case } \mathfrak{b}\\
                      2\, \nu \, &  \nu \in \mathbb{N} & \text{case } \mathfrak{d} \\
                    \end{array} \right.
\end{equation}
has the following very simple and elegant structure
{\large
\begin{eqnarray}
\label{zolotayaformula}
 Z^{\mathfrak{b},\mathfrak{d}}\left(\beta_0,\beta_{i=1,\dots,1
 +\nu}\right) &=& 
c^{\mathfrak{b},\mathfrak{d}} \frac{ (8\pi^2)^{\nu+1} e^{-\beta_0}}{\prod_{i=1}^{\nu+1} (\beta_0^2 - \beta_i^2)}  \\
                   \nonumber\\
  c^{\mathfrak{b}} &=& \frac{2\pi}{\beta_0} \nonumber\\
  c^{\mathfrak{d}} &=& 1 \nonumber
\end{eqnarray}
}
with a small difference in the denominator between the odd-dimensional and even dimensional case of $q$ which, from the Lie algebra theory point of view, corresponds to the distinction between the series $\mathfrak{b}_{\nu}$ and $\mathfrak{d}_{\nu}$.
This means that the formula (\ref{carbonero}) for the Gibbs distributions becomes the
following:
\begin{eqnarray}\label{carbonato}
  \mathrm{G}^{\mathfrak{b}/\mathfrak{d}}\left(\boldsymbol{\beta}_0\, \beta_i \, ; \,  g\in \mathcal{S} \, \mid \, \boldsymbol{\Upsilon} \right) & \equiv & \frac{\exp\left[ - \, \beta_0 \, \boldsymbol{\mathfrak{P}}_0 \, - \, \sum_{i=1}^{\nu+1} \, \beta^{i} \,\boldsymbol{\mathfrak{P}}_i \left(g[\boldsymbol{\Upsilon}]\right) \right]} {Z^{\mathfrak{b}/\mathfrak{d}}(\beta_0,\beta_i)} \nonumber\\
  \boldsymbol{\beta}_0 & = & \text{temperature associated with the $\uu(1)$ Cartan generator and $\boldsymbol{\mathfrak{P}}_0 \left(\boldsymbol{\Upsilon}\right)$ its
  moment map}\nonumber\\
  \boldsymbol{\beta}_i & = & \text{temperatures associated with the $\mathbb{H}^\prime$ Cartan generators and $\boldsymbol{\mathfrak{P}}_i \left(\boldsymbol{\Upsilon}\right)$ their
  moment maps} \nonumber\\
  g & = & \text{any group element in $\mathcal{S}$} \nonumber\\
  {g}[\boldsymbol{\Upsilon}] & = & \text{new solvable parameters after a $g$-transformation}
  \end{eqnarray}
and the convergence conditions of the defining integral, derived with mathematical techqniques briefly summarized in section \ref{sashasecta}, that will be the object of a future purely mathematical publication by the present authors, are the following ones:
\begin{equation}\label{convercondo}
\text{Conv Cone $\boldsymbol{\mathfrak{C}}$} \, =\,  \left\{\beta_0 \, > \, 0 \quad ; \quad \beta_0  > |\beta_i|\, ,\, \forall i = 1,.., \nu+1\right\}
\end{equation}
\par
Correspondingly the stochastic hamiltonian from which we derive the thermodynamical contact metric (\ref{contactmetric}) is:
\begin{eqnarray}
\label{fardellus}
  \boldsymbol{\mathcal{H}}^{\mathfrak{b}}(\boldsymbol{\beta}) &=& \beta_0 \, +\log[\beta_0] +\sum_{i=1}^{\nu+1}\log\left[\beta_0^2-\beta_i^2\right] \, + \, \text{const}\nonumber  \\
  \boldsymbol{\mathcal{H}}^{\mathfrak{d}}(\boldsymbol{\beta}) &=& \beta_0 \, +\sum_{i=1}^{\nu+1}\log\left[\beta_0^2-\beta_i^2\right] \, + \, \text{const}
\end{eqnarray}
We postpone to a future publication the detailed analysis of the thermodynamic
Riemannian metric following from eq.(\ref{fardellus}) but we note here that the convergence conditions define in the temperature space (the dual $\boldsymbol{\mathcal{C}}^\star$ of the compact Cartan subalgebra     $\boldsymbol{\mathcal{C}}\subset\mathbb{H}\subset \mathbb{U}$) a positivity domain that is the convex cone $\boldsymbol{\mathfrak{C}}$ where the symplectic potential,
namely the stochastic hamiltonian (\ref{fardellus}) is real. 
\section{Calabi-Vesentini manifolds}
\label{CVmanifesto}
As we already pointed out in our previous paper \cite{geotermico}, apart from the few sporadic cases based on exceptional Lie algebras, the infinite towers of K\"ahler non-compact symmetric spaces are just four:
\begin{alignat}{4}\label{quaresima}
  &\mathcal{M}_{su}^{[1,1+q]}\quad & \equiv &\quad \frac{\mathrm{SU(1,1+q)}}{\mathrm{U(1)\times SU(1+q)}} & \quad ; \quad & \mathcal{M}_{CV}^{[2,q]}\quad & \equiv &\quad \frac{\mathrm{SO(2,2+q)}}{\mathrm{SO(2)\times SO(2+q)}} \nonumber\\
  & \mathcal{M}_{Sieg}^{[g]}\quad & \equiv & \quad  \frac{\mathrm{Sp(2g,\mathbb{R})}}{\mathrm{U(1)\times SU(g)}}& \quad ; \quad &\mathcal{M}_{so*}^{[n]}\quad  & \equiv  &\quad \frac{\mathrm{SO^\star(2n)}}{\mathrm{U(1) \times SU(n)}}
\end{alignat}
From the point of view of the equivalent solvable Lie group, which is strategic for all our considerations and perspective applications to Machine Learning, the first tower $\mathcal{M}_{su}^{[1,1+q]}$ might be unconvenient since its solvable Lie algebra is exotic (see for instance the book \cite{advancio}) the algebraic Tits Satake projection of the root system not being a smaller root system as instead it happens for all  other cases of simple Lie algebras. For this reason we do not consider, for the moment, the first tower. The Siegel half planes of genus $g$, namely $\mathcal{M}_{Sieg}^{[g]}$ 
present no exotic anomaly but are maximally split, which means that their 
non-compact rank grows with dimensions and there is no Tits Satake projection,  the Paint Group being just the identity. Similarly, in the last tower $\mathcal{M}_{so*}^{[n]}$ the non-compact rank grows with the dimension $2n$  and the tower does not form a single Tits Satake universality class. If we are interested in combining the study of Gibbs statistical distributions with the analysis of the consequences of Paint Group symmetry, then the obligatory choice is that of the Calabi-Vesentini manifolds   
$\mathcal{M}_{CV}^{[2,q]}$. These manifolds have an additional remarkable property.
Their tensor product with a hyperbolic plane $\mathrm{SL(2,\mathbb{R})}/\mathrm{SO(2)}$ constitutes \textbf{a special K\"ahler manifold of the local type} namely $\mathcal{SK}_{3+q}$ already mentioned in eq.(\ref{speckal}).
This implies that there is a symplectic bundle and a symplectic embedding of the group $\mathrm{U}=\mathrm{SO(2,2+q)}$ which acts by linear fractional transformation on a symmetric complex matrix $\mathcal{N}_{\Lambda\Sigma}(p)$ of dimension $(4+q)\times(4+q)$ that is parameterized by the points of the symmetric space $\mathrm{U/H}$, just as in the case of the Siegel plane. We postpone to a later subsection, the discussion of the symplectic embedding: first we analyse the structure of the equivalent solvable Lie group. 
\subsection{The equivalent solvable Lie Group and its structure}
\label{equivosolvo}
The solvable Lie subgroup:
\begin{equation}\label{solsub}
  \mathcal{S}_{[2,2+q]} \, \subset \, \mathrm{SO(2,2+q)}
\end{equation}
whose dimension is equal to that of the symmetric space $\mathcal{M}^{[2,q]}_{CV}$ and explicitly realizes the Cartan-Hadamard diffeomorphism, has a solvable Lie algebra $Solv_{\so(2,2+q)}$, whose structure is best described by
the corresponding Maurer Cartan equations. By introducing  a Paint index  running in the fundamental vector representation of the Paint Group 
$\mathrm{G_{Paint}} \, = \, \mathrm{SO(q)}$, the MC equations have a simple, general, Paint-invariant form holding true for the entire TS universality class,  that is the following one:
\begin{equation}\label{maurocartus}
Solv_{[2,2+q]} \Leftrightarrow\left\{\begin{array}{lcl}
 \mathrm{d}e^1 & = & 0 \\
 \mathrm{d}e^2 & = & 0 \\
 \mathrm{d}e^3+\frac{1}{2} (2 e^{1}\wedge e^{3}-2 e^{2}\wedge e^{3}) & = & 0 \\
 \mathrm{d}e^4+\frac{1}{2} \left(2 e^{1}\wedge e^{4}+2 e^{2}\wedge e^{4}\right)&=& 
 \frac{1}{\sqrt{2}}\sum^{q}_{i=1} \, e^{5,i}\wedge e^{6,i} \\
 \mathrm{d}e^{5,i}+\frac{1}{2} \left(2 e^{1}\wedge e^{5,i}+\sqrt{2} e^{3}\wedge e^{6,i}\right) & = & 0 \\
 \mathrm{d}e^{6,i}+e^{2}\wedge e^{6,i} & = & 0 \\
\end{array}\right.
\end{equation}
In the above MC equations one easily reads off all the properties of the Tits-Satake and sub Tits-Satake projections. As explained in \cite{pgtstheory} and previously reviewed in the book \cite{advancio} and in the original paper \cite{tittusnostro}, the \textbf{sub-Tits Satake} Lie algebra $\mathbb{U}_{\mathrm{subTS}} \subset \mathbb{U}_{\mathrm{TS}} \subset \mathbb{U}$ is the subalgebra that commutes with the compact Paint subalgebra $\mathbb{G}_{\mathrm{Paint}} \subset \mathbb{U}$:
\begin{equation}\label{perdinci1}
  \left [\mathbb{U}_{\mathrm{subTS}}\, , \, \mathbb{G}_{\mathrm{Paint}}  \right] \, = \, 0
\end{equation}
just as the Tits-Satake subalgebra $\mathbb{U}_{\mathrm{TS}} \subset \mathbb{U}$  is that one that commutes with the \textbf{sub-Paint subalgebra} $\mathbb{G}_{\mathrm{subPaint}}\subset \mathbb{G}_{\mathrm{Paint}}$:
\begin{equation}\label{perdinci2}
  \left [\mathbb{U}_{\mathrm{TS}}\, , \, \mathbb{G}_{\mathrm{subPaint}} \right] \, = \, 0
\end{equation}
Furthermore considering the maximal compact subgroups $\mathrm{H}_{\mathrm{subTS}}\subset \mathbb{U}_{\mathrm{subTS}}$ and
$\mathrm{H}_{\mathrm{TS}}\subset \mathbb{U}_{\mathrm{TS}}$, one has that also
the coset manifolds $\mathrm{U}_{\mathrm{subTS}}/\mathrm{H}_{\mathrm{subTS}}$ and
$\mathrm{U}_{\mathrm{TS}}/\mathrm{H}_{\mathrm{TS}}$ are non-compact symmetric spaces,
each metrically equivalent to a solvable Lie group, respectively, $\mathcal{S}_{\mathrm{subTS}}$ and $\mathcal{S}_{\mathrm{TS}}$,  with the inclusions:
\begin{alignat}{4}
\label{fragolino}
& \frac{\mathrm{U}_{\mathrm{subTS}}}{\mathrm{H}_{\mathrm{subTS}}}
&\quad \subset &\quad \frac{\mathrm{U}_{\mathrm{TS}}}{\mathrm{H}_{\mathrm{TS}}}&\quad \subset
&\quad \frac{\mathrm{U}}{\mathrm{H}} \nonumber\\
&\quad\Updownarrow & \quad &\quad \Updownarrow &\quad &\quad \Updownarrow \nonumber\\
& \mathcal{S}_{\mathrm{subTS}}&\quad \subset &\quad \mathcal{S}_{\mathrm{TS}}&\quad \subset &\quad \mathcal{S}_{\mathrm{U}}\nonumber\\
&\quad\downarrow & \quad &\quad \downarrow &\quad &\quad \downarrow \nonumber\\
&\quad \mathrm{U}_{\mathrm{subTS}}&\quad \subset &\quad \mathrm{U}_{\mathrm{TS}}&\quad \subset &\quad \mathrm{U}
\end{alignat}
In the first line of eq.(\ref{fragolino}, the inclusion is like differentiable submanifolds, while in the second and in the third line the inclusion is like  Lie subgroups. The arrows between the first and the second line recall the metric equivalence, namely the
Cartan-Hadamard diffeomorphism of the upper symmetric space with the lower solvable group and the \textit{derivation of the metric} of the upper Riemannian manifold from a \textit{positive definite quadratic form} defined on the solvable Lie algebras  $Solv_{\mathrm{subTS}}$, $Solv_{\mathrm{TS}}$, $Solv_{\mathrm{U}}$ of the lower solvable Lie groups.  
\par
In the case of the Calabi-Vesentini manifolds we have:
\begin{alignat}{5}\label{gromillo}
& \mathbb{U}_{\mathrm{subTS}} &\quad = &\quad \so(2,1) \times \so(2,1) &\quad \supset &\quad Solv_{[2,2]} \nonumber \\
& \quad \downarrow &\quad  &\quad \quad \quad \quad\downarrow &\quad  &\quad \downarrow \nonumber \\
& \quad \mathbb{U}_{\mathrm{TS}} &\quad = &\quad\quad\quad\; \so(2,3)  &\quad \supset &\quad Solv_{[2,3]} \nonumber \\
& \quad \downarrow &\quad  &\quad \quad \quad \quad \downarrow &\quad  &\quad \downarrow \nonumber \\
& \quad \mathbb{U} &\quad = &\quad \quad \;\;\so(2,2+q)  &\quad \supset &\quad Solv_{[2,2+q]} \quad q >1 
\end{alignat}
where the down arrow denotes the inclusion like a Lie subalgebra. Correspondingly we have the following chain of embeddings of  symmetric spaces:
\begin{equation}\label{gronkus}
  \underbrace{\frac{\mathrm{SO(2,1)}}{\mathrm{SO(2)}}\times \frac{\mathrm{SO(2,1)}}{\mathrm{SO(2)}} \simeq \frac{\mathrm{SL(2,\mathbb{R})}}{\mathrm{SO(2)}}\times \frac{\mathrm{SL(2,\mathbb{R})}}{\mathrm{SO(2)}}}_{\text{sub TS manif.}} \, \subset \, 
  \underbrace{\frac{\mathrm{SO(2,3)}}{\mathrm{SO(2)\times SO(3)}} \simeq
  \frac{\mathrm{Sp(4,\mathbb{R})}}{\mathrm{SU(2)\times U(1)}}}_{\text{TS manif.}}\, \subset \,
  \underbrace{\frac{\mathrm{SO(2,2+q)}}{\mathrm{SO(2)\times SO(2+q)}}}_{\text{CV manif.}}
\end{equation}
The relevant geometrical question is the following: what is the differentiable geometrical structure of the including manifold with respect to the included one?
The answer was already given in \cite{tassellandum} where it was shown that the 
symmetric spaces of type 
\begin{equation}\label{romeogiulietta}
  \mathcal{M}^{[r,r+q]} \, \equiv \, \frac{\mathrm{SO(r,r+q)}}{\mathrm{SO(r)\times SO(r+q)}}
\end{equation}
can all be regarded as the total manifolds of as many \textbf{vector bundles} over
the \textbf{TS submanifold} regarded as the \textbf{base-manifold} and structural group:
\begin{equation}\label{cravotto}
  \mathrm{G}_{\mathrm{struc}} \, = \, \mathrm{SO(r)} \times \mathrm{G}_{\mathrm{subPaint}} \, = \, \mathrm{SO(r)} \times \mathrm{SO(q-1)}
\end{equation}
Exactly the same can be said trading the TS base manifold for the sub TS one and enlarging the structural group by upgrading the subPaint factor to the full Paint Group.
\par
Hence the CV manifolds can be seen as vector bundles over a \textit{pair of hyperbolic planes} that play the role of \textit{base manifold}, having  as \textit{standard fibre} a $2\,q$-dimensional vector space $\mathbb{F}_{2q}$ which is in the bi-fundamental representation $(2\mid q)$ of the structural group $\mathrm{SO(2)}\times \mathrm{SO(q)}$. 
\par 
As it was already discussed in general terms in \cite{tassellandum} the above outlined vector bundle structure is quite promising for the construction of convolutionary Cartan Neural Networks adopting the view point that in such networks 
the various layers are to be regarded as different vector bundles over the same base manifold as proposed in \cite{4_GoverH_CNN1,5_GoverH_CNN2,6_G_CNN,Bronstein_2017}.
\par
This vision is particularly relevant in connection with the results of the present paper. The Gibbs distribution that we construct here are statistical distribution over a vector bundle so that one can inspect the probability distribution of points
in the fibres above each base-point.
\subsubsection{Algebraic substructures of the solvable Lie algebras}
In view of the above discussion we can now reinspect the MC equations in (\ref{maurocartus}) and interpret them. 
\paragraph{The subTS subalgebras}
Setting $e^{5,i} = e^{6,i} =0$ we obtain the MC equations of the 4-dimensional solvable Lie algebra  $Solv_{[2,2]}$. We easily see that it is the direct product of two 2-dimensional solvable algebras. Indeed defining
two new generators:
\begin{equation}\label{sollipiccoli}
  e^- \, = \, \ft 12 (e^1 - e^2) \quad ; \quad  e^+ \, = \, \ft 12 (e^1 + e^2)
\end{equation}
we see that the reduced system decouples in two isomorphic solvable algebras, $Solv^I_{[1,2]}$ and $Solv^{II}_{[1,2]}$,
the first with generators $(e^-,e^3)$, the second with generators $(e^+,e^4)$.
These are the solvable Lie algebras  of the two solvable groups, $\mathcal{S}^I_{subTS}$, $\mathcal{S}^{II}_{subTS}$ equivalent to the first and the second of the hyperbolic planes constituting the subTS manifold. Yet the role of the two subalgebras $Solv^{I,II}_{[1,2]}$ is not symmetric within the full solvable Lie algebra $Solv_{[2,2+q]}$. Indeed as we see from the MC the extra generators dual to $e^{5,i}$ and $e^{6,i}$ commute to the generator dual to $e^4$ in $Solv^{II}_{[1,2]}$, but none of their commutators produce elements of $Solv^I_{[1,2]}$. On the other hand the extra generators  $e^{5,i}$ and $e^{6,i}$ 
are in a linear representation of $Solv^I_{[1,2]}\times Solv^{II}_{[1,2]}$ as it appears from the last two MC equations in (\ref{maurocartus}). The role of the Paint-Group as exterior automorphism group of the solvable Lie algebra (see \cite{advancio}) is evident in eq.(\ref{maurocartus}).
\paragraph{The Maximal Heisenberg Ideal.} We note that the solvable Lie algebra contains a maximal Heisenberg ideal $\mathrm{Heis}$ of dimension $2q+1$ composed by the generators
respectively dual to $e^4$ and $e^{5,i},e^{6,i}$. Indeed setting
$e^1=e^2=e^3=0$ we obtain the closed MC equations:
\begin{equation}\label{eisenbergo}
\mathrm{Heis}_{2q+1} \Leftrightarrow\left\{\begin{array}{lcl}
 \mathrm{d}e^4 &=& 
 \frac{1}{\sqrt{2}}\sum^{q}_{i=1} \, e^{5,i}\wedge e^{6,i} \\
 \mathrm{d}e^{5,i} & = & 0 \\
 \mathrm{d}e^{6,i} & = & 0 \\
\end{array}\right.
\end{equation}
\paragraph{The Maximal Abelian Ideal.} On the other hand the maximal abelian ideal $\mathcal{AI}_{max}$ of dimension $q+2$, exactly equal to one-half of the total dimension of the solvable group and, hence, of the differentiable symmetric space $\mathcal{M}^{[2,q]}_{CV}$, can be singled out, directly from eq. (\ref{maurocartus})   by means of the following positions:
\begin{equation}\label{carnoldo}
  e^1\, = \, 0 \quad ; \quad e^2 \, = \, 0 \quad ; \quad e^{6,i} \, = \, 0
\end{equation}
So doing we obtain
\begin{equation}\label{premaximabidea}
\mathcal{AI}_{max} \Leftrightarrow\left\{\begin{array}{lcl}
 \mathrm{d}e^3 & = & 0 \\
 \mathrm{d}e^4 &=& 0\\
 \mathrm{d}e^{5,i} & = & 0 \\
\end{array}\right.
\end{equation}
\subsection{The symplectic embedding}
The symplectic embedding of the group $\mathrm{U=SO(2,2+q)}$ is on one side the consequence of the above mentioned matter of fact that $\mathcal{SK}_{3+q}$ as defined in eq.(\ref{speckal}) is a special K\"ahler manifold and on the other from a general scheme of symplectic embedding of all the groups $\mathrm{ST[m,n]}\equiv\mathrm{SL(2,\mathbb{R})} \times \mathrm{SO(m,n)}$ that was described in the 1995 Trieste Spring lectures given by one of us \cite{mylecture}. 
\par
Let us start from the Special K\"ahler view-point. The manifold $\mathcal{SK}_{3+q}$
has real dimension $2+2(2+q)$ and hence complex dimension $\bar{n}\, = \, 3+q$. Naming $\hat{z}=\hat{\mathbf{z}}^i$ the $\bar{n}$ complex coordinates of the manifold, according with the general theory of Special K\"ahler geometry of the local type (see its review in \cite{pgtstheory} or in the book \cite{advancio}), there necessarily exists a flat symplectic vector-bundle of dimension $2\bar{n}+2\, = \, 8 + 2\,q$  whose holomorphic sections:
\begin{equation}\label{careno}
  \mho(\hat{\mathbf{z}}) \, = \, \left\{X^\Lambda(\hat{\mathbf{z}}) \, , \,F_\Sigma(\hat{\mathbf{z}})\right\} \quad ; \quad \Lambda,\Sigma \, = \, 1,2,\dots, \bar{m} \, = \, \bar{n}+1 \,=\, 4+q
\end{equation}
provide the K\"ahler potential in the form\footnote{In  formulae (\ref{careno}) and (\ref{kallerone}), the coordinates are generically named $\widehat{\mathbf{z}}$ which denote all the complex coordinates of the Special K\"ahler manifold. In the case of interest to us where, as a differentiable manifold, the Special K\"ahler manifold is the direct product of a hyperbolic plane with a Calabi Vesentini manifold, we have $\widehat{\mathbf{z}} = (S, \mathbf{z})$ where $S$ is the complex coordinate on the upper complex plane ($\mathrm{Im}(S) >0$) and $\mathbf{z}$ the complex coordinate of the CV manifold to be discussed in the next pages.}:
\begin{equation}\label{kallerone}
 \mathcal{ K}(\,\widehat{\mathbf{z}},\widehat{\bar{\mathbf{z}}}\,) \, = \, - \, \log \,\left[ \mathit{i} \, \mho^\dagger(\,\widehat{\bar{\mathbf{z}}}\,) \,
 \left(
 \begin{array}{c|c}
  \mathbf{0} & \mathbf{1} \\
  \hline
   -\mathbf{1} & \mathbf{0} \\
   \end{array}
   \right)\, \mho(\mathbf{z}) \right] \, = \, \log \left(\,\bar{X}^\Lambda(\,\widehat{\bar{\mathbf{z}}}\,)\, F_\Lambda(\,\widehat{\mathbf{z}}\,)- \bar{F}_\Sigma(\,\widehat{\bar{\mathbf{z}}}\,)\, X^\Sigma(\,\widehat{\mathbf{z}}\,) \right)
\end{equation}
The K\"ahler manifold $\mathcal{SK}_{3+q}$ admits:
\begin{equation}\label{uskgroup}
  \mathrm{U}_{\mathcal{SK}} \, \equiv \, \mathrm{SL(2,\mathbb{R})} \times \mathrm{SO(2,2+q)}
\end{equation}
as group of isometriees.
Compatibility with the Special Geometry structure requires the existence of a $(8+2q)$-dimensional symplectic 
representation of such a group that is named the $\mathbf{W}$ representation. In other words,   a symplectic embedding  of the isometry group 
$\mathrm{U}_{\mathcal{SK}}$ 
\begin{equation}
  \mathrm{U}_{\mathcal{SK}} \mapsto \mathrm{Sp(8+2\,q, \mathbb{R})}
\label{sympembed}
\end{equation}
necessarily exists such that for each element $\xi \in \mathrm{U}_{\mathcal{SK}}$ we have its representation by means of a suitable real 
symplectic matrix: 
\begin{equation}
  \xi \mapsto \Lambda_\xi \equiv \left( \begin{array}{cc}
     A_\xi & B_\xi \\
     C_\xi & D_\xi \
  \end{array} \right)
\label{embeddusmatra}
\end{equation}
satisfying the defining relation (in terms of the symplectic antisymmetric metric $\mathbb{C}$):
\begin{equation}
  \Lambda_\xi ^T \, \underbrace{\left( \begin{array}{cc}
     \mathbf{0}_{n \times n}  & { \mathbf{1}}_{n \times n} \\
     -{ \mathbf{1}}_{n \times n}  & \mathbf{0}_{n \times n}  \
  \end{array} \right)}_{ \equiv \, \mathbb{C}} \, \Lambda_\xi = \underbrace{\left( \begin{array}{cc}
     \mathbf{0}_{n \times n}  & { \mathbf{1}}_{n \times n} \\
     -{ \mathbf{1}}_{n \times n}  & \mathbf{0}_{n \times n}  \
  \end{array} \right)}_{\mathbb{C}}
\label{definingsympe}
\end{equation}
which implies the following relations on the $(4+q) \times (4+q)$ blocks:
\begin{eqnarray}
A^T_\xi \, C_\xi - C^T_\xi \, A_\xi & = & 0 \nonumber\\
A^T_\xi \, D_\xi - C^T_\xi \, B_\xi& = & \mathbf{1}\nonumber\\
B^T_\xi \, C_\xi - D^T_\xi \, A_\xi& = & - \mathbf{1}\nonumber\\
B^T_\xi \, D_\xi - D^T_\xi \, B_\xi & =  & 0 \label{symplerele}
\end{eqnarray}
Under an element of the isometry group the symplectic section $\mho$ of Special Geometry (\ref{careno}) transforms as follows:
\begin{equation}
\mho\left( \xi \, \cdot \, z\right) \, = \, \Lambda_\xi \, \mho\left ( z \right )
\end{equation} 
As we said the symplectic embedding of the group:
\begin{equation}\label{USKaletto}
  \mathrm{U}_{\mathcal{SK}}\, = \,\mathrm{SL(2,\mathbb{R})} \times \mathrm{SO(2,2+q)}
\end{equation}
follows from the identification of its appropriate $\mathbf{W}$-representation, according with the general scheme outlined in  \cite{mylecture}, namely:
\begin{equation}\label{binasco}
  \mathbf{W}-\text{rep} \, = \, \left(2 \mid 4+q\right)
\end{equation}
is the bi-fundamental of the two groups. Explicitly let:
\begin{equation}\label{etatdefi}
  \eta_t \, = \, \left(
\begin{array}{cc|c|cc}
 0 &  0 & 0 & 0 &  1 \\
 0 &  0 & 0 & 1 &  0 \\
 \hline
 \mathbf{0}_{q\times 1} &  \mathbf{0}_{q\times 1} & \mathbf{1}_{q \times q} & \mathbf{0}_{q\times 1} & \mathbf{0}_{q\times 1}  \\
 \hline
  0 &  1 & 0 & 0 &  0 \\
 1 &  0 & 0 & 0 &  0 \\
\end{array}
\right)
\end{equation}
be the form of the invariant metric defining the group $\mathrm{SO(2,2+q)}$, which is appropriate to the triangular embedding (see section 3 of \cite{pgtstheory} eq.(3.45)):
\begin{equation}\label{invametra4piuq}
  \mathrm{SO(2,2+q)} \, = \, \left\{L \in \mathrm{GL(4+q,\mathbb{R})} \, \mid \, 
  L^T \cdot \eta_t \cdot L \, = \, \eta_t\right\}
\end{equation}
Then the symplectic embedding of a generic $\mathrm{SO(2,2+q)}$-group element $L$ is given by:
\begin{equation}\label{fucecchioA}
  \forall L \in \mathrm{SO(2,2+q)} \, \stackrel{\iota_{sym}}{\hookrightarrow} \, \left(
  \begin{array}{c|c}
  L & \mathbf{0}_{(4+q)\times (4+q)}\\
  \hline
  \mathbf{0}_{(4+q)\times (4+q)} & \left(L^T\right)^{-1}\\
  \end{array}
   \right)\in \mathrm{Sp(8+2q,R)}
\end{equation}
while the symplectic embedding of a generic element of the group
$\mathrm{SL(2,\mathbb{R})}$ is as follows:
\begin{equation}\label{fucecchioB}
  \forall \left( \begin{array}{cc}
                   a & b \\
                   c & d 
                 \end{array}
  \right) \in \mathrm{SL(2,\mathbb{R})} \, \stackrel{\iota_{sym}}{\hookrightarrow} \, \left(
  \begin{array}{c|c}
  a \,\mathbf{1}_{(4+q)\times (4+q)}  & b \, \eta_t \\
  \hline
  c \, \eta_t & d\, \mathbf{1}_{(4+q)\times (4+q)}
  \end{array}
   \right)\in \mathrm{Sp(8+2q,R)}
\end{equation}
\subsubsection{The kinetic matrix $\mathcal{N}$}
Given the symplectic embedding explicitly displayed in eq.s (\ref{fucecchioA}-\ref{fucecchioB}), following the steps reported in \cite{mylecture}
and utilizing the master formula derived by Gaillard and Zumino in 1981 \cite{Gaillard:1981rj} (see eq.(85) of \cite{mylecture}) one obtains the so named 
$\mathcal{N}^{\Lambda\Sigma}(z)$ complex symmetric matrix which in supergravity theories provides the kinetic metric of the vector fields (see eq.(13) of \cite{mylecture}) and transforms, under the action of the group $\mathrm{U}$ by means 
of linear fractional transformations: 
\begin{equation}\label{canobbio}
  \forall \xi \in \mathrm{U} \quad : \quad \mathcal{N}(\xi.z) \, = \
  \left(C_\xi \, + \, D_\xi\,\mathcal{N}(z) \right)\cdot\left(A_\xi \, + \, B_\xi\,\mathcal{N}(z) \right)^{-1} 
\end{equation}
Aside from its physical interpretation in supergravity theories, the matrix $\mathcal{N}(z)$ provides a canonical embedding of the special K\"ahler manifold
$\mathcal{SK}_{3+q}$ into the Siegel half-plane of genus $g= 4+q$:
\begin{equation}\label{fringuello}
  \frac{\mathrm{SL(2,\mathbb{R})}}{\mathrm{SO(2)}}\times \frac{\mathrm{SO(2,2+q)}}{\mathrm{SO(2)\times SO(2+q)}}\, \hookrightarrow \, \frac{\mathrm{Sp(8+2q,\mathbb{R})}}{\mathrm{U(1)\times SU(4+q)}}
\end{equation}
Indeed a point in the Siegel half-plane is a symmetric complex matrix with positive definite imaginary part. In the recent past Barbaresco and Cabanes \cite{ondoso,radarone,barbaresco3} have utilized
the imaginary part of the Siegel plane symmetric matrix as a box where to map 
radar-data in the species of their covariance matrix,  and Gibbs probability distributions {\`a} la Souriau defined over the Siegel plane as a statistical tool to analyse such data\footnote{In one of the quoted publications, Barbaresco et al note also that the covariance matrix datum can be complemented with additional Doppler data which they associate with an additional Poincaré plane. If we correctly interpret such a statement, which is not obvious to us,  it is suggested that the entire Special K\"ahler manifold is utilized to represent such kind of radar data.}. The $\mathcal{N}(z)$ on the special K\"ahler manifold $\mathcal{SK}_{3+q}$ is just such an object, simply not fully general as a $(4+q)\times (4+q)$ matrix rather endowed  with a restricted form parameterized by $4+2q+2$-parameters (the solvable coordinates). The explicit structure of the 
$\mathcal{N}(z)$ matrix (see eq.(106) of \cite{mylecture}) is:
\begin{equation}\label{corleone}
  \mathcal{N} \, = \, \mathit{i} \, \text{Im}S \, \mathbb{L}(\boldsymbol{\Upsilon})\,\mathbb{L}^T(\boldsymbol{\Upsilon})\, 
  + \, \text{Re}S \, \eta_t
\end{equation}
where $S$ is the standard complex coordinate in the upper Poincaré-Lobachevsky plane, corresponding to the first factor in the direct product (\ref{speckal}). If we reduce the manifold to its Calabi-Vesentini component, by setting $S=\mathit{i}$, we reduce $\mathcal{N}$ to its imaginary part which provides a challenging model-type for time-sequence data according to the quoted radar-example and for other data that can be arranged into symmetric matrices. Other uses of the CV-manifolds,  of course, are  not contradictory with the above observations.  
\section{Calabi Vesentini coordinates of type I and type II and moment maps}
\label{baracco}
As we have emphasized in previous sections there are two types of \textbf{abelian structures} we are interested in for different purposes and that coexist on CV manifolds:
\begin{enumerate}
  \item The \textbf{non-compact abelian} structure where the actions are the moment maps of the maximal abelian ideal and the complementary submanifold $\boldsymbol{\mathcal{Q}}^{2+q}\simeq \mathbb{R}^{2+q}$ is the translation group manifold of such an ideal, whose transformations are all non-compact isometries of $\mathcal{M}^{[2,2+q]}$.
  \item The \textbf{compact abelian structure} where, naming $\nu$ the integer part of $q/2$,  we have $2+\nu$ actions identified with the moment maps of the $2+\nu$ generators in a basis $\mathcal{C}_i$ of the compact Cartan subalgebra  $\boldsymbol{\mathcal{C}}\subset\mathbb{H}\subset \mathbb{U}$ inside the isometry Lie algebra, while the  remaining $q-\nu$ actions are non linear functions of the Killing moment-maps. Correspondingly the complementary submanifold $\boldsymbol{\mathcal{Q}}^{2+q}\simeq \boldsymbol{\mathcal{T}}^{2+q}$ is a $(2+q)$-torus which includes the group manifold of the unique $(2+\nu)$-torus contained in $\mathrm{U}$ extended with another $(q-\nu)$-torus whose angles are not isometry parameters.
\end{enumerate}
Correspondingly to these two \textit{radically different abelian structures} one can find two different bases of complex coordinates that are well-adapted to either one of the two abelian structures and that both are entitled to be named \textit{Calabi Vesentini coordinates} since both follow from the complexification scheme introduced in the classical paper by Calabi-Vesentini of 1960 \cite{cvclassic}, summarized in the 1995 Trieste Lectures by one of us \cite{mylecture}.
\par
We shall name \textbf{Type II CV-coordinates} those that are well adapted to the \textbf{non-compact abelian structure} and \textbf{Type I CV-coordinates} those that are well adapted to the \textbf{compact abelian structure}. This numbering is for chronological reasons, since it is the Type I that appears explicitly in some formulae of \cite{cvclassic} (see for instance (3.15), page 497). Furthermore 
the Type I CV coordinates were utilized independently from and previously than by Calabi-Vesentini, also  by the Chinese Mathematician L.K. Hua in his 1946 paper \cite{Hua1946}, later reviewed in his 1963 book \cite{Hua1963}. This early literature 
is anchored to a vision, manifested also in the titles of the publications, related with Cartan's 1935 work on complex bounded domains \cite{Cartan1935} which emphasizes the complex analysis aspects more  than the structural algebraic ones and introduces a corresponding nomenclature strictly tied to the utilized coordinate system which turns out of very difficult immediate interpretation in more intrinsic algebraic approaches based on Lie algebra structures \footnote{For instance the very symmetric space, looked at as a complex domain, is named a \textit{Lie Ball}, the
exponential of the K\"ahler potential is christened the \textit{Bergman denominator}, 
the actions that are moment-maps of compact Cartan generators are named \textit{ azimuthal Cartan moments}, while those additional actions that are non-linear functions of the Killing moment maps are named \textit{structural magnitudes}. Furthermore one has \textit{plane leaves} associated with pairs of Type I CV complex coordinates that define an $\so(2)$ rotation and \textit{orphan axes} that remain after the maximal number of planes has been formed in an odd dimension n = 2m + 1 
} although it is of great value in the singling out and in the construction of the compact abelian structures. 
\subsection{Type II CV coordinates and the maximal abelian ideal}
\label{CVtipodue}
In order to put into evidence the role of the maximal abelian ideal and arrive at a K\"ahler potential that depends only on the imaginary parts of $2+q$ complex coordinates
and, at the same time, is derived from the structural general formulae of Special K\"ahler Geometry we need to go through the following 4 steps:
\begin{enumerate}
  \item Change the definition of the solvable coordinates by performing the exponential $\Sigma$ map:  
  \begin{equation}\label{sigmamapnew}
    \Sigma \, : \, Solv_{[2,2+q]} \, \Leftrightarrow \, \mathcal{S}_{[2,2+q]}
  \end{equation}
  in such a way that the exponential of the maximal abelian ideal sits on the extreme left before the exponentials of the other generators not belonging to the ideal.
  \item Adapt the Calabi-Vesentini procedure, summarized  in section 6.1.2 of \cite{mylecture}, with particular attention at the formulae (212-216), to the case where the $\mathrm{SO(2,2+q)}$-invariant metric is not block diagonal, rather it is $\eta\, = \, \eta_t$, consistent with the upper triangular embedding of the  solvable Lie subgroup. 
 \item Find the corresponding analogue of eq.(217) of \cite{mylecture} for the structure of the upper half of the holomorphic symplectic section $X^\Lambda(\mathbf{z})$, ($\mathbf{z}$ being the complex coordinates of the Calabi Vesentini manifolds) which is recalled above in eq.(\ref{careno}), the lower half being, just as it is standard in most supergravity literature and it is discussed in \cite{mylecture}:     
     \begin{equation}\label{onerac}
       F_\Lambda(S,\mathbf{z}) \, = \, S \, \times \, \eta_{\Lambda\Sigma} \, X^\Sigma(\mathbf{z})
     \end{equation}
     where $S$, the dilaton/Kalb-Ramond field of string/supergravity models, mathematically is nothing else but the complex coordinate in the upper complex plane realizing the symmetric space $\mathrm{SL(2,\mathbb{R})/SO(2)}$.
  \item Find the exact correspondence between the new solvable coordinates $\mathbf{W}$ and the complex coordinates $\mathbf{z}$.   
\end{enumerate}
The above programme has been implemented as follows.
\subsubsection{Well adapted solvable coset representative.} For the Calabi-Vesentini manifolds, the solvable Lie group element is now defined as follows:
\begin{equation}\label{krollus}
  \mathbb{L}\left(\mathbf{W}\right)\, = \, \underbrace{\exp\left[\sum^{q}_{i=1}\,w_{5,i}\, T^{5,i} + w_3\,T^3 + w_4\,  T^4\right]}_{\text{max. ab. id.}} \cdot \underbrace{\exp\left[w_1\,T^1 + w_2 \, T^2\right]}_{\text{non comp. Cartan}}\cdot\underbrace{\exp\left[\sum^{q}_{i=1}\,w_{6,i}\, T^{6,i}\right]}_{\text{compl. ab. subalg.}}
\end{equation}
where the explicit form of the generators for the smallest case with non-trivial TS projection, namely $q=2$, are listed in table \ref{baldovino} in appendix \ref{so2com4}. The explicit form of the coset representative for the case $q=2$ is displayed in eq.(\ref{newsolvab}). By means of an appropriate super-routine named
\textbf{mainCVBseries} of the background 
MATHEMATICA background NoteBook named \textbf{NonCompSymSpacNeuNet26Up2.nb} constructed by one of us [P.F.] all items pertaining to the Lie Algebras $\so(2,2+q)$ can be explicitly constructed for all values of $q$.
\subsubsection{The Calabi-Vesentini complex structure method} Given the coset representative $\mathbb{L}_\Sigma^\Lambda(\mathbf{W})$ in the fundamental representation of $\mathrm{SO(2,2+q)}$, ($\Lambda,\Sigma\, = \, 1,\dots,4+q$), the Calabi-Vesentini complexification method (see \cite{mylecture}) is based on considering the eigenvalues and eigenvectors of the $\mathrm{SO(2)}$ generator $X_c$ which,  independently from the utilized $\eta$ tensor, are always of the following form:
\begin{equation}\label{crisalide}
  \text{Eigenvalues of $X_c$} \, =\, \left\{\mathit{i},-\mathit{i},\underbrace{0,\dots,0}_{2+q}\right\} \quad ;\quad \text{Corresponding Eigenvectors:}\quad\vec{v}_{a} \quad ; \quad a=1,\dots, 4+q
\end{equation}
In the case $q=2$ which we utilize as model for all the others we have:
\begin{equation}\label{vectorini}
  \vec{v}_{a} \, = \, \left(
\begin{array}{cccccc}
 -1 & -i & 0 & 0 & i & 1 \\
 -1 & i & 0 & 0 & -i & 1 \\
 1 & 0 & 0 & 0 & 0 & 1 \\
 0 & 1 & 0 & 0 & 1 & 0 \\
 0 & 0 & 0 & 1 & 0 & 0 \\
 0 & 0 & 1 & 0 & 0 & 0 \\
\end{array}
\right)
\end{equation}
each row of the  matrix (\ref{vectorini}) being the eigenvector corresponding to one of the eigenvalues in the order presented in eq.(\ref{crisalide}). The first two rows correspond to the eigenvalues $\pm \mathit{i}$ while all the others correspond to the $0$-eigenvalues. Following Calabi-Vesentini method, one introduces a complex $(4+q)$-vector $\Psi^\Lambda(\mathbb{W})$ defined as follows:
\begin{equation}\label{posai}
  \Psi^\Lambda(W) \, = \, \sum_{\Sigma} \, \mathbb{L}_\Sigma^\Lambda(\mathbf{W}) \,\vec{v}_1^\Sigma
\end{equation}
and verifies that:
\begin{eqnarray}
 \Psi^\Lambda(\mathbf{W})\,\eta_{\Lambda\Sigma} \Psi^\Sigma(\mathbf{W}) &=& 0 \label{barecco1} \\
  {\overline{\Psi}}^\Lambda(\overline{\mathbf{W}})\,\eta_{\Lambda\Sigma} \Psi^\Sigma(\mathbf{W}) &=& \text{const} \, \neq \, 0 \label{barecco2}
\end{eqnarray}
where, as usual, the bar means complex conjugation. In the case $q=2$, the explicit result is the following:
\begin{equation}\label{rammarico}
  \Psi\, = \, \left(
\begin{array}{l}
 -\frac{1}{8} i e^{w_1} \left(-2 i \left(w_{5,1}^2+2 i w_{6,1}
   w_{5,1}+w_{5,2}^2+2 i w_{5,2} w_{6,2}+4\right)+\sqrt{2} w_3
   \left(w_{6,1}^2+w_{6,2}^2+4\right)-4 i w_3 w_4-4 \sqrt{2}
   w_4\right) \\
 -\frac{1}{4} e^{w_2} \left(2 \sqrt{2} w_4+i
   \left(w_{6,1}^2+w_{6,2}^2+4\right)\right) \\
 -\frac{w_{5,1}+i w_{6,1}}{\sqrt{2}} \\
 -\frac{w_{5,2}+i w_{6,2}}{\sqrt{2}} \\
 -\frac{1}{2} e^{-w_2} \left(\sqrt{2} w_3-2 i\right) \\
 e^{-w_1} \\
\end{array}
\right)
\end{equation}
and it has a structure that immediately reveals its own generalization to all values of $q$ by Paint Group covariance. Indeed treating $w_{5,i}$ and $w_{6,i}$ as the components of two vectors, respectively $\mathbf{w}_5$ and $\mathbf{w}_6$, eq.(\ref{rammarico}) admits the obvious Paint covariant transcription  displayed below:
\begin{equation}\label{cariolarotta}
\Psi^\Lambda(\mathbf{W})\, = \,  \left(
\begin{array}{l}
 -\frac{1}{8} i e^{w_1} \left(\sqrt{2} w_3
   \left(\mathbf{w}_6^2+4\right)-4 i w_3 w_4-4 \sqrt{2} w_4-2 i
   \left(\mathbf{w}_5^2+2 i \mathbf{w}_5.\mathbf{w}_6+4\right)\right)
   \\
 -\frac{1}{4} e^{w_2} \left(2 \sqrt{2} w_4+i
   \left(\mathbf{w}_6^2+4\right)\right) \\
 -\frac{w_{5,i}+i w_{6,i}}{\sqrt{2}} \\
 -\frac{1}{2} e^{-w_2} \left(\sqrt{2} w_3-2 i\right) \\
 e^{-w_1} \\
\end{array}
\right)
\end{equation}
\subsubsection{Parametric solution of the holomorphic constraint in the Type II case.} Following the logic discussed in \cite{mylecture} what remains to be done is partly algorithmic partly 
a matter of educated inventivness. Indeed, recalling  eq.(\ref{barecco1}) one tries to solve such holomorphic constraint in terms of a holomorphic  vector $X^\Lambda(\mathbf{z})$ dependent on $2+q$ complex variables:
\begin{equation}\label{rondella}
  z^i \, = \, u^i\, + \, \mathit{i} \, v^i
\end{equation}
in such a way that:
\begin{enumerate}
\item The constraint (\ref{barecco1} is always satisfied by $X^\Lambda(\mathbf{z})$:  
\begin{equation}\label{fagiolino}
X^\Lambda (\mathbf{z}) \, \eta_{\Lambda\Sigma} \, X^\Sigma (\mathbf{z}) \, = \, 0
\end{equation}
\item The constraint (\ref{barecco2}) is satisfied by:
\begin{equation}\label{cristoforone}
  \Psi^\Lambda(\mathbf{z}) \, \equiv \, \frac{1}{\sqrt{\mathrm{N}(\mathbf{z},\overline{\mathbf{z}})}}\times X^\Lambda(\mathbf{z}) \quad ; \quad 
  \mathrm{N}(\mathbf{z},\overline{\mathbf{z}})\, = \, \overline{X}^\Lambda(\overline{\mathbf{z}})\,\eta_{\Lambda\Sigma}\, {X}^\Sigma(\mathbf{z})
  \end{equation}
\item Upon a suitable change of coordinates:
\begin{equation}\label{wintoz}
  \mathbf{W} \, \to \, \mathbf{W}(\mathbf{u},\mathbf{v})
\end{equation}
$\Psi^\Lambda(\mathbf{z})$ as defined in eq.(\ref{cristoforone}) and as calculated 
in eq.(\ref{posai}) do coincide.
\end{enumerate}
Once the above outlined programme is realized in all of its points, the K\"ahler potential of the $\mathrm{U/H}$ K\"ahler symmetric space is provided by:
\begin{equation}\label{ferdinando}
  \mathcal{K}(\mathbf{z},\overline{\mathbf{z}})\, = \, \textit{const} \times \log\left[\overline{X}^\Lambda(\overline{\mathbf{z}})\,\eta_{\Lambda\Sigma}\, {X}^\Sigma(\mathbf{z})\right]
\end{equation}
We have realized the programme obtaining the following general structure for the holomorphic section in terms of $q+2$ complex coordinates:
\begin{equation}\label{piromane}
  X^\Lambda(\mathbf{z})\, = \, \left(
\begin{array}{c}
 -\sum _{i=1}^q z_i^2-2 z_{q+1} z_{q+2} \\
 \sqrt{2} z_{q+2} \\
 \sqrt{2} z_i \\
 \sqrt{2} z_{q+1} \\
 1 \\
\end{array}
\right)
\end{equation}
and the following coordinate transformation expressing the $4+2q$ solvable coordinates $\mathbf{W}$ in terms of the real $u^i$ and imaginary $v^i$ parts of $q+2$ complex coordinates $z^i$: 
{\large\begin{equation}\label{transoCos} 
\begin{array}{|c|c|}
\hline
\null & \null \\
 w_1\to \log \left(\sqrt{-\sum_{i=1}^q \, v_i^2-2 v_{q+1} v_{q+2}}\right) & w_2\to
   -\log \left(\frac{\sqrt{2} v_{q+1}}{\sqrt{-\sum_{i=1}^q \, v_i^2-2 v_{q+1} v_{q+2}}}\right) \\
   \null & \null \\
 \hline
 \null & \null \\
 w_4\to -2 u_{q+2} & w_3\to -2 u_{q+1}   \\
 \null & \null \\
 \hline
 \null & \null \\
  w_{5,i}\to -2 u_i &
  w_{6,i}\to \mp\frac{2 v_i}{\sqrt{-\sum_{i=1}^q \, v_i^2-2 v_{q+1} v_{q+2}}} \\
  \null & \null \\
 \hline
\end{array}
\end{equation}
}
As one sees from eq.(\ref{transoCos}) the real parts $u^i$ are associated in one-to-one correspondence with the parameters of the maximal abelian ideal $\mathcal{AI}_{max}$.
This goes hand to hand with the structure of the K\"ahler potential, defined by eq. 
(\ref{ferdinando}) which takes the following explicit expression:
\begin{equation}\label{fronellino}
  \mathcal{K}_{CV}(\mathbf{v})\, = \, \text{const} \times \log\left[ - \,\sum_{i=1}^q v_i^2 - 2v_{q+1} \, v_{q+2}   \right]
\end{equation}
 We are therefore in the case discussed in section \ref{metraturacontatta}, in particular from eq.(\ref{complessini}) to eq.(\ref{fradiavolo}). 
 \par
 An obvious question that immediately arises from inspection of equations (\ref{transoCos}-\ref{fronellino}) is the following. The quadratic expression:
 $ - \,\sum_{i=1}^q v_i^2 - 2v_{q+1} \, v_{q+2} $ 
 should be positive in order for the solvable coordinates $\mathbf{W}$ to be real and
 for the K\"ahler potential to have the same property. Indeed as we illustrate in the next subsection \ref{mariosecta} the constraint:
 \begin{equation}\label{boaconstrictor}
   - \,\sum_{i=1}^q v_i^2 - 2v_{q+1} \, v_{q+2} \, > \, 0
 \end{equation}
is always satisfied by having one of the two coordinates $v_{q+1},v_{q+2}$ positive and the other negative and of appropriate magnitude so as to satisfy (\ref{boaconstrictor}). 
\subsubsection{The metric and non-compact abelian structure}
\label{mariosecta}
We start from the parametrization of the coset representative of CV manifolds defined in eq.(\ref{krollus}) and explicitly displayed for the case $q=2$ in eq.(\ref{newsolvab}). As before we use such case whose algebraic structures are explicitly provided in appendix \ref{so2com4}, as a paradigm for all the others. Indeed, just as before, all explicit results we display here are immediately generalized to all values of $q$ by using Paint Group covariance. 
\par
As a first step, from the Lie group element $\mathbb{L}(\mathbf{W})$ shown in eq. (\ref{newsolvab}) we calculate the left-invariant $1$-form:
\begin{equation}\label{canovaccio}
  {\boldsymbol{\Theta}}(\mathbf{W}) \, \equiv \, {\mathbb{L}}^{-1}(\mathbf{W})\, 
  \mathrm{d}{\mathbb{L}}(\mathbf{W})
\end{equation}
and projecting the latter on the coset generators we obtain the vielbein:
\begin{equation}\label{fillinuovi}
  V^A \, = \, \text{Tr} \left[{\boldsymbol{\Theta}}(\mathbf{W})\cdot K^A\right] \quad ; \quad A\, = \, 1,\dots ,4+2\, q
\end{equation}
In our $q=2$ case we obtain:
\begin{equation}\label{8gambe}
  V^A \, = \,\left(
\begin{array}{l}
 \sqrt{2} \,\text{dw}_1 \\
 \sqrt{2} \,\text{dw}_2 \\
 \frac{e^{w_2-w_1}}{\sqrt{2}}\,\text{dw}_3 \\
 \frac{1}{8} e^{-w_1-w_2} \left(-4 e^{w_2} \left(
   w_{6,1}\,\text{dw}_{5,1} \, +\,  w_{6,2} \,\text{dw}_{5,2}\right)-\sqrt{2} e^{2
   w_2} \left(w_{6,1}^2+w_{6,2}^2\right)\, \text{dw}_3\, +\, 4 \sqrt{2}\,
   \text{dw}_4\right) \\
 \frac{1}{2} e^{-w_1} \left( \sqrt{2}
   \, \text{dw}_{5,1} \, +\, e^{w_2} w_{6,1}\, \text{dw}_3 \right) \\
  \frac{1}{2} e^{-w_1} \left( \sqrt{2}
   \, \text{dw}_{5,2} \, +\, e^{w_2} w_{6,2}\, \text{dw}_3 \right)\\
 \frac{1}{\sqrt{2}}\,\left(\text{dw}_{6,1}\, +\,  w_{6,1} \text{dw}_2\right) \\
 \frac{1}{\sqrt{2}}\,\left(\text{dw}_{6,2}\, +\,  w_{6,2} \text{dw}_2\right) \\
\end{array}
\right) 
\end{equation}
and the above result is immediately generalized to all values of $q$ by Paint covariance, namely by writing:
\begin{equation}\label{moltegambe}
  V^A \, = \,\left(
\begin{array}{l}
 \sqrt{2} \,\text{dw}_1 \\
 \sqrt{2} \,\text{dw}_2 \\
 \frac{e^{w_2-w_1}}{\sqrt{2}}\,\text{dw}_3 \\
 \frac{1}{8} e^{-w_1-w_2} \left(-4\, e^{w_2} \,
   \mathbf{w}_{6}\cdot\text{d}\mathbf{w}_{5} \,- \,\sqrt{2}\, e^{2
   w_2} \, \mathbf{w}_{6}\cdot \mathbf{w}_{6}\, \text{dw}_3\, +\, 4 \sqrt{2}\,
   \text{dw}_4\right) \\
 \frac{1}{2} e^{-w_1} \left( \sqrt{2}
   \, \text{dw}_{5,i} \, +\, e^{w_2} w_{6,i}\, \text{dw}_3 \right) \\
 \frac{1}{\sqrt{2}}\,\left(\text{dw}_{6,i}\, +\,  w_{6,i} \text{dw}_2\right) \\
\end{array}
\right)  \quad ; \quad i=1, \dots , q
\end{equation}
The quite different structure of eq.(\ref{moltegambe}) with respect to that of eq.s(\ref{mc1formeA},\ref{mc1formeB}) is due to the different prescription of exponentiation from the solvable Lie algebra to the solvable Lie group. Both the $\boldsymbol{\Upsilon}$ and the $\mathbf{W}$ are solvable coordinates and realize the diffeomorphism of the symmetric space with the solvable Lie group and hence with $\mathbb{R}^{4+2q}$ yet their different structures put into evidence different aspects of the CV geometry. The $\mathbf{W}$ coordinates put into evidence the maximal abelian ideal and lead to an easy derivable Darboux expression of the K\"ahler $2$-form $\boldsymbol{\mathcal{K}}$.
\paragraph{The metric and the K\"ahler potential}
We calculate next the Riemannian metric and the K\"ahler $2$-form and from now on we 
confine ourselves to the $q=2$ case as a simple but complete exemplification.
\par
The line-element takes the following explicit appearance:
\begin{alignat}{3}\label{dsq8}
  &ds^2_{2,2} &\quad = \quad & \frac{1}{2} \left(\text{dw}_2
   w_{6,1}+\text{dw}_{6,1}\right){}^2+\frac{1}{4} e^{-2 w_1}
   \left(\text{dw}_3 e^{w_2} w_{6,1}+\sqrt{2}
   \text{dw}_{5,1}\right){}^2+\frac{1}{2} \left(\text{dw}_2
   w_{6,2}+\text{dw}_{6,2}\right){}^2\nonumber\\
   &\null &\null & +\frac{1}{4} e^{-2 w_1}
   \left(\text{dw}_3 e^{w_2} w_{6,2}+\sqrt{2}
   \text{dw}_{5,2}\right){}^2\nonumber\\
   &\null &\null & +\frac{1}{64} e^{-2 w_1-2 w_2} \left(-4
   e^{w_2} \left(\text{dw}_{5,1} w_{6,1}+\text{dw}_{5,2}
   w_{6,2}\right)-\sqrt{2} \text{dw}_3 e^{2 w_2}
   \left(w_{6,1}^2+w_{6,2}^2\right)+4 \sqrt{2}
   \text{dw}_4\right){}^2\nonumber\\
   &\null &\null & +\frac{1}{2} \text{dw}_3^2 e^{2 w_2-2 w_1}+2\,
   \text{dw}_1^2+2 \,\text{dw}_2^2
\end{alignat}
If we rename the coordinates associated with the maximal abelian ideal as the real parts of the complex coordinates $z^i$ introduced in eq. (\ref{rondella}):
\begin{equation}\label{tadalafil}
  w_3\, \to\,  -2\, u_3,\quad w_4\, \to\,  -2\, u_4,\quad w_{5,1}\, \to\,  -2\, u_1,\quad w_{5,2}\, \to\,  -2\,u_2
\end{equation}
we verify, as it should be the case, that  in the expression (\ref{dsq8}) for the line element, the $u_i$ appear only under derivative as the differentials $\mathrm{d}u_i$. This establishes the isometry under translations $u_i \, \to \, u_i + c_i$. Furthermore the differentials  $\mathrm{d}u_i$ never mix with the differential of the complementary set of coordinates $\{w_{1},w_2,w_{6,1},w_{6,2}\}$. The contribution of the differentials $\mathrm{d}u_i$ to the line element is in the form\footnote{We use the notation $\widehat{\mathcal{F}}_{i,j}$ instead of ${\mathcal{F}}_{i,j}$ because, for simplicity, we omit the factor $1/2$ that appears
in the formula (\ref{kallometrica}).} :
\begin{equation}\label{giraldo}
  ds^2_{2,2} \, = \,  \mathrm{d}u_i \, \widehat{\mathcal{F}}_{i,j}\, \mathrm{d}u_j  \, + \, \text{more}
\end{equation}
where the matrix $\widehat{\mathcal{F}}_{i,j}$ depends only on the remaining coordinates
$\{w_{1},w_2,w_{6,1},w_{6,2}\}$ and has the following explicit expression:
\begin{eqnarray}\label{fufu}
& \widehat{\mathcal{F}}_{i,j} \, = \, &\nonumber\\
&\left(
\begin{array}{cccc}
 e^{-2 w_1} \left(w_{6,1}^2+2\right) & e^{-2 w_1} w_{6,1} w_{6,2} &
   \frac{e^{w_2-2 w_1} w_{6,1} \left(w_{6,1}^2+w_{6,2}^2+4\right)}{2
   \sqrt{2}} & -\sqrt{2} e^{-2 w_1-w_2} w_{6,1} \\
 e^{-2 w_1} w_{6,1} w_{6,2} & e^{-2 w_1} \left(w_{6,2}^2+2\right) &
   \frac{e^{w_2-2 w_1} w_{6,2} \left(w_{6,1}^2+w_{6,2}^2+4\right)}{2
   \sqrt{2}} & -\sqrt{2} e^{-2 w_1-w_2} w_{6,2} \\
 \frac{e^{w_2-2 w_1} w_{6,1} \left(w_{6,1}^2+w_{6,2}^2+4\right)}{2
   \sqrt{2}} & \frac{e^{w_2-2 w_1} w_{6,2}
   \left(w_{6,1}^2+w_{6,2}^2+4\right)}{2 \sqrt{2}} & \frac{1}{8} e^{2
   w_2-2 w_1} \left(w_{6,1}^2+w_{6,2}^2+4\right){}^2 & -\frac{1}{2}
   e^{-2 w_1} \left(w_{6,1}^2+w_{6,2}^2\right) \\
 -\sqrt{2} e^{-2 w_1-w_2} w_{6,1} & -\sqrt{2} e^{-2 w_1-w_2} w_{6,2}
   & -\frac{1}{2} e^{-2 w_1} \left(w_{6,1}^2+w_{6,2}^2\right) & 2
   e^{-2 \left(w_1+w_2\right)} \\
\end{array}
\right)& \nonumber\\
\end{eqnarray}
Next, according with the results of section \ref{baracco}, in particular with eq.(\ref{fronellino}), we assume that the K\"ahler potential is given by:
\begin{equation}\label{kallerofino}
 \mathcal{K}(\mathbf{v}) \, = \,  -\log \left(-v_1^2-v_2^2-2 v_3 v_4\right)
\end{equation}
where $v_i$ are the imaginary parts of the $4$ complex coordinates. According with eq.(\ref{hesso1}) the matrix 
in eq.(\ref{fufu}) should be identical to the Hessian of the K\"ahler potential
(\ref{kallerofino}) that is the following one:
\begin{eqnarray}\label{brandeburgo}
\frac{\partial^2\mathcal{K}}{\partial v_i \partial v_j}  &=&\left(
\begin{array}{cccc}
 \frac{2 \left(v_1^2-v_2^2-2 v_3 v_4\right)}{\left(v_1^2+v_2^2+2 v_3
   v_4\right){}^2} & \frac{4 v_1 v_2}{\left(v_1^2+v_2^2+2 v_3
   v_4\right){}^2} & \frac{4 v_1 v_4}{\left(v_1^2+v_2^2+2 v_3
   v_4\right){}^2} & \frac{4 v_1 v_3}{\left(v_1^2+v_2^2+2 v_3
   v_4\right){}^2} \\
 \frac{4 v_1 v_2}{\left(v_1^2+v_2^2+2 v_3 v_4\right){}^2} & -\frac{2
   \left(v_1^2-v_2^2+2 v_3 v_4\right)}{\left(v_1^2+v_2^2+2 v_3
   v_4\right){}^2} & \frac{4 v_2 v_4}{\left(v_1^2+v_2^2+2 v_3
   v_4\right){}^2} & \frac{4 v_2 v_3}{\left(v_1^2+v_2^2+2 v_3
   v_4\right){}^2} \\
 \frac{4 v_1 v_4}{\left(v_1^2+v_2^2+2 v_3 v_4\right){}^2} & \frac{4
   v_2 v_4}{\left(v_1^2+v_2^2+2 v_3 v_4\right){}^2} & \frac{4
   v_4^2}{\left(v_1^2+v_2^2+2 v_3 v_4\right){}^2} & -\frac{2
   \left(v_1^2+v_2^2\right)}{\left(v_1^2+v_2^2+2 v_3 v_4\right){}^2}
   \\
 \frac{4 v_1 v_3}{\left(v_1^2+v_2^2+2 v_3 v_4\right){}^2} & \frac{4
   v_2 v_3}{\left(v_1^2+v_2^2+2 v_3 v_4\right){}^2} & -\frac{2
   \left(v_1^2+v_2^2\right)}{\left(v_1^2+v_2^2+2 v_3 v_4\right){}^2}
   & \frac{4 v_3^2}{\left(v_1^2+v_2^2+2 v_3 v_4\right){}^2} \\
\end{array}
\right) 
\end{eqnarray}
The quadratic equation:
\begin{equation}\label{quadrequa}
  \frac{\partial^2\mathcal{K}}{\partial v_i \partial v_j} \, = \, \widehat{\mathcal{F}}_{i,j} 
\end{equation}
can be solved in both directions, for $\{w_{1},w_2,w_{6,1},w_{6,2}\}$ in terms of
$\{v_{1},v_2,v_{3},v_{4}\}$, or viceversa. The solution for $\mathbf{W}$ in terms of $\mathbf{v}$ exactly reproduces eq.(\ref{transoCos}), while the solution in the opposite direction is the following one:
{\large
\begin{equation}\label{fraballo}
  \begin{array}{|c|c|}
  \hline
  \null & \null \\
 v_1\to \pm \left(\frac{1}{2} e^{w_1} w_{6,1}\right) & v_2\to \pm
   \left(\frac{1}{2} e^{w_1} w_{6,2}\right) \\
   \null & \null \\
   \hline
   \null & \null \\
 v_3\to \mp \frac{ \, e^{w_1-w_2} }{\sqrt{2}} & v_4\to \pm \frac{e^{w_1+w_2}
   \left(w_{6,1}^2+w_{6,2}^2+4\right)}{4 \sqrt{2}} \\
   \null & \null \\
   \hline
\end{array}
\end{equation}
}
that can be immediately generalized to all values of $q$ by Paint group covariance:
{\large
\begin{equation}\label{fraballone}
  \begin{array}{|c|c|}
  \hline
  \null & \null \\
 v_i\to \pm \left(\frac{1}{2} e^{w_1} w_{6,i}\right) & u_i \to -\ft 12 \, w_{5,i} \\
   \null & \null \\
   \hline
   \null & \null \\
 v_{q+1}\to \mp \frac{ \, e^{w_1-w_2} }{\sqrt{2}} & v_{q+2}\to \pm \frac{e^{w_1+w_2}}{4 \sqrt{2}} \,\left(\mathbf{w}_{6}\cdot\mathbf{w}_{6}+4\right)\\
   \null & \null \\
   \hline
   \null & \null \\
   u_{q+1}\to - \ft 12 \, w_3 & u_{q+2}\to -\ft 12 \, w_4\\
   \null & \null \\
   \hline
\end{array}
\end{equation}
}
Eq.(\ref{fraballone}) is very important since it shows that the entire CV manifold 
represented by the  $4+2q$ solvable coordinates $\mathbf{W}$ spanning $\mathbb{R}^{4+2q}$ is mapped into $\mathbb{C}^{2+q}\simeq \mathbb{C}^2 \times \mathbb{C}^q$ realized by the complex coordinates $z_{i=1,\dots,q}$,$z_{q+1,q+2}$ in two disconnected branches:
\begin{eqnarray}
\label{duebrache}
  \text{Branch 1} &=& \left\{\text{Im}z_{q+1} < 0 \, , \, \text{Im}z_{q+2} >0 \,\mid \,  -2\, \text{Im}z_{q+1} \,\text{Im}z_{q+2} > \sum_{i=1}^q \, (\text{Im}z_i)^2 \right\} \times \mathbb{C}^q \nonumber\\
\text{Branch 2} &=& \left\{\text{Im}z_{q+1} > 0 \, , \, \text{Im}z_{q+2} < 0 \,\mid \,  -2\, \text{Im}z_{q+1} \,\text{Im}z_{q+2} > \sum_{i=1}^q \, (\text{Im}z_i)^2 \right\} \times \mathbb{C}^q 
\end{eqnarray}
This is the analogue of what happens with the Hyperbolic plane where the full manifold $\mathrm{SL(2,\mathbb{R})/SO(2)}$, diffeomorphic to $\mathbb{R}^2$, via the the 2-solvable coordinates, is mapped into either the upper or lower complex plane $\text{Im}z >0$ or $\text{Im}z <0$. 
\par
In any case the argument of the logarithm expressing the K\"ahler potential in eq.(\ref{kallerofino}) is strictly positive.
\paragraph{The K\"ahler 2-form and the moment maps.} In order to complete our programme and arrive at the representation of the metric in the form of eq. (\ref{tantaroba}) we consider the K\"ahler $2$-form. The intrinsic definition of the latter in terms of the Vielbein is given in eq.(\ref{formaggiomagro}) of the appendix. If we utilize the vielbein in the $\mathbf{W}$ coordinate basis displayed in eq.(\ref{8gambe}) and we make the substitution (\ref{tadalafil}) we arrive at the following result:
\begin{equation}\label{farlengo}
  \boldsymbol{\mathcal{K}} \, = \, \sum_{4}^{A=1} \mathrm{d}u_A \wedge \mathrm{d}\mathfrak{m}^A
\end{equation}
where the moment-maps $\mathfrak{m}^i$ are defined as follows:
\begin{equation}\label{craccarino}
  \mathfrak{m}^A \, = \, \left\{e^{-w_1} w_{6,i},\frac{e^{w_2-w_1}
   \left(\mathbf{w}_{6}\cdot\mathbf{w}_{6}+4\right)}{2 \sqrt{2}},-\sqrt{2}
   e^{-w_1-w_2}\right\}
\end{equation}
Replacing in eq.(\ref{craccarino}) the coordinate transformation (\ref{fraballone}) we obtain: 
\begin{equation}\label{craccarino}
  \mathfrak{m}^A \, = \, \left\{-\frac{2 v_i}{\sum_{i=1}^q v_i^2+2 v_{q+1} v_{q+2}},-\frac{2 v_{q+2}}{\sum_{i=1}^q v_i^2+2 v_{q+1} v_{q+2}},-\frac{2 v_{q+1}}{\sum_{i=1}^q v_i^2+2 v_{q+1} v_{q+2}}\right\} \, = \, \frac{\partial\mathcal{K}(\mathbf{v}) }{\partial v_A}
\end{equation}
In this way we see that the coordinates $\mathfrak{m}^A$ are indeed the momenta  in the symplectic geometry sense, since they are the derivatives of the K\"ahler potential with respect to the coordinates $v_A$. The relation (\ref{craccarino}) is easily invertible and we have:
\begin{equation}\label{cannabisonte}
  \left\{v_1\to -\frac{2
   \mathfrak{m}_i}{\sum_{i=1}^q \mathfrak{m}_i^2+2 \mathfrak{m}_{q+1} \mathfrak{m}_{q+2}},v_{q+1}\to -\frac{2
   \mathfrak{m}_{q+2}}{\sum_{i=1}^q \mathfrak{m}_i^2+2 \mathfrak{m}_{q+1} \mathfrak{m}_{q+2}},v_{q+2}\to -\frac{2
   \mathfrak{m}_{q+1}}{\sum_{i=1}^q \mathfrak{m}_i^2+2 \mathfrak{m}_{q+1} \mathfrak{m}_{q+2}}\right\}
\end{equation}
Hence we can easily calculate the Legendre transform of the K\"ahler potential, namely the symplectic potential:
\begin{equation}\label{corgonalo}
  \mathcal{G}(\boldsymbol{\mathfrak{m}})= v_A \, \mathfrak{m}^A \,-\,\mathcal{K}(\mathbf{v})\, = \, \,  -\log \left(-\sum_{i=1}^q \mathfrak{m}_i^2-2 \mathfrak{m}_{q+1} \mathfrak{m}_{q+2}\right)
\end{equation}
and the K\"ahler metric takes exactly the form (\ref{tantaroba}) in terms of the $\boldsymbol{\mathfrak{m}},\mathbf{v}$ variables and of the Hessian of $ \mathcal{G}(\boldsymbol{\mathfrak{m}})$ (apart from the overall factor $1/2$ that we deleted in the above calculations for simplicity).
\subsubsection{Souriau thermodynamics on finite volume symmetric spaces}
\label{finitovolume}
The crucial observation that one should make at this point is that the $\mathfrak{m}_A$
coordinates are not only the moment maps in the symplectic Darboux sense but are also moment maps in the Lie algebra sense. Indeed if we calculate the moment maps of the Killing vectors associated with the generators of the maximal abelian ideal, utilizing the universal formula  (\ref{gelindoelapecora}) and any coordinate basis, upon coordinate change to the $\boldsymbol{\mathfrak{m}},\mathbf{u}$ basis we find:
\begin{alignat}{4}\label{kannone}
  &\mathfrak{P}_{\mathfrak{T}_A} &=& \,\,k \times \mathfrak{m}_A &\quad ; \quad &k\, =\,  \text{const.}\nonumber\\
  &\mathfrak{T}_A & = & \underbrace{\left(T^{5,i}, T^3 , T^4\right)}_{\text{max. abel. ideal}}  &\quad ; \quad & A=1,\dots, q+2
\end{alignat}
One can therefore consider Souriau like unnormalized Gibbs distributions of the type:
\begin{equation}\label{pecoracarolina}
  G_{un}(\boldsymbol{\mathfrak{m}},\mathbf{u}) \, = \, \exp\left[-\beta^A\, \mathfrak{m}_A\right]
\end{equation}
and try to normalize them by calculating the partition function:
\begin{equation}\label{cuccu}
  Z[\pmb{\beta}] \, = \, \int \exp\left[-\beta_A\, \mathfrak{m}^A\right]\, \underbrace{\boldsymbol{\mathcal{K}}\wedge\dots \wedge\boldsymbol{\mathcal{K}}}_{\text{$(q+2)$-times}} 
\end{equation}
The integral is certainly divergent for any choice of $\boldsymbol{\beta}$ because of the 
non-compact coordinates $\mathbf{u}$. It becomes convergent if we mod the CV manifold by an appropriate discrete subgroup of $\mathrm{SO(2,2+q)}$ as those analyzed and presented in \cite{pgtstheory}. Any of these discrete subgroups contain abelian translation subgroups that compactify the $\mathbf{u}$-coordinates but they act also on the moment variables $\mathfrak{m}^A$ so that it is impossible to give a general recipe for the evaluation of the partition function (\ref{cuccu}). Yet it is clear that it can be evaluated by imposing the appropriate discrete subgroup dependent integration limits. Hence a large variety of Souriau like thermodynamics is possible on finite volume Calabi-Vesentini manifolds and we postpone their analysis to a future publication . These thermodynamic models are very much different from those that we analyze in section \ref{sashasecta}, their temperature vectors belonging to the coadjoint orbit of the maximal abelian ideal rather than to the coadjoint orbit (plus extensions) of the compact Cartan subalgebra. We have presented the sketch of such different Souriau thermodynamic models both to anticipate a further interesting research line and to emphasize the appropriate conceptual localization of the results obtained in section \ref{sashasecta}. 
\subsubsection{Embedding into Special K\"ahler Geometry via Type II CV coordinates}  
 Relying on the above preliminaries the embedding of the CV manifolds into the general framework of Special K\"ahler Geometry, as discussed both in \cite{pgtstheory} and in the book \cite{advancio}, is very simple. The symplectic holomorphic section $\mho_{II}(\mathbf{z})$ is given by eq.(\ref{careno}) where $X^\Lambda(\mathbf{z})$ is defined by eq.(\ref{piromane}) and $F_\Sigma(S,\mathbf{z})$ by eq.(\ref{onerac}), the K\"ahler potential defined by eq.
 (\ref{kallerone}) splits into the sum
 \begin{eqnarray}\label{splittometro}
  \mathcal{K}(S,\mathbf{z}) &=& \text{const}^\prime \times \log\left[S-\bar{S}\right] \, + \, \mathcal{K}_{II}(\mathbf{v}) \nonumber\\
  &=& \text{const}^\prime \times \log\left[\mathrm{Im}(S)\right]\, +\, \text{const} \times \log\left[ -\sum_{i=1}^q v_i^2 - 2v_{q+1} \, v_{q+2}   \right]
 \end{eqnarray}
 which explicitly shows the $q+3$ translational symmetries of the Special K\"ahler manifold since the K\"ahler potential depends only on the imaginary parts of all the $q+3$ complex coordinates. In particular the symmetric complex matrix with positive definite imaginary part (\ref{corleone}), upon use of the solvable Lie group element (\ref{krollus}) becomes
 \begin{equation}\label{corleone}
  \mathcal{N}(\text{Im}S,\mathbf{W}) \, = \, \mathit{i} \, \text{Im}S \,\, \mathbb{L}(\boldsymbol{W})\,\mathbb{L}^T(\boldsymbol{W})\, 
  + \, \text{Re}S \,\, \eta_t
\end{equation}
and upon further use of the conversion of the solvable coordinates into the real and imaginary parts of the complex ones given in eq.(\ref{transoCos}) becomes function of the latter. In explicit applications to  Cartan Neural Networks and to geometrical thermodynamics one must use the solvable $\mathbf{W}$ coordinates, yet the important fact is that any chosen coordinate system for the CV manifold parameterizes the imaginary positive definite part of a complex matrix $\mathcal{N}$, whose real part is zero when we freeze the hyperbolic plane factor.
\subsection{Type I CV coordinates and their relation with Type II coordinates}
\label{CVtipouno}
In order to introduce the type I CV coordinates it is convenient to write first the
complete $(8+2q)$-dimensional holomorphic symplectic section of type II as a matter of comparison and as a prerequisite to establish the conversion formulae from I to II. Fixing $S=\mathit{i}$, using the triangular invariant metric (\ref{etatdefi}) and the master formulae (\ref{careno}),(\ref{onerac}) we obtain:
\begin{equation}\label{tipusduemho}
  \mho_{II}(\mathbf{z}) \, = \, 
  \left(\begin{array}{c}
  -\sum_{i=1}^q \,z_i^2 \, - \, 2 \, z_{q+1} \,z_{q+2} \\
  \sqrt{2} z_{q+2} \\
  \sqrt{2} z_{i} \\
  \sqrt{2} z_{q+1}  \\
  1 \\
  \hline
  \mathit{i} \\
  \mathit{i}\, \sqrt{2} z_{q+1} \\
  \mathit{i}\,\sqrt{2} z_{i} \\
  \mathit{i}\, \sqrt{2} z_{q+2} \\
  \mathit{i}\,(-\sum_{i=1}^q \,z_i^2 \, - \, 2 \, z_{q+1} \, z_{q+2}) \\
  \end{array}
  \right)
\end{equation}
The type I Calabi-Vesentini coordinates are instead obtained from a different holomorphic section:
\begin{equation}\label{tiponeunoX}
  \mho_{I}(\mathbf{y}) \, = \, \left( \begin{array}{r}
                                        X^\Lambda_{I}(\mathbf{y}) \\
                                        \mathit{i} \, \eta^{b}_{\Sigma\Delta} \, X^\Delta_{I}(\mathbf{y}) 
                                      \end{array}
  \right)
\end{equation}
where the $\eta^{b}_{\Lambda\Sigma}$ invariant tensor is the traditional block diagonal one:
\begin{equation}\label{etab}
  \eta^b \, = \, \left(\begin{array}{c|c|c}
                        \mathbf{1}_{q\times q} & \mathbf{0}_{q\times 2} &\mathbf{0}_{q\times 2}\\
                        \hline
                        \mathbf{0}_{2\times q} & \mathbf{1}_{2\times 2} & \mathbf{0}_{2\times 2} \\
                        \hline
                        \mathbf{0}_{2\times q} & \mathbf{0}_{2\times 2} & -\mathbf{1}_{2\times 2}
                      \end{array}
  \right)
\end{equation}
and the half holomorphic section $X^\Lambda(\mathbf{y})$ is required to satisfy the holomorphic constraint in the form:
\begin{equation}\label{novoide}
  X^\Lambda(\mathbf{y})\,X^\Sigma(\mathbf{y}) \, \eta^b_{\Lambda\Sigma} \, = \,0
\end{equation}
Since the block diagonal invariant metric $\eta^b$ is easily obtained from the triangular one $\eta_t$ by means of an orthogonal transformation:
\begin{equation}\label{quadrone}
  \eta^b \, = \, \Omega\cdot \eta_t \cdot \Omega^T
\end{equation}
where the $\Omega$-matrix is the following: 
\begin{equation}\label{omegonematro}
  \Omega \, = \, \left(\begin{array}{c|c|c}
                         \mathbf{0}_{q\times 2} & \mathbf{1}_{q\times q} & \mathbf{0}_{q\times 2} \\
                         \hline
                         \frac{1}{\sqrt{2}}\, \left( \begin{array}{cc}
                         1 & 0 \\
                         0 & 1 \\
                         \end{array}\right) &\mathbf{0}_{2\times q} & \frac{1}{\sqrt{2}}\, 
                         \left( \begin{array}{cc}
                         0 & 1 \\
                         1 & 0 \\
                         \end{array}\right)\\
                          \hline
                        \frac{1}{\sqrt{2}}\, \left( \begin{array}{cc}
                         1 & 0 \\
                         0 & 1 \\
                         \end{array}\right) &\mathbf{0}_{2\times q} & -\frac{1}{\sqrt{2}}\, 
                         \left( \begin{array}{cc}
                         0 & 1 \\
                         1 & 0 \\
                         \end{array}\right)\\ 
                       \end{array}
   \right)
\end{equation}
 we might easily obtain a solution of the holomorphic constraint (\ref{novoide}) simply setting:
\begin{equation}\label{carnenonvale}
  X(\mathbf{y})\, = \, \Omega \cdot X(\mathbf{z}) 
\end{equation}
and still using the type II CV coordinates $\mathbf{z}$. This however is not the point. The type I CV coordinates correspond to a different solution of the holomorphic constraint that is aimed at the introduction of complex coordinates $\mathbf{y}$ that, differently from the $\mathbf{z}$, transform linearly under the compact subgroup $\mathrm{SO(2+q)}$.  Such task is fulfilled by the following half section:
\begin{equation}\label{tarrapunto}
  X^\Lambda(\mathbf{y}) \, = \, \left(\begin{array}{c}
                            y_a \\
                            \frac{\mathit{i}}{2}\left(1-\sum_{a=1}^{2+q} y^2_a\right) \\
                            \frac{1}{2}\left(1+\sum_{a=1}^{2+q} y^2_a \right)
                          \end{array}
   \right)
\end{equation}
so that the complete holomorphic section turns out to be the following one:
\begin{equation}\label{tarralinea}
  \mho_{I}(\mathbf{y}) \, = \, \left(
  \begin{array}{c}
  y_a \\
  \frac{\mathit{i}}{2}\left(1-\sum_{a=1}^{2+q} y^2_a\right) \\
  \frac{1}{2}\left(1+\sum_{a=1}^{2+q} y^2_a \right)\\
  \hline
  \mathit{i} \,y_a \\
  \frac{1}{2}\left(1-\sum_{a=1}^{2+q} y^2_a\right) \\
  - \,\frac{\mathit{i}}{2}\left(1+\sum_{a=1}^{2+q} y^2_a \right)
  \end{array}
   \right)
\end{equation}
The main difference with respect to type II coordinates is visually evident. The new $y_a$ form now a $(2+q)$-vector and the structure of the section $\mho_{I}(\mathbf{y})$ is covariant with respect to linear $\mathrm{SO(2+q)}$-transformations, while type II coordinates have such property only with respect to the Paint Group that is the largest possible automorphism group of the solvable Lie algebra. It follows that  Type I CV-coordinates facilitate the task of setting up compact abelian structures, yet at the price of obscuring the solvable Lie algebra structure and hence having a much more involved relation with the solvable coordinates $\mathbf{W}$ that are necessary for Machine Learning algorithms.
Notwithstanding its relative complexity, the transformation from Type I CV coordinates to solvable ones does exist and we are able to construct it in two steps, finding first the expression of the $y$ in terms of the $z$ and then using the expression of the latter in terms of the $\mathbf{W}$ already presented in eq.(\ref{fraballone}). Prior to that let us work out the K\"ahler potential and the metric in Type I coordinates. Utilizing the general formula (\ref{kallerone}) we
obtain:
\begin{eqnarray}\label{KalpotUno}
  \mathcal{K}_{I}(\mathbf{y},\bar{\mathbf{y}}) & = & - \, \log\left[ \mathit{i} \, \mho^\dagger(\bar{\mathbf{y}})\cdot \mathbb{C} \cdot \mho({\mathbf{y}})\right] \, = \,  - \, \log\left[ \underbrace{1 \, - \, 2\, \bar{\mathbf{y}}\cdot \mathbf{y}\, + \, |\mathbf{y}\cdot \mathbf{y}|^2}_{\equiv DB} \right]\nonumber\\
\end{eqnarray}
The conversion from one coordinate basis to the other is performed by solving the following equation:
\begin{equation}\label{bisonteunodue}
  \mho_{II}(\mathbf{z}) \, = \, \mathcal{F}(\mathbf{z}) \, \Sigma[\Omega^T] \cdot \mho_I(\mathbf{y}[\mathbf{z}])
\end{equation}
where $\Sigma[\Omega^T]=\Sigma[\Omega]^T$ is the uplifting to the symplectic group $\mathrm{Sp(8+2q,\mathbb{R})}$ of the orthogonal matrix $\Omega^T$ defined in eq.(\ref{omegonematro}), namely using
\begin{eqnarray}\label{cerulisina}
&&\Sigma \,: \, \mathrm{SO(4+q)} \hookrightarrow \mathrm{Sp(8+2q,\mathbb{R})} \nonumber\\
&&\forall \mathbf{X} \quad  / \quad  \mathbf{X}\cdot \mathbf{X}^T \, = \, \mathbf{1} \quad:\quad \Sigma[\mathbf{X}] \, = \, \left(\begin{array}{c|c}
                                \mathbf{X} & \mathbf{0} \\
                                \hline
                                \mathbf{0} & \mathbf{X} 
                              \end{array}\right)
\end{eqnarray}
we have                             
\begin{eqnarray}\label{tadalafil}
  \Sigma[\Omega^T] & = & \left(\begin{array}{c|c}
                                \Omega^T & \mathbf{0} \\
                                \hline
                                \mathbf{0} & \Omega^T 
                              \end{array}
   \right) \quad ; \quad \Sigma[\Omega]^T \cdot \mathbb{C} \cdot \Sigma[\Omega] \, = \, \mathbb{C} \, = \, \left(\begin{array}{c|c}
                                \mathbf{0} & \mathbf{1} \\
                                \hline
                                -\mathbf{1} & \mathbf{0} 
                              \end{array}
   \right)
\end{eqnarray}
while the transformation $\mathbf{y}[\mathbf{z}]$ and the overall holomorphic function $\mathcal{F}(\mathbf{z})$ are the unknowns to be determined. We find:
\begin{eqnarray}
\label{basilisco}
  y_i &=& -\frac{2 i z_i}{\sum _{i=1}^q z_i^2+2 z_{q+2} z_{q+1}+i \sqrt{2}
   z_{q+1}-i \sqrt{2} z_{q+2}+1} \nonumber \\
 y_{q+1} &=& \frac{i \left(\sum _{i=1}^q z_i^2+2 z_{q+1} z_{q+2}-1\right)}{\sum
   _{i=1}^q z_i^2+2 z_{q+2} z_{q+1}+i \sqrt{2} z_{q+1}-i \sqrt{2}
   z_{q+2}+1} \nonumber\\
  y_{q+2} &=& -\frac{i \sqrt{2} \left(z_{q+1}+z_{q+2}\right)}{\sum _{i=1}^q
   z_i^2+2 z_{q+2} z_{q+1}+i \sqrt{2} z_{q+1}-i \sqrt{2} z_{q+2}+1}\nonumber \\
  \mathcal{F}(\mathbf{z}) &=& \frac{i \left(\sum _{i=1}^q z_i^2+2 z_{q+2} z_{q+1}+i \sqrt{2}
   z_{q+1}-i \sqrt{2} z_{q+2}+1\right)}{\sqrt{2}} 
\end{eqnarray}
Two observations are important at this point. The first, which is a general property of Special K\"ahler geometry is that the two symplectic sections $\mho_{II}(\mathbf{z})$ and 
$\mho_I(\mathbf{y})$ differ by a constant symplectic transformation (the matrix $\Sigma_\Omega$) and by an overall holomorphic function (the factor $\mathcal{F}(\mathbf{z})$). These are precisely the two allowed gauge transformations that do not change the metric because the K\"ahler potential is defined by the general formula (\ref{KalpotUno}), so that:
\begin{equation}\label{kannato}
  \mathcal{K}_{II}(\mathbf{z},\bar{\mathbf{z}}) \, = \, \mathcal{K}_{I}(\mathbf{y}[\mathbf{z}],\bar{\mathbf{y}}[\bar{\mathbf{z}}]) \, -\,\log[\bar{\mathcal{F}}(\bar{\mathbf{z}})\,\mathcal{F}(\mathbf{z})]
\end{equation}
the additional term being the real part of a holomorphic function and as such do not contributing to the metric (K\"ahler transformation).
\par
The second observation is that the transformation (\ref{basilisco}) explicitly breaks the $\mathrm{SO(2+q)}$ linear covariance of  Type I Calabi Vesentini coordinates $y$, preserving only the Paint Group linear covariance $\mathrm{SO(q)}$. The transformations lying in the Grassmannian $\mathcal{F}_T \, \equiv \,\frac{\mathrm{SO(2+q)}}{\mathrm{SO(2)\times SO(q)}}$ (see \cite{pgtstheory}) act linearly on the $\mathbf{y}.s$ but act non linearly on the $\bar{z}.s$ and a fortiori with heavier non linearity on the solvable coordinates $\mathbf{W}$. In Cartan Neural Networks data are mapped linearly into the $\mathbf{W}$, so that as described in
\cite{TSnaviga,naviga}, the transformations associated with the above quoted Grassmannian $\mathcal{F}_T$ are important contributors to the non-linearity of the maps between layers. On the other hand the construction of the compact abelian structure is centered on the compact subalgebra $\mathbb{H}\subset \mathbb{U}$ and for that task the Type I CV coordinates are better suited.
\section{Moment Maps Revisited and the Moment Map Matrix}
\label{momentissimi}
In section \ref{deagostini} we provided the general formula (\ref{gelindoelapecora}) for the moment-maps $\mathfrak{P}_\Lambda$  of the Killing vector fields $\boldsymbol{\mathfrak{k}}_\Lambda$ associated with a  $J_\Lambda$ generator basis of the Lie algebra $\mathbb{U}$. In
(\ref{gelindoelapecora}) the $\mathfrak{P}_\Lambda$ are computed in terms of the coset representative, for instance $\mathbb{L}(\mathbf{W})$ of eq.(\ref{krollus})
and they turn out to be functions of the coordinates used to parameterize the coset representative $\mathbb{L}$, in this case the solvable ones. In order to obtain the same moment maps directly expressed in terms of complex coordinates (Calabi-Vesentini of type I or II, for instance) there is another completely general formula of Special K\"ahler Geometry, explained and utilized for instance in \cite{stabillo} that is the following:
\begin{eqnarray}\label{francione}
  \mathfrak{P}_A(\mathbf{t},\bar{\mathbf{t}}) & = &
  \exp[\mathcal{K}(\mathbf{t},\bar{\mathbf{t}})] \times < \bar{\mho}(\bar{\mathbf{t}})\cdot \Sigma[J^{\eta}_A] \mid {\mho}({\mathbf{t}})> \nonumber\\
  &=& \exp[\mathcal{K}(\mathbf{t},\bar{\mathbf{t}})] \times \left( \bar{\mho}(\bar{\mathbf{t}})
    \cdot \Sigma[J^{\eta}_A] \cdot \mathbb{C} \cdot \mho({\mathbf{t}})\right)
\end{eqnarray}
where $\mathbf{t}$ is any set of complex coordinates and $\Sigma[J^{\eta}_A]$ is the symplectic uplifting
\begin{equation}\label{maialerosso}
  \Sigma[J^{\eta}_A] \, = \, \left(
                                     \begin{array}{c|c}
                                       J^{\eta}_A & 0 \\
                                       \hline
                                       0 &-\,\left( J^{\eta}_A\right)^T \\
                                     \end{array}
                                   \right)
\end{equation}
 of the $\mathbb{U}=\so(2,2+q)$ Lie algebra generators $J^{\eta}_A$ in the basis used by the symplectic section namely such that:
\begin{equation}\label{cartongesso}
  \left( J^{\eta}_A\right)^T\cdot\eta +\eta \cdot J^{\eta}_A \, = \, 0 \quad ; \quad X^\Lambda(\mathbf{t}) \, X^\Sigma(\mathbf{t}) \eta_{\Lambda\Sigma} \, = \,  0 \quad ; \quad F_\Sigma(\mathbf{t}) \, = \, \mathit{i} \, \eta_{\Sigma\Delta} \, X^\Delta(\mathbf{t})
\end{equation}
\par
Note that using accurately all the definitions the Killing moment maps calculated via the Special K\"ahler Geometry general formula (\ref{francione}) are the same in any coordinate basis. For instance one has
\begin{equation}\label{identitapesante}
  \mathfrak{P}_A(\mathbf{z},\bar{\mathbf{z}}) \, = \, \mathfrak{P}_A(\mathbf{y},\bar{\mathbf{y}})
\end{equation}
This follows from eq. (\ref{bisonteunodue}) that relates the two symplectic sections, from eq. (\ref{kannato}) that relates the two K\"ahler potentials and from eq. (\ref{quadrone}) that relates the two eta-matrices and is uplifted to a symplectic transformation. Altogether to go from one coordinate basis to the other one utilizes the intrinsic setup of Special K\"ahler Geometry that encodes the metric and all of its properties in the flat symplectic bundle of dimension $8+2q$ times the Hodge line bundle (see \cite{pgtstheory} for a review). Without the powerful scheme of Special Geometry the coordinate changes relevant to our purposes could not be easily found and this is an important signal that ML algorithms and Souriau-like Thermodynamics are best tuned not only to K\"ahler Geometry but even better to Special K\"ahler Geometry.
\subsection{The moment map matrix and the spectrum of its sublock $\mathfrak{D}$}
\label{momappamatra}
As it was discovered by one of us (A.S.) that is mostly responsible for finding the clue to the explicit construction of the compact abelian structures and the
consequent explicit calculation of the integrals defining the partition functions (see the general result anticipated in section \ref{lodicoprima}),  a very useful tool of analysis is the \textbf{moment-map} defined as follows:
\begin{equation}\label{gongolando}
  \mathfrak{MP} \, \equiv \, \kappa^{AB} \,\mathfrak{P}_A \, J_B
\end{equation}
where the constant matrix $\kappa^{AB}$ is the invariant Killing metric\footnote{As we observed above, we always utilize a basis of generators normalized in such a way that $\kappa^{AB}$ is diagonal with eigenvalue $+1$ for non compact generators and eigenvalue $-1$ for compact ones.} on the full Lie algebra $\mathbb{U} \, = \, \so(2,2+q)$. The object $\mathfrak{MP} \in \mathbb{U}$ is just an element of the Lie algebra $\mathbb{U}$ parameterized by the Killing moment maps. It can exhibited in different ways depending on the coordinates utilized to calculate the moment-maps, solvable $\mathbf{W}$, Type II CV $(\mathbf{z},\bar{\mathbf{z}})$ or Type I CV $(\mathbf{y},\bar{\mathbf{y}})$. Finally it can also be exhibited in abstract form simply by naming the moment-maps with their symbols. To this effect we introduce the following convention. According with the standard order of the Lie algebra generators
utilized in previous articles and in the MATHEMATICA Code \textbf{NonCompSymSpacNeuNet26Up2.nb} that enables all our computations, the first $4+2q$ generators are the non-compact ones $K_a$ belonging to the subspace $\mathbb{K}$ (the coset generators) while the remaining $2+\ft12 q(q+3)$ are the generators of the compact subalgebra $\mathbb{H} = \so(2)\oplus \mathbb{H}^\prime$.
For visual convenience we introduce the following convention. We name we straight Latin  letters $P$ the non-compact moment maps:
\begin{equation}\label{noncompiti}
  P_a \, = \, \mathfrak{P}_a \quad ; \quad a=1,\dots, 4+2q
\end{equation}
and with calligraphic $\mathcal{P}$ the compact starting with $0$ for the $\so(2)$ generator associated with the K\"ahler $2$-form: 
\begin{equation}\label{compiti}
  \mathcal{P}_i \, = \, \mathfrak{P}_{5+2q+i} \quad ; \quad i=0,\dots, 1+\ft12 q(q+3)
\end{equation}
Furthermore in eq.(\ref{gongolando}) we utilize the generators $J_A$ in the $\eta^{b}$ block-diagonal basis. To illustrate the result we utilize the case $q=3$.
\begin{equation}\label{cavernus}
\mathfrak{MP} \, = \,  \left(
\begin{array}{ccccc||cc}
 0 & -\frac{\mathcal{P}_4}{\sqrt{2}} &
   -\frac{\mathcal{P}_3}{\sqrt{2}} &
   \frac{\mathcal{P}_{10}}{\sqrt{2}} &
   \frac{\mathcal{P}_7}{\sqrt{2}} & \frac{P_5}{\sqrt{2}} &
   \frac{P_8}{\sqrt{2}} \\
 \frac{\mathcal{P}_4}{\sqrt{2}} & 0 &
   -\frac{\mathcal{P}_2}{\sqrt{2}} & \frac{\mathcal{P}_9}{\sqrt{2}}
   & \frac{\mathcal{P}_6}{\sqrt{2}} & \frac{P_6}{\sqrt{2}} &
   \frac{P_9}{\sqrt{2}} \\
 \frac{\mathcal{P}_3}{\sqrt{2}} & \frac{\mathcal{P}_2}{\sqrt{2}} & 0
   & \frac{\mathcal{P}_8}{\sqrt{2}} & \frac{\mathcal{P}_5}{\sqrt{2}}
   & \frac{P_7}{\sqrt{2}} & \frac{P_{10}}{\sqrt{2}} \\
 -\frac{\mathcal{P}_{10}}{\sqrt{2}} &
   -\frac{\mathcal{P}_9}{\sqrt{2}} & -\frac{\mathcal{P}_8}{\sqrt{2}}
   & 0 & -\frac{\mathcal{P}_1}{\sqrt{2}} & \frac{P_1}{\sqrt{2}} &
   \frac{P_3}{2}-\frac{P_4}{2} \\
 -\frac{\mathcal{P}_7}{\sqrt{2}} & -\frac{\mathcal{P}_6}{\sqrt{2}} &
   -\frac{\mathcal{P}_5}{\sqrt{2}} & \frac{\mathcal{P}_1}{\sqrt{2}}
   & 0 & \frac{P_3}{2}+\frac{P_4}{2} & \frac{P_2}{\sqrt{2}} \\
   \hline\hline
 \frac{P_5}{\sqrt{2}} & \frac{P_6}{\sqrt{2}} & \frac{P_7}{\sqrt{2}}
   & \frac{P_1}{\sqrt{2}} & \frac{P_3}{2}+\frac{P_4}{2} & 0 &
   -\frac{\mathcal{P}_0}{\sqrt{2}} \\
 \frac{P_8}{\sqrt{2}} & \frac{P_9}{\sqrt{2}} &
   \frac{P_{10}}{\sqrt{2}} & \frac{P_3}{2}-\frac{P_4}{2} &
   \frac{P_2}{\sqrt{2}} & \frac{\mathcal{P}_0}{\sqrt{2}} & 0 \\
\end{array}
\right) \, = \,
\left(
  \begin{array}{c||c}
    \mathfrak{D} & \mathfrak{B}^T \\
    \hline\hline
    \mathfrak{B} & \mathfrak{A} \\
  \end{array}
\right)
\end{equation}
In the above equation we have put into evidence the block-structure of the moment map matrix that turns out very useful in order to understand the explicit coordinate expression  of the actions and of the angles. 
\par
A quite strategic observation is the following one. The moment maps are in a number much larger than the dimensions of the manifold, hence they cannot be all functionally independent. If they were functionally independent then we would have a larger number of coordinates on the manifold than necessary. The surplus is precisely given by the compact moment maps $\mathcal{P}_i$. By construction the non-compact moment maps are instead as many as the dimensions of the manifold and we can expect that one might invert the relation between the coordinates and the non-compact $P.s$.
This is indeed what we are going to illustrate explicitly below. As for the compact 
moment maps $P_i$ we expect instead that their functional relations should come out in an invariant group-theoretical way.
\par
To clarify such matters we extract the three relevant blocks and we write them in three versions, the formal one, and in the two explicit CV coordinate bases of Type I and Type II, respectively.
\paragraph{Block $\mathfrak{D}$} 
We begin with the formal expression of the $\mathfrak{D}$-block
\begin{equation}\label{DBlockF}
  \mathfrak{D} \, = \, \left(
\begin{array}{ccccc}
 0 & -\frac{\mathcal{P}_4}{\sqrt{2}} &
   -\frac{\mathcal{P}_3}{\sqrt{2}} &
   \frac{\mathcal{P}_{10}}{\sqrt{2}} &
   \frac{\mathcal{P}_7}{\sqrt{2}} \\
 \frac{\mathcal{P}_4}{\sqrt{2}} & 0 &
   -\frac{\mathcal{P}_2}{\sqrt{2}} & \frac{\mathcal{P}_9}{\sqrt{2}}
   & \frac{\mathcal{P}_6}{\sqrt{2}} \\
 \frac{\mathcal{P}_3}{\sqrt{2}} & \frac{\mathcal{P}_2}{\sqrt{2}} & 0
   & \frac{\mathcal{P}_8}{\sqrt{2}} & \frac{\mathcal{P}_5}{\sqrt{2}}
   \\
 -\frac{\mathcal{P}_{10}}{\sqrt{2}} &
   -\frac{\mathcal{P}_9}{\sqrt{2}} & -\frac{\mathcal{P}_8}{\sqrt{2}}
   & 0 & -\frac{\mathcal{P}_1}{\sqrt{2}} \\
 -\frac{\mathcal{P}_7}{\sqrt{2}} & -\frac{\mathcal{P}_6}{\sqrt{2}} &
   -\frac{\mathcal{P}_5}{\sqrt{2}} & \frac{\mathcal{P}_1}{\sqrt{2}}
   & 0 \\
\end{array}
\right)
\end{equation}
As you see the $\mathfrak{D}$-block is just an antisymmetric $(q+2)\times(q+2)$ matrix ($q=3$ in our example) parameterized by $\ft 12 (q+2)(q+1)$ parameters, namely the compact moment maps of the $\mathbb{H}^\prime \equiv \so(2+q)$ Lie algebra. As such its generic rank, namely if no relation existed among the moment maps $\mathcal{P}_1,\dots, \mathcal{P}_{\ft 12 (q+2)(q+1)}$ should be:
\begin{equation}\label{rankoformalo}
  \text{formal rank}[\mathfrak{D}] \, = \, q+1
\end{equation}
Actually, as we are going to see, independently from the value of $q$ the actual rank
of the matrix $\mathfrak{D}$ is always $2$:
\begin{equation}\label{rankoattualo}
  \text{actual rank}[\mathfrak{D}] \, = \, 2
\end{equation}
This is an intrinsic property of the Lie algebra that can be verified explicitly in every chosen coordinate basis and it is a manifestation of the $\mathrm{SO(q-1)}$ \textbf{subPaint-group} invariance of the geometry (see \cite{pgtstheory}) 
and of the grouping of all CV manifolds into a single Tits Satake Universality class. Indeed the rank $2$ is the natural rank of the case $q=1$ that corresponds to the Tits Satake submanifold for the entire class. It means that calculating the pair-wise
conjugate formal eigenvalues of the matrix $\mathfrak{D}$ the only non vanishing ones
are the two that exists in the Tits Satake case and are \textbf{subPaint singlets}. All the other eigenvalues that are \textbf{subPaint non-singlets} do vanish. 
\par
In order to see this effect and in preparation of our discussion of abelian structures in section \ref{sashasecta} let us write explicitly the $\mathfrak{D}$ in
Type I and Type II CV-coordinates. We begin with Type I, the classical Calabi-Vesentini coordinates \cite{cvclassic}, previously and independently introduced by Hua \cite{Hua1946}, as we already remarked. Here we have:
\begin{equation}\label{DBlockCVT1}
\mathfrak{D}[\mathbf{y},\bar{\mathbf{y}}]\, = \,  \left(
\begin{array}{ccccc}
 0 & \frac{i \left(y_1 \bar{y}_2-y_2 \bar{y}_1\right)}{DB} &
   \frac{i \left(y_1 \bar{y}_3-y_3 \bar{y}_1\right)}{DB} &
   \frac{i \left(y_1 \bar{y}_4-y_4 \bar{y}_1\right)}{DB} &
   \frac{i \left(y_1 \bar{y}_5-y_5 \bar{y}_1\right)}{DB} \\
 \frac{i \left(y_2 \bar{y}_1-y_1 \bar{y}_2\right)}{DB} & 0 &
   \frac{i \left(y_2 \bar{y}_3-y_3 \bar{y}_2\right)}{DB} &
   \frac{i \left(y_2 \bar{y}_4-y_4 \bar{y}_2\right)}{DB} &
   \frac{i \left(y_2 \bar{y}_5-y_5 \bar{y}_2\right)}{DB} \\
 \frac{i \left(y_3 \bar{y}_1-y_1 \bar{y}_3\right)}{DB} &
   \frac{i \left(y_3 \bar{y}_2-y_2 \bar{y}_3\right)}{DB} & 0
   & \frac{i \left(y_3 \bar{y}_4-y_4 \bar{y}_3\right)}{DB} &
   \frac{i \left(y_3 \bar{y}_5-y_5 \bar{y}_3\right)}{DB} \\
 \frac{i \left(y_4 \bar{y}_1-y_1 \bar{y}_4\right)}{DB} &
   \frac{i \left(y_4 \bar{y}_2-y_2 \bar{y}_4\right)}{DB} &
   \frac{i \left(y_4 \bar{y}_3-y_3 \bar{y}_4\right)}{DB} & 0
   & \frac{i \left(y_4 \bar{y}_5-y_5 \bar{y}_4\right)}{DB} \\
 \frac{i \left(y_5 \bar{y}_1-y_1 \bar{y}_5\right)}{DB} &
   \frac{i \left(y_5 \bar{y}_2-y_2 \bar{y}_5\right)}{DB} &
   \frac{i \left(y_5 \bar{y}_3-y_3 \bar{y}_5\right)}{DB} &
   \frac{i \left(y_5 \bar{y}_4-y_4 \bar{y}_5\right)}{DB} & 0
   \\
\end{array}
\right)
\end{equation}
where the denominator $DB$ is the object displayed in equation (\ref{KalpotUno}) and
produced by the exponential of the K\"ahler potential appearing in eq.(\ref{francione}).
The important remarks about the matrix (\ref{DBlockCVT1}) are
\begin{enumerate}
  \item its structure is immediately evident and generalizable to all values of $q$ since the upper triangular matrix elements are of the form:      
      \begin{equation}\label{barlume}
        \mathfrak{D}_{ab}[\mathbf{y},\bar{\mathbf{y}}] \, = \,\frac{i \left(y_a \bar{y}_b-y_b \bar{y}_a\right)}{DB} \, = \, - \,\mathfrak{D}_{ba}[\mathbf{y},\bar{\mathbf{y}}] \quad ; \quad b\, > \,a
      \end{equation}
  \item The action of the $\mathrm{SO(2+q)}$ is manifestly linear on the $y$.s and the denominator (\textit{i.e.} the exponential of the K\"ahler potential is manifestly invariant under such action.
\end{enumerate}
Notwithstanding these appearances, imposed by the choice of the coordinates, the inner working of the not maximally split $\mathbb{U}$ Lie algebra implying a 
non-trivial Paint Group manifests itself in the spectral property of the matrix $\mathfrak{D}[\mathbf{y},\bar{\mathbf{y}}]$. Indeed if we calculate the eigenvalues
of  $\mathfrak{D}[\mathbf{y},\bar{\mathbf{y}}]$ we find that there are always 
$q-1$ vanishing eigenvalues and only two non vanishing pair-conjugate eigenvalues:
\begin{equation}\label{pairlambdi}
  \lambda^\pm_1 \, = \, \pm \,\mathit{i} \, \frac{\sqrt{\left(\sum_{a=1}^{q+2} y_i\,\bar{y}_i\right)^2-\left(\sum_{a=1}^{q+2} y_i^2\right)\left(\sum_{a=1}^{q+2} \bar{y}_i^2\right)}}{DB}\, \equiv \, \pm \,\mathit{i}\, \underbrace{\sqrt{\frac{(\bar{\mathbf{y}}\cdot \mathbf{y})^2 \,-\,|\mathbf{y}\cdot \mathbf{y}|^2}{\left(1 \, - \, 2\, \bar{\mathbf{y}}\cdot \mathbf{y}\, + \, |\mathbf{y}\cdot \mathbf{y}|^2\right)^2}}}_{\sqrt{\boldsymbol{\mathfrak{C}}[\mathbb{H}_1]}}
\end{equation}
where we have posed $\mathbb{H}_1\, \equiv\,\mathbb{H}^\prime$.
The most important thing is the intrinsic algebraic interpretation of the non vanishing eigenvalues. Indeed by using the explicit expression of the compact moment maps in terms of the Type I CV coordinates we easily verify that the following identity holds true:
\begin{equation}\label{kurchatone}
  \frac{(\bar{\mathbf{y}}\cdot \mathbf{y})^2 \,-\,|\mathbf{y}\cdot \mathbf{y}|^2}{\left(1 \, - \, 2\, \bar{\mathbf{y}}\cdot \mathbf{y}\, + \, |\mathbf{y}\cdot \mathbf{y}|^2\right)^2} \, = \, \boldsymbol{\mathfrak{C}}[\mathbb{H}_1] \, \equiv \, \frac{1}{2}\,\sum_{i=1}^{\ft 12 (q+2)(q+1)} \, \mathcal{P}_i^2 
\end{equation}
namely, apart from their phase, the modulus of the two non-vanishing  $\mathfrak{D}$-matrix  eigenvalues is the square root of the quadratic Casimir of the $\mathbb{H}_1=\mathbb{H}^\prime$ subalgebra, \textit{i.e.} the sum of squares of the moment-maps of all its generators (remember that the Killing metric of the Lie algebra $\mathbb{U}$ and, hence, also of its compact subalgebra $\mathbb{H}^\prime$ is diagonal). By construction the function $\boldsymbol{\mathfrak{C}}[\mathbb{H}^\prime]$ has vanishing Poisson bracket with all the moment maps of the compact Cartan subalgebra $\mathbb{H}=\so(2)\oplus \mathbb{H}_1$:
\begin{equation}\label{cardenson}
  \left\{\mathcal{P}_i\, , \, \boldsymbol{\mathfrak{C}}[\mathbb{H}_1]\right\}_{PB} \, = \, 0 \quad ; \quad i=0,1,\dots, \ft 12 (q+2)(q+1)
\end{equation}
Such property obviously extends also to the square root of the same object:
\begin{equation}\label{cadrega}
  \left\{\mathcal{P}_i\, , \, \sqrt{\boldsymbol{\mathfrak{C}}[\mathbb{H}_1]}\right\}_{PB} \, = \, 0 \quad ; \quad i=0,1,\dots, \ft 12 (q+2)(q+1)
\end{equation}
and in particular includes the  vanishing Poisson bracket with $\sqrt{\boldsymbol{\mathfrak{C}}[\mathbb{H}_1]}$ of each one in any basis $\mathcal{C}_i$ ($i=0,1,\dots,\nu+1$) of Cartan generators  spanning the compact Cartan subalgebra $\boldsymbol{\mathcal{C}}\subset \mathbb{H}$ (having denoted by $\nu +1$ the rank of $\so(2+q)$).
\par 
Let us consider, in the  formulae below, the $\mathfrak{D}$-matrix (for our illustrative example $q=3$) where we have marked with a symbol $\times$ and delimited between $2$-lines two columns and two rows that have to be deleted (or better replaced with zeros) in order to extract a $3\times 3$ minor:
\begin{eqnarray}\label{sdildoy}
  \widehat{\mathfrak{D}}[\mathbf{y},\bar{\mathbf{y}}] & = & \left(
\begin{array}{|c|cc|c|c}
\hline
\times&\times&\times&\times&\times\\
\hline
\times& 0 & \frac{i \left(y_2 \bar{y}_3-y_3
   \bar{y}_2\right)}{DB} &\times& \frac{i \left(y_2
   \bar{y}_5-y_5 \bar{y}_2\right)}{DB} \\
\times& \frac{i \left(y_3 \bar{y}_2-y_2 \bar{y}_3\right)}{DB} &
   0 &\times& \frac{i \left(y_3 \bar{y}_5-y_5
   \bar{y}_3\right)}{DB} \\
   \hline
\times&\times&\times&\times&\times\\
\hline
\times& \frac{i \left(y_5 \bar{y}_2-y_2 \bar{y}_5\right)}{DB} &
   \frac{i \left(y_5 \bar{y}_3-y_3 \bar{y}_5\right)}{DB} &\times   & 0 \\
\end{array}
\right)\, = \, \left(
\begin{array}{|c|cc|c|c}
\hline
\times&\times&\times&\times&\times\\
 \hline
\times& 0 & -\frac{\mathcal{P}_2}{\sqrt{2}} &\times&
   \frac{\mathcal{P}_6}{\sqrt{2}} \\
\times& \frac{\mathcal{P}_2}{\sqrt{2}} & 0 &\times&
   \frac{\mathcal{P}_5}{\sqrt{2}} \\
   \hline
\times&\times&\times&\times&\times\\
 \hline
\times& -\frac{\mathcal{P}_6}{\sqrt{2}} &
   -\frac{\mathcal{P}_5}{\sqrt{2}} &\times& 0 \\
\end{array}
\right)\nonumber\\
\end{eqnarray}
The minor has two pair-conjugate eigenvalues:
\begin{equation}\label{frosinone}
  \lambda^\pm_2\, = \, \pm \mathit{i} \sqrt{-\frac{ \left(y_3 \bar{y}_2-y_2 \bar{y}_3\right){}^2+
   \left(y_5 \bar{y}_2-y_2 \bar{y}_5\right){}^2+ \left(y_5
   \bar{y}_3-y_3 \bar{y}_5\right){}^2}{\left(1 \, - \, 2\, \bar{\mathbf{y}}\cdot \mathbf{y}\, + \, |\mathbf{y}\cdot \mathbf{y}|^2\right)^2} }\, =\, \pm \mathit{i}
   \,\sqrt{ \frac{\mathcal{P}_2^2+\mathcal{P}_5^2+\mathcal{P}_6^2}{
   2} }\, = \, \pm \mathit{i} \, \sqrt{\boldsymbol{\mathfrak{C}}[\mathbb{H}_2]} 
\end{equation}
and the object $\boldsymbol{\mathfrak{C}}[\mathbb{H}_2]$ is just the Casimir of a subalgebra $\so(3) \simeq \mathbb{H}_2 \subset \mathbb{H}_1\subset \mathbb{H}$ that is generated, in the Poisson representation  by $\mathcal{P}_2,\mathcal{P}_5,\mathcal{P}_6$, namely by the generators $J_{13},J_{16},J_{17}$ in the standard matrix representation of $\so(2,2+q)$. How is this subalgebra determined? It is simply the normalizer of the first Cartan generator of $\mathbb{H}_1$. There is nothing of intrinsic in the choice of minors of the $\mathfrak{D}$ matrix. This latter is simply an element of the $\so(2+q)$ Lie algebra in the fundamental $(2+q)$ representation parameterized by the moment maps associated with a basis of its generators. The correct interpretation of the minor is to put all the parameters in the marked column and rows to zero so as to restrict the matrix to a subalgebra;  what is intrinsic is the definition of the subalgebra as the normalizer of the first Cartan generator. Should we change the Cartan basis the normalizer would migrate to some different minor yet it would be in any case an $\so(2+q-2)=\so(q)$ Lie algebra. As we explain in section \ref{sashasecta}, in the general $q$ case one has to repeat the same operation as many times as it is the rank of $\mathbb{H}^\prime$ and this generates as many subalgebras and corresponding Casimirs $\boldsymbol{\mathfrak{C}}[\mathbb{H}_i]$, whose square roots can play the role of the missing actions, in order to complete the action list of a compact abelian structure. Indeed, by their very construction, all these Casimirs and their square roots commute among themselves and with the $\mathcal{P}_{\mathcal{C}_i}$ associated with compact Cartan generators. The explicit expression of both the Casimirs and of the Cartan moment maps in Type I CV coordinates is useful, in particular to single out the conjugate angles, yet by no mean it is essential. Indeed all the actions can be written abstractly in terms of the Killing moment-maps and the latter can be evaluated in whatever coordinates one prefers (solvable $\mathbb{W}$, Type II CV or Type I). We come back to this point in section \ref{sashasecta}. 
Now in order to better appreciate the intrinsic character of what we did so far we display the structure of the $\mathfrak{D}$-matrix in Type II CV coordinates that is the following one:
{\scriptsize{
\begin{eqnarray}
\label{DBlockCVT2}
&& \mathfrak{D}[\mathbf{z},\bar{\mathbf{z}}]\, =\, \nonumber\\
&& \left(
\begin{array}{ccc|cc}
 0 & \frac{i \left(z_1 \bar{z}_2-z_2 \bar{z}_1\right)}{DC} &
   \frac{i \left(z_1 \bar{z}_3-z_3 \bar{z}_1\right)}{DC} &
   \frac{i \left(\bar{z}_1 \mathcal{Q}(z)-z_1
   \bar{\mathcal{Q}}\left(\bar{z}\right)\right)}{2 DC} &
   \frac{i \left(z_1
   \left(\bar{z}_4+\bar{z}_5\right)-\left(z_4+z_5\right)
   \bar{z}_1\right)}{\sqrt{2} DC} \\
 \frac{i \left(z_2 \bar{z}_1-z_1 \bar{z}_2\right)}{DC} & 0 &
   \frac{i \left(z_2 \bar{z}_3-z_3 \bar{z}_2\right)}{DC} &
   \frac{i \left(\bar{z}_2 \mathcal{Q}(z)-z_2
   \bar{\mathcal{Q}}\left(\bar{z}\right)\right)}{2 DC} &
   \frac{i \left(z_2
   \left(\bar{z}_4+\bar{z}_5\right)-\left(z_4+z_5\right)
   \bar{z}_2\right)}{\sqrt{2} DC} \\
 \frac{i \left(z_3 \bar{z}_1-z_1 \bar{z}_3\right)}{DC} &
   \frac{i \left(z_3 \bar{z}_2-z_2 \bar{z}_3\right)}{DC} & 0
   & \frac{i \left(\bar{z}_3 \mathcal{Q}(z)-z_3
   \bar{\mathcal{Q}}\left(\bar{z}\right)\right)}{2 DC} &
   \frac{i \left(z_3
   \left(\bar{z}_4+\bar{z}_5\right)-\left(z_4+z_5\right)
   \bar{z}_3\right)}{\sqrt{2} DC} \\
   \hline 
 \frac{i \left(z_1 \bar{\mathcal{Q}}\left(\bar{z}\right)-\bar{z}_1
   \mathcal{Q}(z)\right)}{2 DC} & \frac{i \left(z_2
   \bar{\mathcal{Q}}\left(\bar{z}\right)-\bar{z}_2
   \mathcal{Q}(z)\right)}{2 DC} & \frac{i \left(z_3
   \bar{\mathcal{Q}}\left(\bar{z}\right)-\bar{z}_3
   \mathcal{Q}(z)\right)}{2 DC} & 0 & \frac{i
   \left(z_4+z_5\right) \bar{\mathcal{Q}}\left(\bar{z}\right)-i
   \left(\bar{z}_4+\bar{z}_5\right) \mathcal{Q}(z)}{2 \sqrt{2}
   DC} \\
 \frac{i \left(\left(z_4+z_5\right) \bar{z}_1-z_1
   \left(\bar{z}_4+\bar{z}_5\right)\right)}{\sqrt{2} DC} &
   \frac{i \left(\left(z_4+z_5\right) \bar{z}_2-z_2
   \left(\bar{z}_4+\bar{z}_5\right)\right)}{\sqrt{2} DC} &
   \frac{i \left(\left(z_4+z_5\right) \bar{z}_3-z_3
   \left(\bar{z}_4+\bar{z}_5\right)\right)}{\sqrt{2} DC} &
   \frac{i \left(\left(\bar{z}_4+\bar{z}_5\right)
   \mathcal{Q}(z)-\left(z_4+z_5\right)
   \bar{\mathcal{Q}}\left(\bar{z}\right)\right)}{2 \sqrt{2}
   DC} & 0 \\
\end{array}
\right)\nonumber \\
\end{eqnarray}
}}
where we have:
\begin{eqnarray}
\label{spiegone}
  \mathcal{Q}(z) &=& z_1^2+z_2^2+z_3^2+2 z_4 z_5-1  \nonumber\\
  DC &=& \left(z_1-\bar{z}_1\right){}^2+\left(z_2-\bar{z}_2\right){}^2+\left(
   z_3-\bar{z}_3\right){}^2+2 \left(z_4-\bar{z}_4\right)
   \left(z_5-\bar{z}_5\right) 
\end{eqnarray}
As it is evident from eq.(\ref{DBlockCVT2}) in Type II coordinates the action of the $\mathrm{SO(2+q)}$ is linear only for the $\mathrm{SO(q)=G_{Paint}}$ subgroup (in our example the upper $3\times 3$ diagonal block). As we already stressed the remaining transformations act non linearly and this is an important ingredient in Cartan Neural Networks. Yet the rank of the $\mathfrak{D}$-matrix remains obviously equal to two and the pair conjugate eigenvalues are expressed in terms of the square root of the Casimir. The expression of this latter in Type II coordinates is extremely complicated and it is not worth  being displayed. On the other hand the expression of the Casimirs directly calculated in solvable coordinates is manageable, as we will see in section  \ref{sashasecta} and this is what matters for applications to Machine Learning. The complicated structure of compact objects in Type II coordinates is the dual of the highly complex structure of 
non-compact objects in Type I coordinates, in particular of the vital solvable group transformations.    
We have dwelled on the $\mathfrak{D}$-block of the Moment-map matrix because of its relevance in determining the missing actions. 
\section{Compact abelian structures on CV manifolds and the generalized Souriau temperatures and partition functions}
\label{sashasecta}
In this section, which is the principal core of the present paper, we present the construction of \textbf{abelian structures} on Calabi-Vesentini manifolds that include the moment-maps of the compact Cartan subalgebra $\boldsymbol{\mathcal{C}} \subset \mathbb{H} \subset \mathbb{U}$ as starting point. The tools for this construction were already presented in the previous section \ref{momentissimi}, and the result for Souriau partition functions was anticipated in section \ref{lodicoprima}.
This means that here we consider and generalize the Souriau-like thermodynamics on CV manifolds as it is defined in section \ref{topocaldo} and we develop in a full-fledged manner the programme presented in sect.\ref{abelprogram} (see in particular eq.s(\ref{carbonero}-\ref{cardioloso}). This is stressed at the very beginning of the present section in order to help  reader's understanding and mark the difference between  the Souriau-like thermodynamics based on the compact abelian structures and that based on the non-compact abelian structure associated with the maximal abelian ideal discussed instead in section
\ref{mariosecta} and in particular in section \ref{finitovolume}. We also stress at the very beginning of this section that all the various integration architectures based on the compact abelian structure that here are simply skipped and instead will be analyzed and classified in a forthcoming paper \cite{abelstrutture}  yield different generalization of Souriau compact thermodynamics, yet they all collapse to the same general result anticipated in eq.(\ref{zolotayaformula}) if the considered temperatures are limited to the coadjoint orbit of the compact subalgebra $\boldsymbol{\mathcal{C}}$. In the present section we expose the general method and we illustrate it alternatively with the example $q=1$ (the Tits Satake manifold with vanishing Paint Group) and with the already above quoted example $q=3$, that is the smallest one with a non-trivial, non abelian Paint Group.
\subsection{The Souriau Partition Functions for CV manifolds}
A statistical probability distribution defined over a CV space that incorporates the large symmetry $\mathrm{U}$ of its K\"ahler metric, is of the general Gibbs type as defined in eq.(\ref{ciacolone}), where the observable functions $X(q)$ measured on the CV microscopic manifold are chosen to be the moment maps of the Killing vectors generating the $\mathrm{U}$-isometries, as we explained already in section \ref{topocaldo} and we made explicit in eq.(\ref{cerusico}).
Denoting by $\boldsymbol{\Upsilon}$ any type of solvable coordinates (in particular the $\mathbf{W}$ defined by the integration scheme in eq.(\ref{krollus})), 
the moment maps $\boldsymbol{\mathfrak{P}_\Lambda}(\boldsymbol{\Upsilon})$ satisfy, with respect to the Poisson bracket introduced in eq.(\ref{corallino}), the $\mathbb{U}$ Lie algebra in its full fledged form as displayed in eq.(\ref{quirito}).
Hence assuming:
\begin{description}
  \item[A)] that the generalized temperature vector $\boldsymbol{\beta}$ is such that the partition function converges,
  \item[B)] that the generalized temperature vector $\boldsymbol{\beta}$ lies in the $\mathrm{U}$-adjoint orbit\footnote{We do not distinguish among adjoint and coadjoint orbits since the Lie algebras we utilize are simple and the Killing metric is not degenerate (altrough of indefinite signature) and furthermore always standardized to be diagonal.} of the compact subalgebra $\mathbb{H}\subset \mathbb{U}$. 
\end{description}
we can always rotate the temperature vector, by an adjoint transformation to the Cartan subalgebra of the compact subalgebra $\boldsymbol{\mathcal{C}}\subset \mathbb{H}$. Then, after this formal reduction to $\boldsymbol{\mathcal{C}}$:
\begin{equation}\label{reduzia}
 \mathbb{U} \ni\boldsymbol{\beta} \, \stackrel{\text{Adj(g)}}{\Longrightarrow} \boldsymbol{\beta}_c \, \in \, \boldsymbol{\mathcal{C}} \subset \mathbb{H} \subset \mathbb{U}
\end{equation}
we can study the conditions on $\boldsymbol{\beta}_c$ for the integral convergence and the possible $\boldsymbol{\beta} \in \mathbb{U}$ guaranteeing convergence will be those that are in the adjoint orbit of the $\boldsymbol{\beta}_c$ satisfying the found conditions.
As we  show in the later section \ref{theTSmanifoldo} (with the calculation details provided in appendix \ref{integso23}), even in the simplest case of the Tits Satake submanifold $\mathrm{SO(2,3)/SO(2)\times SO(3)}$ of the entire Tits Satake universality class of CV manifolds, the direct integration of the unnormalized Gibbs-distribution in order to obtain its normalizing denominator, \textit{i.e.} \textbf{the partition function} $Z(\boldsymbol{\beta}_c)$, anticipated in the general formulae (\ref{zolotayaformula}), is a very hard task in whatever coordinate system one utilizes, both real, like the solvable coordinates for the various choices of their definition, or complex like the Calabi-Vesentini coordinates discussed in section \ref{baracco}. The key to obtain the general result displayed in eq.s(\ref{zolotayaformula}) is the completion of the Abelian Structure programme outlined above in section \ref{abelprogram}. 
\par
The real dimension of the microscopic CV manifold is always $2n$ where $n=2+q$ but we have to distinguish the two cases $q=\mathrm{odd}$ and $q=\mathrm{even}$ since this decides whether the compact Lie algebra $\so(2+q)$ belongs to the $\boldsymbol{\mathfrak{b}}$-series or to the $\boldsymbol{\mathfrak{d}}$-series.
We introduce therefore the following notations:
\paragraph{Case $\boldsymbol{\mathfrak{b}}$:  $q=2\nu+1$, $\nu\in \mathbb{N}$; $n=3+2\nu$}
\begin{alignat}{8}\label{pagnaccob}
  &\boldsymbol{\lambda}_c^a  &\quad = \quad & \left\{\underbrace{\boldsymbol{\beta}_c^i}_{i=0,1,\dots, \nu+1} \right.\, , \,\left. \underbrace{\boldsymbol{h}^j}_{j=1,\dots,\nu+1}\right\} &\quad ; \quad & (a=1,\dots ,2\nu +3=n) \nonumber \\
  &\boldsymbol{\mathfrak{p}}_a(\boldsymbol{\Upsilon}) &\quad = \quad & \left\{\underbrace{{\mathcal{P}}_{c|i}(\boldsymbol{\Upsilon})}_{i=0,1,\dots, \nu+1} \right.\, , \,\left. \underbrace{\sqrt{\boldsymbol{\mathfrak{C}}^j(\boldsymbol{\Upsilon})}}_{j=1,\dots,
  \nu+1}\right\}&\quad ; \quad & (a=1,\dots ,2\nu +3=n)
\end{alignat}
and
\begin{alignat}{8}\label{pagnacco2b}
  &\boldsymbol{\mathfrak{q}}^a(\boldsymbol{\Upsilon}) \quad &\quad = \quad & \left\{\underbrace{\boldsymbol{\mathfrak{\theta}}^i(\boldsymbol{\Upsilon})}_{i=0,
  1,\dots, \nu+1} \right.\, , \,\left. \underbrace{\boldsymbol{\mathfrak{\psi}}^j(\boldsymbol{\Upsilon})}_{j=1,\dots,\nu+1}\right\}
  &\quad ; \quad & (a=1,\dots ,2\nu +3=n)
\end{alignat}
\paragraph{Case $\boldsymbol{\mathfrak{d}}$:  $q=2\nu$, $\nu\in \mathbb{N}$; $n=2+2\nu$}
\begin{alignat}{8}\label{pagnaccod}
  &\boldsymbol{\lambda}_c^a  &\quad = \quad & \left\{\underbrace{\boldsymbol{\beta}_c^i}_{i=0,1,\dots, \nu+1} \right.\, , \,\left. \underbrace{\boldsymbol{h}^j}_{j=1,\dots,\nu}\right\} &\quad ; \quad & (a=1,\dots ,2\nu +2=n) \nonumber \\
  &\boldsymbol{\mathfrak{p}}_a(\boldsymbol{\Upsilon}) &\quad = \quad & \left\{\underbrace{\mathcal{P}_{c|i}(\boldsymbol{\Upsilon})}_{i=0,1,\dots, \nu} \right.\, , \,\left. \underbrace{\sqrt{\boldsymbol{\mathfrak{C}}_j(\boldsymbol{\Upsilon})}}_{j=1,\dots,
  \nu}\right\}&\quad ; \quad & (a=1,\dots ,2\nu +2=n)
\end{alignat}
and
\begin{alignat}{8}\label{pagnacco2d}
  &\boldsymbol{\mathfrak{q}}^a(\boldsymbol{\Upsilon}) \quad &\quad = \quad & \left\{\underbrace{\boldsymbol{\mathfrak{\theta}}^i(\boldsymbol{\Upsilon})}_{i=0,
  1,\dots, \nu+1} \right.\, , \,\left. \underbrace{\boldsymbol{\mathfrak{\psi}}^j(\boldsymbol{\Upsilon})}_{j=1,\dots,\nu}
  \right\}
  &\quad ; \quad & (a=1,\dots ,2\nu +2=n)
\end{alignat}
where the $\boldsymbol{\mathfrak{p}}_a(\boldsymbol{\Upsilon})$ are the action functions of the solvable coordinates and $\boldsymbol{\mathfrak{q}}^a(\boldsymbol{\Upsilon})$ the corresponding conjugate angle variables that reduce the K\"ahler $2$-form at the Darboux structure of equations (\ref{cromatosko}-\ref{coricidino2}). 
\par
As we explained above in eq.(
\ref{compiti}), $\mathcal{P}_i$ with $(i=1,\dots,1+\ft12 q(q+3)$ is our notation for the moment maps of all $\mathbb{H}$-generators. For brevity we denote instead $\mathcal{P}_{c|i}$ ($i=0,1,\dots,\nu+1$) the $2+\nu$ among them that are associated with the Cartan subalgebra of $\mathbb{H}\, = \,\so(2)+\so(2+q)$.
\par
The objects denoted by the symbol $\boldsymbol{\mathfrak{C}}_j(\boldsymbol{\Upsilon})$ in eq.s(\ref{pagnaccob}-\ref{pagnacco2d}) are the Casimirs of the 
stabilizer subalgebras of the compact Cartan generators, already introduced above
in the discussion of the $\mathfrak{D}$-block of the moment map matrix (see text between eq.(\ref{frosinone}) and eq.(\ref{DBlockCVT2})). The rank of the algebra
$\so(2+q)$ is $\nu+1$ in both cases $q=2\nu+1$ and $q=2\nu$, so in both cases $\boldsymbol{\mathfrak{b}}$ and $\boldsymbol{\mathfrak{d}}$, the number of Cartan generators of $\mathbb{H}_1$ is $\nu+1$ and correspondingly there are $\nu+1$ stabilizer subalgebras with $\nu+1$ Casimirs. The question is why in the $\boldsymbol{\mathfrak{b}}$ case the number of utilized Casimirs is $\nu+1$, while in the $\boldsymbol{\mathfrak{d}}$ case, only $\nu$ Casimirs enter the game. The answer is extremely simple. As we are going to show below, in the even case the sequential embedding of  subalgebras does not end not with an $\so(3)$ rather with a one-dimensional $\so(2)$. In other words with an algebra spanned by a single generator. Such generator that, by definition, commutes with the $\mathcal{C}_{\nu}$ Cartan generator is nothing else but the $\mathcal{C}_{\nu+1}$ Cartan generator whose moment map is already included in the list. Therefore the last Casimir action $\sqrt{\mathcal{P}_{{c|\nu+1}}^2}=\mathcal{P}_{{c|\nu+1}}$ is just a repetition and should not be counted twice.
\par
Once the abelian structure is found, the integral defining the partition function can, according with eq.s (\ref{cardioloso}-\ref{cerebroleso}), be rewritten as:
\begin{eqnarray}\label{maggiorana}
  Z^{\boldsymbol{\mathfrak{b}},\boldsymbol{\mathfrak{d}}}
  (\boldsymbol{\lambda}) &=&\int_{\mathcal{P}_\nu^{\boldsymbol{\mathfrak{b}},\boldsymbol{\mathfrak{d}}}}\,
  \exp\left[-H_{\nu}^{\boldsymbol{\mathfrak{b}},\boldsymbol{\mathfrak{d}}}
  \left(\boldsymbol{\mathfrak{p}},\boldsymbol{\lambda}\right)\right]
  \mathrm{d}^n \boldsymbol{\mathfrak{p}} \times \int_{\boldsymbol{\mathcal{T}}^{n}} \,\mathrm{d}^n \boldsymbol{\mathfrak{q}} 
\end{eqnarray}
where $\boldsymbol{\mathcal{T}}^n$ is the $n$-torus advocated by Arnol'd theorem and spanned by the angles $\boldsymbol{\theta}^i(\boldsymbol{\Upsilon})$ and $\boldsymbol{\psi}^j(\boldsymbol{\Upsilon})$ while the argument of the exponential is the generalized hamiltonian displayed below
\begin{equation} \label{eq:extended_Hamiltonian}
H_{\nu}^{\boldsymbol{\mathfrak{b}},\boldsymbol{\mathfrak{d}}}, \left(\boldsymbol{\lambda}\right) \,=\, 
\left\{
\begin{array}{ccc}
               H_{\nu}^{\boldsymbol{\mathfrak{b}}}\left(\boldsymbol{\mathfrak{p}},\boldsymbol{\lambda}\right)
                & = & \sum_{a=1}^{2\nu+3}
\boldsymbol{\lambda}^a \, \boldsymbol{\mathfrak{p}}_a(\boldsymbol{\boldsymbol{\Upsilon}}) \\
                H_{\nu}^{\boldsymbol{\mathfrak{d}}}
                \left(\boldsymbol{\mathfrak{p}},\boldsymbol{\lambda}\right) & = & \sum_{a=1}^{2\nu+2}
\boldsymbol{\lambda}^a \, \boldsymbol{\mathfrak{p}}_a(\boldsymbol{\boldsymbol{\Upsilon}}) \\
             \end{array}
\right.
\end{equation}
the Liouville integration measure:
\begin{equation}\label{cagnone}
  \mathrm{d}\boldsymbol{\mu} \, = \, \boldsymbol{\mathcal{K}}\wedge \, \dots \wedge \boldsymbol{\mathcal{K}}\, \simeq \, \text{const} \times \mathrm{d}^n \boldsymbol{\mathfrak{p}} \, \times \,\mathrm{d}^n \boldsymbol{\mathfrak{p}} 
\end{equation}
is flat both in the angle and in the action space. 
Before proceeding we need to stress a very important point. The actions $\boldsymbol{\mathfrak{p}}_a$ and the angles $\boldsymbol{\mathfrak{q}}^a$, ($a=1,dots,n$) can be regarded as a new coordinate frame for the entire manifold, yet one should stress that there is a fundamental difference between the angles $\boldsymbol{\theta}^i$ and the angles $\boldsymbol{\psi}^j$. The first are cyclic coordinates from which the metric coefficient cannot depend since they are the parameters of $\nu+1$ a $\mathrm{U(1)}^{\nu+1} \subset \mathrm{U}$ isometry subgroup.
On the other hand the second set of angles are the parameters of a $\mathrm{U(1)}^{\nu+1}$ or $\mathrm{U(1)}^{\nu+1}$ group of transformations in the symplectic manifold that are not isometries of its K\"ahler metric. Hence the 
$\boldsymbol{\psi}^j$ are not cyclic variables and the metric coefficients should explicitly depend on them \footnote{It is here the appropriate point to remind the reader that, if we were to consider the Special K\"ahlerian spaces $\frac{\mathrm{SU(1,n)}}{\mathrm{U(1) \times SU(n)}}$ (the so named minimal model spaces) whose real dimension is $2\times n$, then, for such manifolds
we would have a natural compact abelian structure since the compact subgroup contains a $\mathrm{U(1)}^n$ subgroup. Using the non-compact analogue of Fubini-Study complex coordinates the angles are just the phases of such coordinates. The price one pays for such simplicity of the Thermodynamics is that the equivalent solvable group exists but it is much more complicated and not associated with a Lie algebra root system. The well known case of exotic solvable coordinates whose use in Machine Learning is much more problematic.}
This fundamental property allows the partition functions to be evaluated as an integral over the action polytope $\mathcal{P}_\nu^{\boldsymbol{\mathfrak{b}},\boldsymbol{\mathfrak{d}}}$.
The evaluation of the integral (\ref{maggiorana}) is performed through a recursive algorithm that reflects the sub-algebra inclusion chain described in section \ref{cannastorta}.
\par
The partition functions (\ref{maggiorana}) depending on the extended temperature vector $\boldsymbol{\lambda}$ are the normalization denominators of Gibbs distributions of extended Souriau type as it follows:
\begin{eqnarray} \label{extendedGibbs}
\mathbf{G}^{\boldsymbol{\mathfrak{b}}}_\nu\left(\boldsymbol{\lambda},
\boldsymbol{\Upsilon}\right)\, & =&\frac{1}{ Z^{\boldsymbol{\mathfrak{b}}}(\boldsymbol{\lambda})} \times
\exp\left[- \,\beta^0 \mathcal{P}_{c|0}(\boldsymbol{\Upsilon})- \sum_{i=1}^{\nu+1} \beta^i \mathcal{P}_{c|i}(\boldsymbol{\boldsymbol{\Upsilon}}) - \sum_{j=1}^{\nu+1} 
h^j \sqrt{\boldsymbol{\mathfrak{C}}_j(\boldsymbol{\boldsymbol{\Upsilon}})}\right]\nonumber\\
\mathbf{G}^{\boldsymbol{\mathfrak{d}}}_\nu\left(\boldsymbol{\lambda},\boldsymbol{\Upsilon}\right)\, & =&\frac{1}{ Z^{\boldsymbol{\mathfrak{d}}}(\boldsymbol{\lambda})} \times
\exp\left[-  \,\beta^0 \mathcal{P}_{c|0}(\boldsymbol{\Upsilon})- \sum_{i=1}^{\nu+1} \beta^i \mathcal{P}_{c|i}(\boldsymbol{\boldsymbol{\Upsilon}})- \sum_{j=1}^{\nu} 
h^j \sqrt{\boldsymbol{\mathfrak{C}}_j(\boldsymbol{\boldsymbol{\Upsilon}})}\right]
\end{eqnarray}
Eq. (\ref{extendedGibbs}) becomes meaningful as soon the additional $\boldsymbol{\mathfrak{C}}_j(\boldsymbol{\Upsilon})$ functions are determined. We present their algorithmic construction in the next subsection but before launching into that presentation we stress the following relevant issue. From the point of view of a generalized physical theory of thermodynamics where the main goal is just the calculation of the partition function, the choice of the coordinates utilized for the microscopic K\"ahler manifold is irrelevant and those by means of which the integrals (\ref{maggiorana}) are more easily calculated correspond to the best choice. If on the other hand one has in mind applications to machine learning, the useful coordinates are only the solvable ones, in one of their possible versions, because of the strategic use, in Cartan Neural Networks, of the metric equivalence of the symmetric space with its corresponding solvable Lie group. For this reason 
the Gibbs probability distribution must be expressed in terms of
the solvable coordinates, as in eq.(\ref{extendedGibbs}). The usefulness of the new extended Souriau Gibbs distributions critically depends on the structure of the extra objects
$\boldsymbol{\mathfrak{C}}_j(\boldsymbol{\Upsilon})$
that we describe next. Before addressing such description we mention that the principle temporal-like temperature $\beta_0$ acts as a global regularizer. For the Souriau-Gibbs state to remain normalized and stable, $\beta_0$ must satisfy appropriate conditions, ensuring that the non-compact boost components of the coadjoint orbit be exponentially suppressed at the boundary, as we are going to show in a next coming detailed mathematical paper \cite{abelstrutture}.
\subsection{The extra actions completing the abelian structure}
\label{cannastorta}
We finally come to the determination of the additional actions completing the Abelian Structure that includes the moment maps
of the compact Cartan subalgebra:
\begin{equation}\label{cartolone}
  \mathbb{H} \, = \, \uu(1) \oplus \mathbb{H}^\prime \, = \, \so(2) \oplus \so(2+q)
\end{equation}
What we are going to do is the explicit algorithmic implementation of what we already sketched  in section \ref{momappamatra}. 
Since for our purposes in Cartan Neural Networks we need to use the so called triangular basis (see \cite{pgtstheory,TSnaviga,tassellandum}) where the invariant $\eta_t$ metric for the fundamental representation of the group $\mathrm{SO(2,2+q)}$ is the one displayed in eq.(\ref{etatdefi}) that we repeat below for reader's convenience while reading the present fundamental section:
\begin{equation}\label{etatbona}
  \eta_t \, = \, \left(
                         \begin{array}{cc|c|cc}
                           0 & 0 & \mathbf{0}_{1\times q} & 0 & 1 \\
                           0 & 0 & \mathbf{0}_{1\times q} & 1 & 0 \\
                           \hline
                           \mathbf{0}_{q\times 1} & \mathbf{0}_{q\times 1} & \mathbf{1}_{q\times q} & \mathbf{0}_{q\times 1} & \mathbf{0}_{q\times 1} \\
                           \hline
                           0 & 1 & \mathbf{0}_{1\times q} & 0 & 0 \\
                           1 & 0 & \mathbf{0}_{1\times q} & 0 & 0 \\
                         \end{array}
                       \right)
\end{equation}
and since, moreover, we need to put into evidence, at all steps, the role of the Paint Group $\mathrm{SO(q)} \subset \mathrm{SO(2+q)}$, we have a general form for the $X_c$, $\uu(1)$-generator:
\begin{equation}\label{Xcione}
\mathcal{C}_0 \, = \, X_c \, = \,  \left(
                         \begin{array}{cc||c||cc}
                           0 & \frac{1}{2\sqrt{2}} & \mathbf{0}_{1\times q} & -\frac{1}{2\sqrt{2}} & 0 \\
                           -\frac{1}{2\sqrt{2}} & 0 & \mathbf{0}_{1\times q} & 0 & \frac{1}{2\sqrt{2}} \\
                           \hline
                           \hline
                           \mathbf{0}_{q\times 1} & \mathbf{0}_{q\times 1} & \mathbf{0}_{q\times q} & \mathbf{0}_{q\times 1} & \mathbf{0}_{q\times 1} \\
                           \hline
                           \hline
                           \frac{1}{2\sqrt{2}} & 0 & \mathbf{0}_{1\times q} & 0 & -\frac{1}{2\sqrt{2}} \\
                           0 & -\frac{1}{2\sqrt{2}} & \mathbf{0}_{1\times q} & \frac{1}{2\sqrt{2}} & 0 \\
                         \end{array}
                       \right)
\end{equation}
and a general form also for the first Cartan generator of the subgroup $\mathrm{SO(q+2)}$:
\begin{equation}\label{CHp1}
\mathcal{C}_1 \, = \,  \left(
                         \begin{array}{cc||c|c||cc}
                           0 & 0 & \frac{1}{2}&\mathbf{0}_{1\times (q-1)} & 0 & 0 \\
                          0 & 0 & 0&\mathbf{0}_{1\times (q-1)} & 0 & 0 \\
                          \hline
                          \hline
                          -\frac{1}{2}& 0 & 0&\mathbf{0}_{1\times (q-1)} & 0 & -\frac{1}{2} \\ 
                           \hline
                          0 & 0 & 0& \mathbf{0}_{(q-1)\times (q-1)}  & 0 &0\\ 
                           0 & 0 & 0&\mathbf{0}_{1\times (q-1)} & 0 & 0 \\
                           \hline\hline
                           0 & 0 & 0&\mathbf{0}_{1\times (q-1)} & 0 & 0 \\
                           0 & 0 & \frac{1}{2}&\mathbf{0}_{1\times (q-1)} & 0 & 0 \\
                         \end{array}
                       \right)
\end{equation}
which exists also when $q=1$, namely when the Paint Group $\mathrm{SO(q)}$ vanishes, $q=1$ corresponding to the maximally split Tits Satake case. All the other $\nu$ Cartan generators are canonical Cartan generators of the Paint Group $\mathrm{SO(q)}$, like, for instance, for the case $\nu=1$ (q=3) of the odd $\boldsymbol{\mathfrak{b}}$-series the following one:
\begin{equation}
\mathcal{C}_2 \, = \,  \left(                         \begin{array}{cc||c|cc||cc}
                         0 & 0 & 0& 0 & 0&0&0 \\
                         0 & 0 & 0& 0 & 0&0&0  \\
                          \hline
                          \hline
                          0 & 0 & 0& 0 & 0&0&0 \\ 
                           \hline
                          0 & 0 & 0&0 &\frac{1}{\sqrt{2}}&0&0\\ 
                          0 & 0 & 0&-\frac{1}{\sqrt{2}} & 0 & 0 &0\\
                           \hline\hline
                           0 & 0 & 0&0 & 0 & 0 &0\\
                            0 & 0 & 0&0 & 0 & 0 &0\\ 
                         \end{array}
                       \right)
\end{equation}
or for the case $\nu=2$ of the same odd $\boldsymbol{\mathfrak{b}}$-series the following two ones:
\begin{eqnarray}
\label{bringo}
  \mathcal{C}_2 &=& \left(\begin{array}{cc||c|cccc||cc}
                        0 & 0 & 0& 0 & 0&0&0&0&0 \\
                        0 & 0 & 0& 0 & 0&0&0&0&0 \\
                          \hline
                          \hline
                         0 & 0 & 0& 0 & 0&0&0&0&0 \\ 
                           \hline
                          0 & 0 & 0&0 &\frac{1}{\sqrt{2}}&0&0&0&0\\ 
                          0 & 0 & 0&-\frac{1}{\sqrt{2}}&0&0 & 0 & 0 &0\\
                          0 & 0 & 0& 0 & 0&0&0&0&0 \\
                          0 & 0 & 0& 0 & 0&0&0&0&0 \\
                           \hline\hline
                           0 & 0 & 0& 0 & 0&0&0&0&0\\
                            0 & 0 & 0& 0 & 0&0&0&0&0\\ 
                         \end{array}
                       \right) \\
  \mathcal{C}_3 &=& \left(\begin{array}{cc||c|cccc||cc}
                        0 & 0 & 0& 0 & 0&0&0&0&0 \\
                        0 & 0 & 0& 0 & 0&0&0&0&0 \\
                          \hline
                          \hline
                         0 & 0 & 0& 0 & 0&0&0&0&0 \\ 
                           \hline
                          0 & 0 & 0&0 &0&0&0&0&0\\ 
                          0 & 0 & 0&0&0&0 & 0 & 0 &0\\
                          0 & 0 & 0& 0 & 0&0&\frac{1}{\sqrt{2}}&0&0 \\
                          0 & 0 & 0& 0 & 0&-\frac{1}{\sqrt{2}}&0&0&0 \\
                           \hline\hline
                           0 & 0 & 0& 0 & 0&0&0&0&0\\
                            0 & 0 & 0& 0 & 0&0&0&0&0\\ 
                         \end{array}
                       \right) \\
\end{eqnarray}
This produces an embedding chain of subalgebras and hence of subgroups that are defined by the choice of the Cartan generators of the compact subalgebra. The two series are the following ones 
\paragraph{The $\boldsymbol{\mathfrak{b}}_\nu$ embedding chain}
\begin{alignat}{9}\label{gianni1}
&\quad\quad\mathbb{H}_1 &\supset &\quad\quad\mathbb{H}_2 &\supset &\quad\quad\mathbb{H}_3&\supset &\dots &\supset &\quad\mathbb{H}_{\nu+1} \nonumber\\
&\,\mathfrak{so}(2\nu+3) &\supset &\,\mathfrak{so}(2\nu+1) &\supset &\,\mathfrak{so}(2\nu-1)&\supset &\dots &\supset &\quad\mathfrak{so}(3) 
\end{alignat}
\paragraph{The $\boldsymbol{\mathfrak{d}}_\nu$ embedding chain}
\begin{alignat}{10}\label{gianni2}
&\quad\quad\mathbb{H}_1 &\supset &\quad\mathbb{H}_2 &\supset &\quad\quad\mathbb{H}_3&\supset &\dots &\supset &\quad\mathbb{H}_{\nu}\,\,&\supset &\quad\mathbb{H}_{\nu+1} \nonumber\\
&\,\mathfrak{so}(2\nu+2) &\supset &\,\mathfrak{so}(2\nu) &\supset &\,\mathfrak{so}(2\nu-2)&\supset &\dots &\supset &\,\mathfrak{so}(4)&\supset &\quad\mathfrak{so}(2) 
\end{alignat}
and they are intrinsically defined, without the use of any explicit matrix index, in any specific coordinate basis by the following intrinsic rule:
\begin{equation}\label{carambolina}
  \mathbb{H}_i\,= \, \mathrm{N}[\mathcal{C}_{i-1},\mathbb{H}_{i-1}] \, - \,\{\mathcal{C}_{i-1}\}
\end{equation}
where $\mathrm{N}[\mathcal{C}_{i-1},\mathbb{H}_{i-1}]$ denotes the normalizer of the Cartan generator $\mathcal{C}_{i-1}$ in the previous subalgebra:
\begin{equation}\label{craniotto}
  \mathrm{N}[\mathcal{C}_{i-1},\mathbb{H}_{i-1}] \, = \,
  \left\{Y\in\mathbb{H}_{i-1} \, \mid \, \left[\mathcal{C}_{i-1}\, , \, Y \right]\, = \, 0  \right\}
\end{equation}
and the subtraction in eq.(\ref{carambolina}) is in the set-theoretic sense. We have already remarked that in the case of the $\boldsymbol{\mathfrak{d}}$-series the last group in the sequence is $\so(2)$ made of a single generator that is the already counted last Cartan generator.
\par
\par
Given the embedding series described above and the complete set of $\mathbb{U}$-generators $J_\Lambda$ normalized as in eq.(\ref{gelindoelapecora}) that defines the basic moment-maps of the $\mathbb{U}$ Lie algebra, the extra actions completing the abelian structure are obtained in terms of the casimirs of the subalgebras $\mathbb{H}_j$, namely:
and one obtains the extra actions completing the abelian structure by setting
\begin{equation}\label{gulini2}
 \sqrt{ \boldsymbol{\mathfrak{C}}_{j}(\boldsymbol{\Upsilon})}  \,\equiv \, \sqrt{\sum_{J_\Lambda\in\mathbb{H}_j} \mathcal{P}^2_\Lambda(\boldsymbol{\Upsilon})}
 \quad ; \quad \quad j=\left\{\begin{array}{cccc}
                                1, & \dots , & \nu+1 & \text{case $\boldsymbol{\mathfrak{b}}$} \\
                                1, & \dots , & \nu & \text{case $\boldsymbol{\mathfrak{d}}$}
                              \end{array}\right.
\end{equation}
Indeed by the very definition of the chosen construction all the introduced actions 
are in involution as they should, namely commute with eachother in the Poisson bracket. 
\par
As we are going to show in the detailed mathematical derivation to be presented in a forthcoming paper \cite{abelstrutture}, once rewritten in terms of the action angle variables $\boldsymbol{\mathfrak{p}}^a,\boldsymbol{\mathfrak{q}}_a$ the integral on the actions $\boldsymbol{\mathfrak{p}}^a$ is extended to certain convex polytopes whose boundaries  provide the positivity constraint on the generalized temperatures $\boldsymbol{\lambda}$. In this paper we skip the description of polytopes and we just report the result of the integration that, knowing the integration limits, is a straightforward linear exponential integral. 
\subsubsection{Generalized Souriau Partition Functions}
The final outcome of the above quoted integration yields the generalized Souriau partition functions.
In order to write their explicit form  in dependence of  the components of the generalized temperature vector $\boldsymbol{\lambda}$ it is convenient to introduce the following auxiliary variables 
\begin{eqnarray} \label{eq:Omega_def_analysis}
\mu_k :&=& \beta_0 + \sum_{j=k}^{\nu+1} h_j 
\quad k=1, \dots, \nu+1 \quad \quad \,\, \, ; \quad  \text{case $\boldsymbol{\mathfrak{b}}$} \nonumber\\
\hat{\mu}_k :&=& \beta_0 + \sum_{j=k}^{\nu} h_j \,, \quad \hat{\mu}_0 := \hat{\mu}_1\,,
\quad k=1, \dots, \nu \quad ; \quad \text{case $\boldsymbol{\mathfrak{d}}$} 
\end{eqnarray}
\paragraph{Partition function for the $\boldsymbol{\mathfrak{b}}$-series.} In this case the partition functions with the extended temperature vector $\boldsymbol{\lambda}$
\begin{equation} \label{eq:final_Z_odd_sequential}
Z_{\nu}^{\boldsymbol{\mathfrak{b}}}(\boldsymbol{\lambda}) =  \frac{2^{\nu+1} (2\pi)^{2\nu+3} e^{-\beta_0} }{\beta_0\prod_{k=1}^{\nu+1} [\mu_k^2 - \beta_k^2]}
\end{equation}
\par
The largest possible domain of the parameters $\boldsymbol{\lambda}$ for which the partition function $Z_{\nu}^{\boldsymbol{\mathfrak{b}}}(\boldsymbol{\lambda})$ is convergent and positive definite is the intersection of $\nu+2$  linear half-spaces:
\begin{equation} \label{eq:D_Theta_Odd}
D^{\boldsymbol{\mathfrak{b}}}_{\nu}(\boldsymbol{\lambda}) = \left\{ (\boldsymbol{\beta}, \boldsymbol{h}) \in \mathbb{R}^{2\nu+3} \mid \beta_0 > 0, \quad \mu_k > |\beta_k| \text{ for } k=1, \dots, \nu+1 \right\}
\end{equation}
\paragraph{Partition function for the $\boldsymbol{\mathfrak{d}}$-series.}
Integrating on the appropriate  polytopes one obtains\footnote{In the case q=even we use $q=2s$ ($s\in \mathbb{N}$). This allows to distinguish the even and odd case visually.}: 
\begin{equation} \label{eq:final_Z_even_sequential}
    Z_{s}^{\boldsymbol{\mathfrak{d}}}(\boldsymbol{\mathfrak{\lambda}}) = \frac{2^{s+1} (2\pi)^{2(s+1)}  \hat \mu_1 e^{-\beta_0}}{\beta_0 \prod_{k=1}^{s+1} [\hat \mu_{k-1}^2 - \beta_k^2]}
\end{equation}
The largest possible domain $D_{\boldsymbol{\mathfrak{d}}} \subset \mathbb{R}^{2s+1}$ of the parameters $\boldsymbol{\mathfrak{d}}$ for which the partition function $Z_{s}^{\boldsymbol{\mathfrak{d}}}(\boldsymbol{\mathfrak{\lambda}}) $ is convergent and positive definite is the intersection of the following linear half-spaces:
\begin{equation} \label{eq:D_Theta_Even}
\hspace{-0.2cm} \hat D_{\boldsymbol{\mathfrak{d}}} = \left\{ (\boldsymbol{\beta}, \boldsymbol{h}) \in \mathbb{R}^{2s+1} \mid \beta_0 > 0, \quad \hat \mu_1 > \max(|\beta_1|, |\beta_2|), \quad \hat{\mu}_{k-1} > |\beta_k| \text{ for } k=2, \dots, s \right\}
\end{equation}
\paragraph{The Adjoint $\mathrm{U}$-orbit limit of the compact Cartan subalgebra (Souriau)}
By setting to zero all the extra temperatures (those not associated with the Cartan subalgebra of $\mathbb{H}\subset \mathbb{U}$), namely by setting ${h}^i = 0$ in $ Z_{\nu}^{\boldsymbol{\mathfrak{b}},\boldsymbol{\mathfrak{d}}}(\boldsymbol{\mathfrak{\lambda}})$ we find that $\mu_k = \beta_0$ and we retrieve the result of
eq.(\ref{zolotayaformula}) 
All the mathematical details on the derivation of the above general results will be presented in a forthcoming dedicated paper \cite{abelstrutture}.
\subsection{The explicit example  $\nu=0$ in the
$\boldsymbol{\mathfrak{b}}$-series}
In order to illustrate the above explained general results we concentrate on the lowest lying example ($\nu=0$) of the $\boldsymbol{\mathfrak{b}}$-series. The choice of the latter has a very simple reason:  the case $q=1$ corresponds to the unique Tits Satake submanifold of the entire TS-universality class, namely $\mathcal{M}^{[2,3]}\, = \,\mathrm{SO(2,3)/SO(2)\times SO(3)}$ which, by low rank Lie Algebra homomorphisms, is also locally isomorphic to the Siegel plane of genus $2$, namely $\mathcal{M}^{[2,3]}\, \simeq \, \mathrm{Sp(4,\mathbb{R})/U(1)\times SU(2)}$. In section \ref{theTSmanifoldo} we are also going to show how the partition function of this lowest lying case can be alternatively obtained by explicit integration in solvable coordinates, reproducing in this way, at least in the Souriau limit, the result obtained with the above  exposed general methods. The comparison with the complexity of the direct integral provide an additional evidence of the relevance of the general methods based on the compact abelian structure that allow the calculation of the partition function in all cases. 
\par
In the case $\nu = 0$, the partition function, apart from irrelevant constant factors, including the unique extra temperature $h_1$ has the following form:
\begin{equation}\label{nuzeropartfun}
  Z_0^{\boldsymbol{\mathfrak{b}}}(\beta_0,\beta_1,h_1) \, \propto \, \frac{\exp[-\beta^0]}{\beta_0\left[\left(\beta_0+h_1\right)^2\, - \,\beta_1^2\right]}
\end{equation}
\par
Utilizing the solvable coset representative in the $\mathbf{W}$-basis (\ref{krollus})
using the generators later displayed in table (\ref{v10genni}) and applying the general formula (\ref{gelindoelapecora}) we obtain the following  moment-maps 
\begin{eqnarray}\label{biliardino}
&&  \begin{array}{|lcl|}
  \hline
 \mathfrak{P}^{\so(2,3)}_1\, \equiv\, P_1 & = & \frac{1}{8} e^{-w_1-w_2} \left(2 w_4-2 e^{w_2} w_5 w_6-e^{2 w_2} w_3
   \left(w_6^2+2\right)\right)    \\
 \mathfrak{P}^{\so(2,3)}_2\, \equiv\, P_2 & = & \frac{1}{8} e^{-w_1-w_2} \left(2 w_4+e^{2 w_2} w_3
   \left(w_6^2+2\right)\right)   \\
 \mathfrak{P}^{\so(2,3)}_3\, \equiv\, P_3 & = & \frac{e^{-w_1-w_2} \left(w_5^2+2 e^{w_2} w_3 w_6 w_5+2 e^{2
   w_1}+e^{2 w_2} \left(w_3^2-1\right)
   \left(w_6^2+2\right)\right)}{8 \sqrt{2}}  \\
 \mathfrak{P}^{\so(2,3)}_4\, \equiv\, P_4 & = & \frac{e^{-w_1-w_2} \left(-4 w_4^2+4 e^{w_2} w_5 w_6 w_4-2 e^{2
   \left(w_1+w_2\right)} \left(w_6^2+2\right)-e^{2 w_2} w_5^2
   \left(w_6^2+2\right)+4\right)}{16 \sqrt{2}}  \\
 \mathfrak{P}^{\so(2,3)}_5\, \equiv\, P_5 & = & \frac{e^{-w_1-w_2} \left(-2 w_4 w_5+e^{2 w_2} w_3
   \left(w_6^2+2\right) w_5-2 e^{2 w_1+w_2} w_6-e^{w_2}
   \left(-w_5^2+2 w_3 w_4+2\right) w_6\right)}{8 \sqrt{2}}   \\
 \mathfrak{P}^{\so(2,3)}_6\, =\, P_6 & = & \frac{e^{-w_1-w_2} \left(2 e^{w_2} \left(w_3-w_4\right) w_6+w_5
   \left(e^{2 w_2} \left(w_6^2+2\right)+2\right)\right)}{8 \sqrt{2}}   \\
   \hline
 \mathfrak{P}^{\so(2,3)}_7\, \equiv\, \mathcal{P}_3 & = & \frac{e^{-w_1-w_2} \left(-e^{2 w_2} \left(w_6^2+2\right) w_5+2 w_5+2
   e^{w_2} \left(w_3+w_4\right) w_6\right)}{8 \sqrt{2}}   \\
 \mathfrak{P}^{\so(2,3)}_8\, \equiv\, \mathcal{P}_1 & = & \frac{e^{-w_1-w_2} \left(2 w_4 w_5-e^{2 w_2} w_3
   \left(w_6^2+2\right) w_5+2 e^{2 w_1+w_2} w_6+e^{w_2}
   \left(-w_5^2+2 w_3 w_4-2\right) w_6\right)}{8 \sqrt{2}}  \\
 \mathfrak{P}^{\so(2,3)}_9\, \equiv\, \mathcal{P}_2 & = & \frac{1}{32} e^{-w_1-w_2} \left(4 w_4^2-4 e^{2 w_1}-2 w_5^2-4
   e^{w_2} \left(w_3+w_4\right) w_5 w_6+2 e^{2 \left(w_1+w_2\right)}
   \left(w_6^2+2\right)\right.\\
   \null & \null &\left. -e^{2 w_2} \left(2 w_3^2-w_5^2+2\right)
   \left(w_6^2+2\right)+4\right)   \\
   \hline
 \mathfrak{P}^{\so(2,3)}_{10}\, \equiv\, \mathcal{P}_0 & = & \frac{1}{32} e^{-w_1-w_2} \left(-2 \left(2 w_4^2+w_5^2+2\right)-4
   e^{2 w_1}-4 e^{w_2} \left(w_3-w_4\right) w_5 w_6\right.\\
   \null&\null &\left.-2 e^{2
   \left(w_1+w_2\right)} \left(w_6^2+2\right)-e^{2 w_2} \left(2
   w_3^2+w_5^2+2\right) \left(w_6^2+2\right)\right) \\
   \hline
\end{array} 
\end{eqnarray}
that should be compared with their analogues in the different solvable coordinate basis $\boldsymbol{\Upsilon}$ displayed instead in eq.(\ref{carambola}).
\par
According with the general theory discussed above the Darboux basis for the 6-dimensional manifold is composed of three actions $\boldsymbol{\mathfrak{p}}_{a}(\mathbf{W})$ and three angles $\boldsymbol{\mathfrak{q}}^{a}(\mathbf{W})$. The actions are equal to:
\begin{eqnarray}
  \boldsymbol{\mathfrak{p}}_{0}(\mathbf{W}) &=& \mathcal{P}_0 \nonumber \\
  \boldsymbol{\mathfrak{p}}_{1}(\mathbf{W}) &=& \mathcal{P}_2 \\
  \boldsymbol{\mathfrak{p}}_{2}(\mathbf{W}) &=& \sqrt{\boldsymbol{\mathfrak{C}}[\mathbb{H}_1]} \, = \, \sqrt{\mathcal{P}_1^2+\mathcal{P}_2^2+\mathcal{P}_3^2}
\end{eqnarray}
and their explicit form in terms of solvable coordinates is precisely defined by eq.s(\ref{biliardino}).
In particular with simplifications the Casimir function has the following appearance:
\begin{eqnarray}\label{casimiro23}
  \boldsymbol{\mathfrak{C}}[\mathbb{H}_1] & = & \frac{e^{-2 \left(w_1+w_2\right)}}{1024}\times \left( 
  e^{4 w_2} \left(\left(2 w_3^2+w_5^2+2\right){}^2+4 e^{4 w_1}-4 e^{2
   w_1} \left(2 w_3^2-w_5^2+2\right)\right)
   \left(w_6^2+2\right){}^2\right.\nonumber\\
   &&\left.+8 e^{3 w_2} w_5
   \left(\left(w_3-w_4\right) \left(2 w_3^2+w_5^2+2\right)-2 e^{2
   w_1} \left(3 w_3+w_4\right)\right) w_6 \left(w_6^2+2\right)+4
   \left(\left(2 w_4^2+w_5^2+2\right){}^2+4 e^{4 w_1}\right.\right.\nonumber\\
   &&\left.\left.
   -4 e^{2 w_1}
   \left(2 w_4^2-w_5^2+2\right)\right)+16 e^{w_2} w_5 \left(2 e^{2
   w_1} \left(w_3+3 w_4\right)+\left(w_3-w_4\right) \left(2
   w_4^2+w_5^2+2\right)\right) w_6\right.\nonumber\\
   &&\left.+4 e^{2 w_2} \left(\left(4
   \left(w_6^2-2\right) w_4^2+w_5^2 \left(6 w_6^2+4\right)+4
   \left(w_6^2+e^{2 w_1} \left(w_6^2+2\right)-2\right)\right)
   w_3^2\right.\right.\nonumber\\
   &&\left.\left.-8 w_4 \left(w_5^2 \left(w_6^2+2\right)-2 e^{2 w_1}
   w_6^2\right) w_3\right.\right.\nonumber\\
   &&\left.\left.+4 e^{4 w_1}
   \left(w_6^2-2\right)+\left(w_5^2+2\right){}^2
   \left(w_6^2-2\right)-4 e^{2 w_1} \left(\left(3 w_6^2+2\right)
   w_5^2+2 \left(w_6^2-2\right)\right)+w_4^2 \left(\left(6
   w_6^2+4\right) w_5^2\right.\right.\right.\nonumber\\
   &&\left.\left.\left.+4 \left(w_6^2+e^{2 w_1}
   \left(w_6^2+2\right)-2\right)\right)\right)\right)
\end{eqnarray}
The relevant point is that, thanks to the intrinsic definition of all the objects and to the exact evaluation of the partition function in eq.(\ref{nuzeropartfun}), the Gibbs distribution over the genus 2 Siegel plane:
\begin{equation}\label{fachiro}
  \mathrm{G}_{Siegel}(\beta_0,\beta_1,h_1,\zeta)\, = \, \beta_0\left[\left(\beta_0+h_1\right)^2\, - \,\beta_1^2\right]\times
  \exp\left[\beta_0 \left(1-\boldsymbol{\mathfrak{p}}_0(\zeta)\right)-\beta_1 \boldsymbol{\mathfrak{p}}_1(\zeta) \, - \,h_1 \boldsymbol{\mathfrak{p}}_2(\zeta) \right]
\end{equation}
has a universal form in terms of the actions $\boldsymbol{\mathfrak{p}}_{a}(\zeta)$ where by $\zeta \in \mathbb{SH}_2 \, = \,\frac{\mathrm{SO(2,3)}}{\mathrm{SO(2) \times SO(3)}}$ we denote a point on the Siegel plane, in whatever coordinates one prefers to utilize. For instance, instead of the solvable coordinates $\mathbf{W}$, one can use the solvable coordinates $\boldsymbol{\Upsilon}$, defined in \cite{pgtstheory} and utilized in the next section \ref{theTSmanifoldo}, by inserting in  eq.(\ref{fachiro}) the expression of the moment maps presented in eq.(\ref{carambola}) rather than that exhibited in eq. (\ref{biliardino}). In particular referring to the solvable coordinate frame $\boldsymbol{\Upsilon}$ in eq.(7.63) of \cite{pgtstheory}) one finds the correspondence between the solvable coordinates $\Upsilon$ and the entries of complex symmetric $2\times 2$ matrix $Z$ with positive definite imaginary part which is the traditional parameterization of Siegel half plane\footnote{In eq.(7.63) of \cite{pgtstheory} the solvable coordinates are named $w_i$, yet they are those that in the present article we name $\Upsilon_i$.
In \cite{pgtstheory} we used various letters for the solvable coordinates depending on the occasion, always utilizing the same exponentiation scheme. It is only in the  present article that we felt the need to introduce the distinction because we introduced a second  exponentiation scheme finalized to put into evidence the maximal abelian ideal.}. 
Furthermore, if needed, one can use the expression of the corresponding moment maps in Type I or Type II CV coordinates that we have discussed in previous sections in particular in section \ref{momentissimi}.
\subsubsection{The angles}
As we stressed above, once the result for the partition function is established, the Gibbs distribution and the thermodynamical geometry are fixed and there is no need to consider Calabi-Vesentini coordinates of type I any longer. One can freely utilize the solvable coordinates that are those relevant for Machine Learning applications
(see \cite{pgtstheory},\cite{TSnavigation},\cite{tassellandum}). Yet the very result for the partition function relies on the existence of compact angles conjugate to each of the actions, namely on the compactness of the $\boldsymbol{\mathcal{Q}^n}$ manifold mentioned in the introduction (see text in between eq.s(\ref{cartolina}) and (\ref{ntorus}). In order to show how such angles can be found, we consider their construction in the present case and for this purpose it is convenient to resort to 
Type I Calabi-Vesentini coordinates. Hence we recall the expression of the three actions $\boldsymbol{\mathfrak{p}}_{a}$ in such coordinate basis. Relying on the results of section \ref{momappamatra} and in particular on eq.s (\ref{barlume}) and  (\ref{kurchatone}) we have:
\begin{eqnarray}
  \boldsymbol{\mathfrak{p}}_{0}(\mathbf{y},\bar{\mathbf{y}}) &=& \mathcal{P}_0 \, = \,\frac{\left(y_1^2+y_2^2+y_3^2\right)
   \left(\bar{y}_1^2+\bar{y}_2^2+\bar{y}_3^2\right)-1}{\sqrt{2}
   \left(-2 \left(y_1 \bar{y}_1+y_2 \bar{y}_2+y_3
   \bar{y}_3\right)+\left(y_1^2+y_2^2+y_3^2\right)
   \left(\bar{y}_1^2+\bar{y}_2^2+\bar{y}_3^2\right)+1\right)} \nonumber \\
  \boldsymbol{\mathfrak{p}}_{1}(\mathbf{y},\bar{\mathbf{y}}) &=& \mathcal{P}_2 \, = \, \frac{i \sqrt{2} \left(y_3 \bar{y}_2-y_2 \bar{y}_3\right)}{-2
   \left(y_1 \bar{y}_1+y_2 \bar{y}_2+y_3
   \bar{y}_3\right)+\left(y_1^2+y_2^2+y_3^2\right)
   \left(\bar{y}_1^2+\bar{y}_2^2+\bar{y}_3^2\right)+1} \nonumber\\
  \boldsymbol{\mathfrak{p}}_{2}(\mathbf{y},\bar{\mathbf{y}}) &=& \sqrt{\mathcal{P}_1^2+\mathcal{P}_2^2+\mathcal{P}_3^2} \, = \, \frac{\sqrt{-2 \left(y_2 \bar{y}_1-y_1 \bar{y}_2\right){}^2-2 \left(y_3
   \bar{y}_1-y_1 \bar{y}_3\right){}^2-2 \left(y_3 \bar{y}_2-y_2
   \bar{y}_3\right){}^2}}{-2
   \left(y_1 \bar{y}_1+y_2 \bar{y}_2+y_3
   \bar{y}_3\right)+\left(y_1^2+y_2^2+y_3^2\right)
   \left(\bar{y}_1^2+\bar{y}_2^2+\bar{y}_3^2\right)+1}
\end{eqnarray}
The construction of the conjugate angles is a little bit of an art. What is required is to construct, in terms of the complex coordinate $\mathbf{y}$ and $\bar{\mathbf{y}}$ a complex one-dimensional function $\mathit{f}(\mathbf{y},\bar{\mathbf{y}})$ whose argument (or phase) is the one that is shifted by the infinitesimal transformations generated by the hamiltonian vector field of which the considered action is the moment map. This means
\begin{eqnarray}\label{crunco}
  \boldsymbol{\mathfrak{k}}[\boldsymbol{\mathfrak{p}}] & \equiv & g^{ij^\star} \, \frac{\partial \boldsymbol{\mathfrak{p}} }{\partial \bar{y}^{j^\star}} \, \frac{\partial}{\partial y^i}\nonumber\\
  \boldsymbol{\mathfrak{k}}[\boldsymbol{\mathfrak{p}}] \, \text{ArcTan}\,\left[\,- \,\mathit{i} \,\frac{f-\bar{f}}{f+\bar{f}}\right] & = &  1 \nonumber\\
  &\Downarrow & \nonumber\\
  \left\{\boldsymbol{\mathfrak{p}} \, , \, \underbrace{\text{ArcTan}\,\left[\,- \,\mathit{i} \,\frac{f-\bar{f}}{f+\bar{f}}\right]}_{=\theta} \right\}_{PB} & = & 1
  \end{eqnarray}
The first two angles are relatively simple to be found since the hamiltonian vector field associated with the action is a Killing vector. 
\paragraph{The angle $\theta_0$}
The angle conjugate to the moment-map of the K\"ahler Killing vector (the generator of the fundamental $\so(2)$ subgroup of isometry) is the phase of the holomorphic function $f=\mathbf{y}^2=y_1^2+y_2^2+y_3^2$ which is general result for all Calabi-Vesentini manifolds:
\begin{equation}\label{tettazero}
  \theta_0 \, = \, -2  \text{ArcTan}\left(i\frac{y^2-\bar{y}^2}{\bar{y}^2+y^2}\right)
\end{equation}
Indeed we easily verify that:
\begin{equation}\label{formaggino}
  \left\{\boldsymbol{\mathfrak{p}}_{0}\, , \, \theta_0 \right\} \, = \, 1 \quad ; \quad  
  \left\{\boldsymbol{\mathfrak{p}}_{1}\, , \, \theta_0\right\} \, = \, \left\{\boldsymbol{\mathfrak{p}}_{2}\, , \,\theta_0 \right\} \, = \, 0
\end{equation}
\paragraph{The angles $\theta_{1,2}$}
In order to construct the remaining two angles we need two additional complex functions that we describe below:
\begin{eqnarray}
  \chi(\mathbf{y},\bar{\mathbf{\mathbf{y}}}) &=& \left(y_2+\text{iy}_3\right) \left(\sqrt{\left(y\cdot
   \bar{y}\right)^2-y^2 \bar{y}^2}+y\cdot \bar{y}\right)-y^2
   \left(\bar{y}_2+i \bar{y}_3\right) \label{chistu} \\
  \zeta(\mathbf{y},\bar{\mathbf{\mathbf{y}}}) &=& y_1 \left(\sqrt{\left(y\cdot \bar{y}\right)^2-y^2 \bar{y}^2}+y\cdot
   \bar{y}\right)-y^2 \bar{y}_1 \label{zetusto}
\end{eqnarray}
we set:
\begin{eqnarray}\label{unodue}
  \theta_2 &=& \theta _0\,+\,4\,  \text{ArcTan}\left(\mathit{i}\frac{\zeta -\bar{\zeta }}{\bar{\zeta
   }+\zeta }\right) \\
\theta_1 &=& 4 \, \text{ArcTan}\left(\mathit{i}\frac{\chi -\bar{\chi }}{\bar{\chi }+\chi
   }\right)-4 \, \text{ArcTan}\left(\mathit{i}\frac{\zeta -\bar{\zeta
   }}{\bar{\zeta }+\zeta }\right) 
\end{eqnarray}
and we can verify that:
\begin{eqnarray}\label{frenesia}
   \left\{\boldsymbol{\mathfrak{p}}_{1}\, , \, \theta_1 \right\} & = & 1 \quad ; \quad  
  \left\{\boldsymbol{\mathfrak{p}}_{0}\, , \, \theta_1\right\} \, = \, \left\{\boldsymbol{\mathfrak{p}}_{2}\, , \,\theta_1 \right\} \, = \, 0 \nonumber\\
  \left\{\boldsymbol{\mathfrak{p}}_{2}\, , \, \theta_2 \right\} & = & 1 \quad ; \quad  
  \left\{\boldsymbol{\mathfrak{p}}_{0}\, , \, \theta_2\right\} \, = \, \left\{\boldsymbol{\mathfrak{p}}_{1}\, , \,\theta_2 \right\} \, = \, 0
\end{eqnarray}
The same type of constructions can be implemented in all cases, yet as we already emphasized the explicit form of the angles is an irrelevant information for all applications and thermodynamic constructions. It suffices to know that they exist.
\section{The case of $\mathrm{SO(2,3)/SO(2)\times SO(3)}$ in solvable coordinates $\boldsymbol{\Upsilon}$}
\label{theTSmanifoldo}
In this section we discuss once again the case $q=1$ of Calabi-Vesentini manifold, that is the Tits Satake submanifold of the entire Tits-Satake universality class and coincides with the Siegel half-plane $\mathbb{SH}_2$. For $q=1$ we illustrate the derivation of the partition function (reduced to pure Souriau case) via an explicit integration in the solvable coordinate basis. Such a task had already been addressed in our previous paper \cite{geotermico}, yet there the complete analytic form of the partition function was not found and we had to stop at a numerical definition as a compiled function. With a strategic change of the order of integrations  we show here (details being provided in appendix \ref{integso23}) how we can arrive at an explicit analytic closed form which exactly coincides with the general form obtained by other techniques in section \ref{sashasecta}.  This provides a very important consistency check of the various approaches. Furthermore, by means of a Ward Identity differential equation, we provide a check of the transcription of our result in terms of $\mathrm{U}$-group invariants.  This is a most significant  basic confirmation of the main argument on which this and our previous paper \cite{geotermico} do rely, namely the $\mathrm{U}$-invariance of the partition function, when the temperature vector is restricted to adjoint $\mathbb{U}$-orbit the compact Cartan subalgebra. 
When the extra temperatures $\gamma$.s are switched on the $\mathbb{U}$ symmetry is broken. We postpone to a future publication the investigation of all the possible symmetry breaking patterns and of the associated phase-transitions.
\subsection{The generator basis of the Lie algebra $\mathbb{U}=\so(2,3)$ and the moment maps in the $\boldsymbol{\Upsilon}$-solvable coordinate frame}
Although we mostly rely on our previous paper \cite{geotermico} for general concepts
and for a lot of background material, in order to make the present paper independently readable and self-consistent, we report here once again some essential definitions and explicit formulae. In particular we need the explicit list of the standardized generators of the full Lie algebra $\mathbb{U}=\so(2,3)$ in the vector representation
(displayed in table \ref{v10genni}) and the ensuing explicit form of the moment maps.
\par
The $10$ generators, $J_i$ are ordered in the following way:
\begin{equation}\label{cosettico}
  J_{1,\dots,6} \,=\,K_{i}
\end{equation}
are the $6$ non-compact coset generators spanning the vector subspace $\mathbb{K}$ in the orthogonal decomposition
\begin{equation}\label{trappolatura}
  \mathbb{U} \, = \, \mathbb{H} \, \oplus \, \mathbb{K}
\end{equation}
Furthermore the first two generators $J_{1,2}$ are the two non-compact Cartan generators.
\par
The generators:
\begin{equation}\label{hattico}
  J_{7,8,9} \,=\,H_{1,2,3}
\end{equation}
are the generators of the $\su(2)\simeq\so(3)$ subalgebra of the compact subalgebra $\mathbb{H}\, = \, \su(2)\oplus \uu(1)$.
Finally
\begin{equation}\label{konditorei}
 J_{10} \, \equiv \, X_c 
\end{equation}
is the $\uu(1)\simeq \so(2)$ generator responsible for the K\"ahler structure.
\paragraph{The coset representative in $\boldsymbol{\Upsilon}$ solvable coordinates} was already displayed in
\cite{pgtstheory} and \cite{geotermico}. We repeat it here for convenience. In the whole literature we have recently produced on the CaNNs \cite{pgtstheory,naviga,TSnaviga,tassellandum,geotermico}, just as in the original literature on the metric equivalence and on the Tits Satake projection \cite{noipainted,cmappotto,noiconsasha,sashaebog,arrowtime,pancetta2,pancetta1} and also in  the monographic book \cite{advancio}, the solvable coset representative is
denoted by the letter $\mathbb{L}$, while the solvable coordinates from which it depends, are alternatively denoted by the $\boldsymbol{\Upsilon}$ or by $\mathbf{W}$. The coset representative that we display here below is written according with the $\Sigma$ exponential map utilized in our previous papers \cite{pgtstheory,naviga,TSnaviga,tassellandum,geotermico} and not according with the $\Sigma^\prime$ exponential map of eq.(\ref{krollus}) that we utilized above in order to introduce complex coordinates whose real parts correspond to the maximal abelian ideal of translations. In order to mark the difference and to avoid confusions, while we used the $(\mathbf{W},w_A)$-letters for the solvable coordinates generated by the $\Sigma^\prime$ form of the exponential map, we utilize the notation $(\boldsymbol{\Upsilon},\Upsilon_A)$-letters for the solvable coordinates generated by the $\Sigma$ exponential map. In both cases we utilize the letter $\mathbb{L}$ for the coset representative expressed as a solvable Lie group element and we add  a subscript with the name of the group $\mathrm{U}$ to which it is related  since in the present paper we discuss different $\mathrm{U}$-groups of the Calabi Vesentini series. Hence for $\mathrm{SO(2,3)}$ we have:
\begin{equation}\label{so23cosetrep}
  \mathbb{L}_{\so(2,3)}\left(\boldsymbol{\Upsilon}\right)=\left(
\begin{array}{ccccc}
 e^{\Upsilon_1} & \frac{e^{\Upsilon_1} \Upsilon_3}{\sqrt{2}} & \frac{1}{2} e^{\Upsilon_1} \left(\sqrt{2} \Upsilon_5+\Upsilon_3 \Upsilon_6\right) & \frac{1}{8} e^{\Upsilon_1} \left(4 \sqrt{2} \Upsilon_4-4 \Upsilon_5
\Upsilon_6-\sqrt{2} \Upsilon_3 \Upsilon_6^2\right) & -\frac{1}{4} e^{\Upsilon_1} \left(2 \Upsilon_3 \Upsilon_4+\Upsilon_5^2\right) \\
 0 & e^{\Upsilon_2} & \frac{e^{\Upsilon_2} \Upsilon_6}{\sqrt{2}} & -\frac{1}{4} e^{\Upsilon_2} \Upsilon_6^2 & -\frac{e^{\Upsilon_2} \Upsilon_4}{\sqrt{2}} \\
 0 & 0 & 1 & -\frac{\Upsilon_6}{\sqrt{2}} & -\frac{\Upsilon_5}{\sqrt{2}} \\
 0 & 0 & 0 & e^{-\Upsilon_2} & -\frac{e^{-\Upsilon_2} \Upsilon_3}{\sqrt{2}} \\
 0 & 0 & 0 & 0 & e^{-\Upsilon_1} \\
\end{array}
\right)
\end{equation}
which is manifestly upper triangular and constitutes an element of  the $6$-dimensional solvable Lie group $\mathcal{S}_{[2,3]}\subset \mathrm{SO(2,3)}$
\par
Utilizing the general formula in eq.(\ref{gelindoelapecora}) one obtains the explicit result for the 10 moment maps associated with the Killing vectors  that are displayed below:
\begin{eqnarray}\label{carambola}
&&  \begin{array}{|lcl|}
  \hline
 \mathfrak{P}^{\so(2,3)}_1 & = & \frac{1}{16} \left(4 \sqrt{2} \Upsilon_4-4 \Upsilon_5 \Upsilon_6-\sqrt{2}
   \Upsilon_3 \left(\Upsilon_6^2+4\right)\right)  \\
 \mathfrak{P}^{\so(2,3)}_2 & = & \frac{4 \Upsilon_4+\Upsilon_3 \left(\Upsilon_6^2+4\right)}{8 \sqrt{2}} \\
 \mathfrak{P}^{\so(2,3)}_3 & = & \frac{1}{32} \left(e^{\Upsilon_1-\Upsilon_2} \left(\sqrt{2}
   \left(\Upsilon_6^2+4\right) \Upsilon_3^2+4 \Upsilon_5 \Upsilon_6 \Upsilon_3+2 \sqrt{2} \left(\Upsilon_5^2+4\right)\right)-2
   \sqrt{2} e^{\Upsilon_2-\Upsilon_1} \left(\Upsilon_6^2+4\right)\right) \nonumber \\
 \mathfrak{P}^{\so(2,3)}_4 & = & \frac{1}{64} e^{-\Upsilon_1-\Upsilon_2} \left(16 \sqrt{2}-e^{2
   \left(\Upsilon_1+\Upsilon_2\right)} \left(8 \sqrt{2} \Upsilon_4^2-8 \Upsilon_5 \Upsilon_6 \Upsilon_4+\sqrt{2}
   \left(\Upsilon_5^2+4\right) \left(\Upsilon_6^2+4\right)\right)\right) \nonumber \\
 \mathfrak{P}^{\so(2,3)}_5 & = & \frac{1}{32} e^{\Upsilon_1} \left(-4 \sqrt{2} \Upsilon_4 \Upsilon_5+\sqrt{2}
   \Upsilon_3 \left(\Upsilon_6^2+4\right) \Upsilon_5-4 \Upsilon_3 \Upsilon_4 \Upsilon_6+2 \left(\Upsilon_5^2-4\right)
   \Upsilon_6\right)-\frac{1}{4} e^{-\Upsilon_1} \Upsilon_6 \nonumber \\
 \mathfrak{P}^{\so(2,3)}_6 & = & \frac{1}{16} e^{-\Upsilon_2} \left(2 \sqrt{2} \left(\Upsilon_3-e^{2
   \Upsilon_2} \Upsilon_4\right) \Upsilon_6+\Upsilon_5 \left(e^{2 \Upsilon_2} \left(\Upsilon_6^2+4\right)+4\right)\right)  \\
   \hline
 \mathfrak{P}^{\so(2,3)}_7 & = & \frac{1}{16} e^{-\Upsilon_2} \left(2 \sqrt{2} \left(\Upsilon_3+e^{2
   \Upsilon_2} \Upsilon_4\right) \Upsilon_6+\Upsilon_5 \left(4-e^{2 \Upsilon_2} \left(\Upsilon_6^2+4\right)\right)\right)  \\
 \mathfrak{P}^{\so(2,3)}_8 & = & -\frac{1}{4} e^{-\Upsilon_1} \Upsilon_6-\frac{1}{32} e^{\Upsilon_1} \left(-4
   \sqrt{2} \Upsilon_4 \Upsilon_5+\sqrt{2} \Upsilon_3 \left(\Upsilon_6^2+4\right) \Upsilon_5-4 \Upsilon_3 \Upsilon_4 \Upsilon_6+2
   \left(\Upsilon_5^2-4\right) \Upsilon_6\right)  \\
 \mathfrak{P}^{\so(2,3)}_9 & = & \frac{1}{64} e^{-\Upsilon_1-\Upsilon_2} \left[-4 e^{2 \Upsilon_2}
   \left(\Upsilon_6^2+4\right)-2 e^{2 \Upsilon_1} \left(\left(\Upsilon_6^2+4\right) \Upsilon_3^2+2 \sqrt{2} \Upsilon_5 \Upsilon_6
   \Upsilon_3+2 \left(\Upsilon_5^2+4\right)\right)\right. \\
   \null & \null & \left.+e^{2 \left(\Upsilon_1+\Upsilon_2\right)} \left(8 \Upsilon_4^2-4 \sqrt{2}
   \Upsilon_5 \Upsilon_6 \Upsilon_4+\left(\Upsilon_5^2+4\right) \left(\Upsilon_6^2+4\right)\right)+16\right]  \\
   \hline
 \mathfrak{P}^{\so(2,3)}_{10} & = & \frac{1}{64} e^{-\Upsilon_1-\Upsilon_2} \left[-4 e^{2 \Upsilon_2}
   \left(\Upsilon_6^2+4\right)-2 e^{2 \Upsilon_1} \left(\left(\Upsilon_6^2+4\right) \Upsilon_3^2+2 \sqrt{2} \Upsilon_5 \Upsilon_6
   \Upsilon_3+2 \left(\Upsilon_5^2+4\right)\right)\right.\\
   \null & \null & \left.-e^{2 \left(\Upsilon_1+\Upsilon_2\right)} \left(8 \Upsilon_4^2-4 \sqrt{2}
   \Upsilon_5 \Upsilon_6 \Upsilon_4+\left(\Upsilon_5^2+4\right) \left(\Upsilon_6^2+4\right)\right)-16\right] \\
   \hline
\end{array}\nonumber\\
&&
\end{eqnarray}
\begin{table}[htb]
$$
\begin{array}{||lclcl||lclcl||}
\hline\hline
 J_1 & = & \sqrt{2}\,K_1 & = &  \left(
\begin{array}{ccccc}
 1 & 0 & 0 & 0 & 0 \\
 0 & 0 & 0 & 0 & 0 \\
 0 & 0 & 0 & 0 & 0 \\
 0 & 0 & 0 & 0 & 0 \\
 0 & 0 & 0 & 0 & -1 \\
\end{array}
\right) & J_2 & = & \sqrt{2}\,K_2 & = & \left(
\begin{array}{ccccc}
 0 & 0 & 0 & 0 & 0 \\
 0 & 1 & 0 & 0 & 0 \\
 0 & 0 & 0 & 0 & 0 \\
 0 & 0 & 0 & -1 & 0 \\
 0 & 0 & 0 & 0 & 0 \\
\end{array}
\right)\\
\hline
 J_3 & = & \sqrt{2}\,K_3 & = & \left(
\begin{array}{ccccc}
 0 & \frac{1}{\sqrt{2}} & 0 & 0 & 0 \\
 \frac{1}{\sqrt{2}} & 0 & 0 & 0 & 0 \\
 0 & 0 & 0 & 0 & 0 \\
 0 & 0 & 0 & 0 & -\frac{1}{\sqrt{2}} \\
 0 & 0 & 0 & -\frac{1}{\sqrt{2}} & 0 \\
\end{array}
\right) & J_4 & = & \sqrt{2}\,K_4 & = & \left(
\begin{array}{ccccc}
 0 & 0 & 0 & \frac{1}{\sqrt{2}} & 0 \\
 0 & 0 & 0 & 0 & -\frac{1}{\sqrt{2}} \\
 0 & 0 & 0 & 0 & 0 \\
 \frac{1}{\sqrt{2}} & 0 & 0 & 0 & 0 \\
 0 & -\frac{1}{\sqrt{2}} & 0 & 0 & 0 \\
\end{array}
\right) \\
\hline
 J_5 & = & \sqrt{2}\,K_5 & = & \left(
\begin{array}{ccccc}
 0 & 0 & \frac{1}{\sqrt{2}} & 0 & 0 \\
 0 & 0 & 0 & 0 & 0 \\
 \frac{1}{\sqrt{2}} & 0 & 0 & 0 & -\frac{1}{\sqrt{2}} \\
 0 & 0 & 0 & 0 & 0 \\
 0 & 0 & -\frac{1}{\sqrt{2}} & 0 & 0 \\
\end{array}
\right) & J_6 & = & \sqrt{2}\,K_6 & = & \left(
\begin{array}{ccccc}
 0 & 0 & 0 & 0 & 0 \\
 0 & 0 & \frac{1}{\sqrt{2}} & 0 & 0 \\
 0 & \frac{1}{\sqrt{2}} & 0 & -\frac{1}{\sqrt{2}} & 0 \\
 0 & 0 & -\frac{1}{\sqrt{2}} & 0 & 0 \\
 0 & 0 & 0 & 0 & 0 \\
\end{array}
\right)\\
\hline
 J_7 & = & H_1 & = & \left(
\begin{array}{ccccc}
 0 & 0 & 0 & 0 & 0 \\
 0 & 0 & \frac{1}{\sqrt{2}} & 0 & 0 \\
 0 & -\frac{1}{\sqrt{2}} & 0 & -\frac{1}{\sqrt{2}} & 0 \\
 0 & 0 & \frac{1}{\sqrt{2}} & 0 & 0 \\
 0 & 0 & 0 & 0 & 0 \\
\end{array}
\right) & J_8 & = & H_2 & = &  \left(
\begin{array}{ccccc}
 0 & 0 & \frac{1}{\sqrt{2}} & 0 & 0 \\
 0 & 0 & 0 & 0 & 0 \\
 -\frac{1}{\sqrt{2}} & 0 & 0 & 0 & -\frac{1}{\sqrt{2}} \\
 0 & 0 & 0 & 0 & 0 \\
 0 & 0 & \frac{1}{\sqrt{2}} & 0 & 0 \\
\end{array}
\right)\\
\hline
 J_9 & = & H_3 & = &  \left(
\begin{array}{ccccc}
 0 & \frac{1}{2} & 0 & \frac{1}{2} & 0 \\
 -\frac{1}{2} & 0 & 0 & 0 & -\frac{1}{2} \\
 0 & 0 & 0 & 0 & 0 \\
 -\frac{1}{2} & 0 & 0 & 0 & -\frac{1}{2} \\
 0 & \frac{1}{2} & 0 & \frac{1}{2} & 0 \\
\end{array}
\right) & J_{10} & = & H_0 & = &  \left(
\begin{array}{ccccc}
 0 & \frac{1}{2} & 0 & -\frac{1}{2} & 0 \\
 -\frac{1}{2} & 0 & 0 & 0 & \frac{1}{2} \\
 0 & 0 & 0 & 0 & 0 \\
 \frac{1}{2} & 0 & 0 & 0 & -\frac{1}{2} \\
 0 & -\frac{1}{2} & 0 & \frac{1}{2} & 0 \\
\end{array}
\right)\\
\hline\hline
\end{array}
$$
\caption{\it In this table we display the complete list of the 10 generators  of the $\so(2,3)$ Lie algebra, expressed in the 5-dimensional fundamental representaion and in the trianagular basis. \label{v10genni}}
\end{table}
In eq.(\ref{carambola}) we have separated the moment-maps in the three groups. The first group of six are the moment maps of the $\mathbb{K}$ translations. The subsequent group of three yields the moment maps of the $\su(2)$ compact generators, while the last group of just one is the moment map of the $\uu(1)$ generator associated with the K\"ahler structure. Eq.s(\ref{carambola}) should be compared with
eq.s(\ref{biliardino}) in order to appreciate the difference between the two types of
solvable coordinate frames.
\subsection{Construction of the invariant tensors of the Lie algebra $\mathbb{U}$}
The Lie algebra $\mathbb{U}\,=\,\so(2,3)$ has total rank $r=2$ and therefor we expect only two independent invariant tensors corresponding to the two independent Casimir operators admitted by the enveloping algebras. Such tensors are respectively the 2-index tensor providing the Killing metric and a 4-index tensor. Both are easily defined in our chosen basis of table \ref{v10genni} by means of the following traces:
\newpage
\begin{equation}\label{kalingoso23}
  \mathcal{I}\mathit{q}_{AB }\, \equiv \, \frac{1}{2} \text{Tr}\left[J_{A }\cdot J_{B }\right] \, = \, \left(
\begin{array}{cccccccccc}
 1 & 0 & 0 & 0 & 0 & 0 & 0 & 0 & 0 & 0 \\
 0 & 1 & 0 & 0 & 0 & 0 & 0 & 0 & 0 & 0 \\
 0 & 0 & 1 & 0 & 0 & 0 & 0 & 0 & 0 & 0 \\
 0 & 0 & 0 & 1 & 0 & 0 & 0 & 0 & 0 & 0 \\
 0 & 0 & 0 & 0 & 1 & 0 & 0 & 0 & 0 & 0 \\
 0 & 0 & 0 & 0 & 0 & 1 & 0 & 0 & 0 & 0 \\
 0 & 0 & 0 & 0 & 0 & 0 & -1 & 0 & 0 & 0 \\
 0 & 0 & 0 & 0 & 0 & 0 & 0 & -1 & 0 & 0 \\
 0 & 0 & 0 & 0 & 0 & 0 & 0 & 0 & -1 & 0 \\
 0 & 0 & 0 & 0 & 0 & 0 & 0 & 0 & 0 & -1 \\
\end{array}
\right)
\end{equation}
\begin{equation}\label{quartirolo}
  \mathcal{Q}_{A B C D }^{[\mathit{f},\mathit{h}]}
  \, \equiv \, \mathit{f} \, \text{Tr}\left[J_{A }\cdot J_{B }\cdot J_{C
   }\cdot J_{D }\right]+\mathit{h}\, \text{Tr}\left[J_{C
   }\cdot J_{D }\right] \text{Tr}\left[J_{A }\cdot
   J_{B }\right]
\end{equation}
As one sees the second invariant tensor as defined in eq.(\ref{quartirolo}) is actually a two parameter family of 4-index invariant tensors, the component proportional to the coefficient $\mathit{h}$ being the tensor product of the quadratic tensor of equation (\ref{kalingoso23}). The reason for that is that a basis of independent invariant tensors is anyhow provided by the unique 2-index one and by any choice a quartic one not simply proportional to the tensor product of the 2-index one. In order to facilitate the invariant reinterpretation of the partition functions calculated by direct integration it is convenient to choose a basis where the quartic invariant tensor is a special combination that we indicate below, by specializing the parameters $\mathit{f},\mathit{h}$. This implies no loss of generality but it is useful to simplify the final expressions.
\par
\paragraph{The two parameters quartic invariant} has the following explicit form
\noindent
\begin{eqnarray}\label{quartotto}
\mathcal{Q}^{[\mathit{f},\mathit{h}]}&=& 4 \mathit{h} \left(\beta _1^2+\beta _2^2+\beta _3^2+\beta _4^2+\beta _5^2+\beta _6^2-\beta _7^2-\beta
_8^2-\beta _9^2-\beta _{10}^2\right){}^2\nonumber\\&&
+\mathit{f} \left(2 \beta _1^4+2 \beta _2^4+4 \beta _2^2 \beta _3^2+\beta _3^4+4 \beta _2^2 \beta _4^2+6
\beta _3^2 \beta _4^2+\beta _4^4+2 \beta _3^2 \beta _5^2+4 \beta _3 \beta _4 \beta _5^2+2 \beta _4^2 \beta _5^2+2 \beta _5^4\right.\nonumber\\&&
\left. +4 \sqrt{2} \beta _2
\beta _3 \beta _5 \beta _6+4 \sqrt{2} \beta _2 \beta _4 \beta _5 \beta _6+4 \beta _2^2 \beta _6^2+2 \beta _3^2 \beta _6^2-4 \beta _3 \beta _4 \beta
_6^2+2 \beta _4^2 \beta _6^2+4 \beta _5^2 \beta _6^2+2 \beta _6^4-4 \beta _2^2 \beta _7^2-2 \beta _3^2 \beta _7^2 \right. \nonumber\\&&
\left. -4 \beta _3 \beta _4 \beta _7^2-2
\beta _4^2 \beta _7^2-4 \beta _5^2 \beta _7^2-4 \beta _6^2 \beta _7^2+2 \beta _7^4-4 \sqrt{2} \beta _2 \beta _3 \beta _7 \beta _8+4 \sqrt{2} \beta
_2 \beta _4 \beta _7 \beta _8\right.\nonumber\\&&
\left.-2 \beta _3^2 \beta _8^2+4 \beta _3 \beta _4 \beta _8^2-2 \beta _4^2 \beta _8^2-4 \beta _5^2 \beta _8^2-4 \beta _6^2
\beta _8^2+4 \beta _7^2 \beta _8^2+2 \beta _8^4\right.\nonumber\\&&
\left.+4 \sqrt{2} \beta _3 \beta _6 \beta _7 \beta _9-4 \sqrt{2} \beta _4 \beta _6 \beta _7 \beta _9-4 \sqrt{2}
\beta _3 \beta _5 \beta _8 \beta _9-4 \sqrt{2} \beta _4 \beta _5 \beta _8 \beta _9-8 \beta _2 \beta _6 \beta _8 \beta _9-4 \beta _2^2 \beta _9^2\right.\nonumber\\&&
\left.-4
\beta _3^2 \beta _9^2-4 \beta _4^2 \beta _9^2+4 \beta _7^2 \beta _9^2+4 \beta _8^2 \beta _9^2+2 \beta _9^4\right.\nonumber\\&&
\left.+4 \left(2 \beta _2 \beta _5 \beta _7+\sqrt{2}
\beta _3 \left(-\beta _6 \beta _7+\beta _5 \beta _8\right)+\beta _3^2 \beta _9-\beta _4 \left(\sqrt{2} \beta _6 \beta _7+\sqrt{2} \beta _5 \beta
_8+\beta _4 \beta _9\right)\right) \beta _{10}\right.\nonumber\\&&
\left.-4 \left(\beta _2^2+\beta _3^2+\beta _4^2+\beta _5^2+\beta _6^2\right) \beta _{10}^2+2 \beta _{10}^4-4
\beta _1^2 \left(-\beta _3^2-\beta _4^2-\beta _5^2+\beta _8^2+\beta _9^2+\beta _{10}^2\right)\right.\nonumber\\&&
\left.+4 \beta _1 \left(-\sqrt{2} \beta _4 \beta _5 \beta
_6-\sqrt{2} \beta _4 \beta _7 \beta _8+\sqrt{2} \beta _3 \left(\beta _5 \beta _6-\beta _7 \beta _8\right)+2 \beta _5 \beta _7 \beta _9-2 \beta _6
\beta _8 \beta _{10}+\beta _2 \left(\beta _3^2-\beta _4^2-2 \beta _9 \beta _{10}\right)\right)\right)\nonumber\\
\end{eqnarray}
\subsubsection{Choice of the compact Cartan subalgebra}
In the case under study that constitutes the common Tits Satake submanifold for all
Calabi-Vesentini manifolds, the compact subalgebra is the following one and we have no Paint-subalgebra since the Tits Satake submanifold is by definition maximally split:
\begin{equation}\label{coniglioarrosto}
  \so(2,3) \equiv \mathbb{U} \, \supset \, \mathbb{H} \equiv \so(2)\times \so(3) \, = \, \text{span}\,\left\{J_7,J_8,J_9,J_{10}\right\}
\end{equation}
The generators  $J_{7,8,9}$ constitute a basis for the $\so(3)$ Lie algebra while $J_{10}=X_c$ is the unique $\so(2)$ generator responsible for the K\"ahler structure. The Cartan subalgebra of $\mathbb{H}$ can be obtained by adding to $J_{10}$ any generator of $\so(3)$. We have chosen $J_{9}$ so that:
\begin{equation}\label{kumpatsubal}
  \mathbb{H} \,  \supset \, \boldsymbol{\mathcal{C}_c} \, \equiv \, \text{span}\,\left\{J_9,J_{10}\right\}
\end{equation}
\subsubsection{Anticipation on the form of the partition function}
According with the general scheme described in sect.\ref{topocaldo} and  
as we prove in appendix \ref{integso23},  the explicit integration of the Gibbs distribution yielding the partition function, once reduced to the two above chosen compact temperatures produces the following result that agrees with the general scheme derived in section \ref{sashasecta} and with the ensuing general formula
(\ref{zolotayaformula}) anticipated in section \ref{lodicoprima}:
\begin{equation}\label{partfunzina}
  Z\left(\beta _{10},\beta _9\right)=\frac{256 e^{-\beta _{10}} \pi ^3}{\beta _{10}{}^3-\beta _{10} \beta _9{}^2}
\end{equation}
The only invariant way to interpret the explicit result (\ref{partfunzina}) with the two compact temperatures is provided by the following identities
\begin{eqnarray}\label{interpret1}
  \left.\sqrt{\frac{\sqrt{Q_4}+Q_2}{2}}\right| _{\boldsymbol{\mathcal{C}_c}}&=&\beta _{10}\nonumber\\
\sqrt{\frac{\sqrt{Q_4}+Q_2}{2}}\sqrt{Q_4}=
\left.\sqrt{\frac{Q_4\left(\sqrt{Q_4}+Q_2\right)}{2}}\right| _{\boldsymbol{\mathcal{C}_c}} &=& \beta _{10}{}^3-\beta _{10} \beta _9{}^2
\end{eqnarray}
Hence the invariant way of rewriting the result of the explicit calculation of the partition function is the following one
\begin{equation}\label{invaria23parfunzia}
  Z[\boldsymbol{\beta} ]=256\pi ^3\frac{\text{Exp}\left[- \sqrt{\frac{\sqrt{Q_4}+Q_2}{2}}\right]}{\sqrt{\frac{\sqrt{Q_4}+Q_2}{2}}\sqrt{Q_4}}
\end{equation}
We will study in the next subsection the Ward differential equation satisfied by such a partition function, yet the general result of the direct
calculation of integrals should always be interpretable  in an analogous  invariant way as we did in the present case (when the only introduced temperatures are the $\boldsymbol{\beta}$.s). The explicit derivation via integration of the result in eq.(\ref{partfunzina}) is provided in appendix \ref{integso23}.
\subsubsection{The Ward differential Identity}
\label{guardarobaide}
We assume that the partition function $Z(\boldsymbol{\beta})$ depends from the temperatures $\beta_i$ only through the two available invariants that in our chosen basis
of invariants are:
\begin{eqnarray}
\label{invariantnames}
  \Xi &\equiv& Q_4(\boldsymbol{\beta}) \, =  \,\mathcal{Q}^{[1,-1/4]}(\boldsymbol{\beta}) \nonumber \\
  \omega &\equiv & Q_2(\boldsymbol{\beta})  
\end{eqnarray}
Hence we pose:
\begin{equation}\label{posiziazetale}
  Z(\boldsymbol{\beta}) \, =\, \Psi [\Xi ,\omega ]
\end{equation}
where $\Psi [\Xi ,\omega ]$ is an unknown function of the two variables $\Xi,\omega$. 
\par
From the very fact that $\Xi,\omega$ are invariants we can deduce a highly non-trivial differential constraint that should be obeyed by the function $\Psi [\Xi ,\omega ]$ that we name a \textit{differential Ward Identity}. Imposing it on the function calculated by explicit integration and reinterpreted in terms of invariants is a formidable test both on the calculation of the integrals and on the validity of the invariant reinterpretation. Let us derive such a differential constraint.   
\paragraph{The first derivative}
Utilizing the standard notation $\partial_i \, \equiv \, \partial/\partial\beta^i$ where 
$\beta^i$ are the temperature components with their index raised with the invariant metric $\mathcal{I}\mathit{q}^\mathit{ij}$, we immediately obtain:
\begin{equation}\label{comasco}
  \partial _i\Psi [\Xi ,\omega ]=\frac{\partial \Psi [\Xi ,\omega ]}{\partial \Xi }\partial _i\Xi +\frac{\partial \Psi [\Xi ,\omega
]}{\partial \omega }\partial _i\omega
\end{equation}
\paragraph{The second derivative}
Performing next the second derivative we obtain:
\begin{equation}\label{ramodellago}
  \partial _i\partial _j\Psi = \frac{\partial ^2\Psi }{\partial \Xi ^2}\partial _i\Xi \partial _j\Xi +\frac{\partial \Psi }{\partial \Xi
}\partial _i\partial _j\Xi +\frac{\partial ^2\Psi }{\partial \Xi \partial \omega }\left(\partial _i\Xi \partial _j\omega +\partial _j\Xi
\partial _i\omega \right) +\frac{\partial ^2\Psi }{\partial \omega ^2}\partial _i\omega \partial _j\omega +\frac{\partial \Psi }{\partial \omega
}\partial _i\partial _j\omega\
\end{equation}
\paragraph{On the other hand} performing the second derivative directly from the definition of the partition function as integral of the unnormalized Gibbs distribution
(see eq.(\ref{cerusico}) we get:
\begin{equation}\label{comancho}
  \partial _i\partial _j\Psi  = \int \left(\mathfrak{P}^m(\mathbf{\Upsilon})\mathcal{I}\mathit{q}_{\text{jm}}\mathfrak{P}^p(\mathbf{\Upsilon})\mathcal{I}\mathit{q}_{\text{ip}}\right)\text{Exp}\left[-\beta
^r\mathfrak{P}^s(\mathbf{\Upsilon})\mathcal{I}_{\text{rs}}\right]d\mathbf{\Upsilon}
\end{equation}
\paragraph{Next we contract the indices} with the invariant metric and we use the identities satisfied by the moment maps and the invariants.  {From}  {the}  {last} {line} {we}  {have}:
\begin{equation}\label{cromatone1}
  \mathcal{I}\mathit{q}^{\text{ij}} \partial _i\partial _j\Psi =-\int \text{Exp}\left[-\beta ^r\mathfrak{P}^s(\mathbf{\Upsilon})\mathcal{I}_{\text{rs}}\right] \, d\mathbf{\Upsilon}=-\Psi\
\end{equation}
while in the previous lines we utilize the easily verifiable identities:
\begin{eqnarray}
\label{carneatore}
&&\mathcal{I}\mathit{q}^{\text{ij}} \partial _i\Xi \partial _j\Xi = -16\,\Xi \, \omega\nonumber\\
&& \mathcal{I}\mathit{q}^{\text{ij}} \partial _i\partial _j\Xi = -24 \,  \omega\nonumber\\
&&\mathcal{I}\mathit{q}^{\text{ij}}\left(\partial _i\Xi \partial _j\omega +\partial _j\Xi \partial _i\omega \right)= -32 \,\Xi \nonumber\\
&&\mathcal{I}\mathit{q}^{\text{ij}}\partial _i\omega \partial _j\omega =-20
\end{eqnarray}
hence from eq.(\ref{ramodellago}) we obtain the differential Ward Identity
\begin{equation}\label{wardaida}
  -16\, \Xi \, \omega \, \frac{\partial ^2\Psi }{\partial \Xi ^2}-24\,\omega \, \frac{\partial \Psi }{\partial \Xi }-16\,\Xi \, \frac{\partial
^2\Psi }{\partial \Xi \partial \omega }-4 \,\omega \,\frac{\partial ^2\Psi }{\partial \omega ^2}-20\, \frac{\partial \Psi }{\partial \omega }+\Psi  =
0
\end{equation}
which is identically satisfied by the function
\begin{equation}\label{Psione}
  \Psi[\Xi,\omega] \, = \, \frac{256 \sqrt{2} \pi ^3 e^{-\frac{\sqrt{\omega +\sqrt{\Xi
   }}}{\sqrt{2}}}}{\sqrt{\Xi } \sqrt{\omega +\sqrt{\Xi }}}
\end{equation}
namely by the invariant reinterpretation of the partition function calculated by means of integrals.
\section{Conclusions}
In this paper we have tried to tame so named \textit{Information Geometry} \cite{raone,cenzone,amarone}, connecting it (at list for probability distributions of Gibbs type) to its conceptual bases that 
belong to a,  historically long and variegated, development within the framework of theoretical physics, whose starting point can be located in the 1957 foundational papers by Jaynes \cite{gianno1,gianno2}. 
\paragraph{Thermodynamic K\"ahler metric.}
Following the basic ideas of Geometrical Thermodynamics of Lychagin, Roop and collaborators \cite{lychaginlecture}, \cite{Kushner_2020,Lychagin_2020,ludaed,ludaed2}, we have introduced a contact structure as basic feature of the macroscopic thermodynamical space $\mho$, yet, up to our knowledge, \underline{\textit{we have been the first}} to introduce  on the contact manifold also \underline{\textit{the unique general metric}} compatible with the contact structure,  showing that its reduction to the submanifolds transverse to Reeb field, which is the entropy gradient, is K\"ahlerian. Such K\"ahler metric has a generic \textbf{
non-compact translation group} of isometries that correspond to the shifts of the average values $\mathbf{x}$ of the observable functions $\mathbf{X}(\mathbf{q})$ defined over the microscopic manifold of events $\Omega$. It is precisely this translation group, naturally associated with the Thermodynamics à la Jaynes, what provides the mathematical basis for the \textbf{non-compact abelian structure} characterizing   Geometric Thermodynamics. The identification of \textbf{the Hessian of the stochastic hamiltonian} (the Fisher matrix) with a \textit{metric in Gibbs distribution parameter space} rigorously follows as the \textit{pull-back of the Thermodynamical K\"ahler metric} on the \textbf{equilibrium states} that are, by definition, \textbf{Lagrangian submanifolds} where the thermodynamic K\"ahler 2-form vanishes. 
\paragraph{Thermodynamics à la Souriau.}    
It is within such a general geometric framework of thermodynamics that we have located the Gibbs distributions à la Souriau, defined as the choice, in the capacity of microscopic event manifold $\Omega$, of a \textbf{homogeneous K\"ahler manifold} $\mathcal{M}$ with a non-trivial isometry group $\mathrm{U}$ having transitive action on $\mathcal{M}$, and, in the capacity of observable functions $\mathbf{X}(\mathbf{p})$ on $\Omega$, the moment-maps $\boldsymbol{\mathfrak{P}}_A(\mathbf{p})$, $\mathbf{p}\in\mathcal{M}$  of the Killing vectors $\boldsymbol{\mathfrak{k}}_A$ closing the isometry Lie algebra $\mathbb{U}$.
\par
Motivated by the perspective applications of such Gibbs distributions to Machine Learning and in particular to \textbf{Cartan Neural Networks}\cite{pgtstheory,naviga,TSnaviga,tassellandum}, as already foreseen in \cite{geotermico}, we concentrated on the case of Calabi-Vesentini manifolds $\mathcal{M}_{CV}^{[2,q]}$, defined in eq.(\ref{quaresima}) that, it is most convenient and inspiring to look at, as \textit{complex codimension one submanifolds} of the Special K\"ahler manifolds $\mathcal{SK}_{3+q}$, recalled in eq.(\ref{speckal}). From the point of view of \textbf{Cartan Neural Networks}, these manifolds constitute the ideal next case, after the Hyperbolic Planes, utilized in \cite{naviga}, for the implementation of the network algorithm theorized in \cite{TSnaviga} and based on homomorphisms of the solvable Lie algebras corresponding to each layer of the network. Their K\"ahler structure, that allows the parallel construction of thermodynamics, makes them an almost unique obligatory choice among the non-compact symmetric spaces. 
\paragraph{Exact Partition Functions.}
Strongly relying on the  encoding of CV manifolds within the framework of Special K\"ahler Geometry and on the techniques of \textbf{symplectic/poissonian geometry} for integrable dynamical systems in order to construct \textbf{action/angle bases} and Darboux local coordinates corresponding to \textbf{compact abelian structures}, we were able to perform the explicit integrations involved in the calculation of the partition functions, arriving at the general and very simple formulae presented in eq.s (\ref{zolotayaformula}) and (\ref{eq:final_Z_odd_sequential}-\ref{eq:final_Z_even_sequential}). 
This very general result allows the study of the macroscopic thermodynamic K\"ahler geometry for all such microscopic manifolds of events whose relevance for Cartan Neural Networks has already been outlined. 
\paragraph{Generalization of Souriau thermodynamics and symmetry breaking.}
The winning token that allowed the calculation of partition functions was the setting up of \textbf{compact abelian structures} whose \textbf{actions} cannot all be moment-maps of Killing vectors, since the isometry algebra $\mathbb{U}$ does not include a sufficient number of compact Cartan generators. The \textbf{missing actions} turn out to be \textit{square roots of quadratic polynomials in the Killing moment-maps} whose coefficients are \textit{invariant tensors} for a \textbf{nested sequence of principal subalgebras} $\mathbb{H}_i \subset \mathbb{H}_{i-1}$ of the compact subalgebra $\mathbb{H}\subset \mathbb{U}$. As long as the generalized temperatures $\boldsymbol{\beta}$ are in the (co)-adjoint representation of $\mathbb{U}$ and in the orbit of the compact Cartan subalgebra, the partition function $Z(\boldsymbol{\beta})$ is invariant with respect to the full isometry group $\mathrm{U}$. A natural emerging idea, however, is that of including new generalized temperatures $\boldsymbol{h}$ \footnote{we name them magnetic fields for reasons that will become apparent below} associated with each of the missing actions. This certainly breaks the $\mathbb{U}$ symmetry to a smaller subgroup. We postpone to a future publication the detailed analysis of the symmetry breaking patterns associated with the switching on of such new temperatures. We simply observe that when the $\boldsymbol{\gamma}$ do not vanish, we can calculate the non vanishing \textbf{mean-values} of the new observable functions defined on the microscopic event manifold, namely the square roots of the $\mathbb{H}_i$-subalgebra Casimir functions, by setting $\boldsymbol{\mathfrak{C}}[\mathbb{H}_i](p)$,\,$\forall p\in \Omega$, by setting:
\begin{equation}\label{magnamagna}
  0 \neq m_i \, = \, < \sqrt{\boldsymbol{\mathfrak{C}}[\mathbb{H}_i]}> \, = \, - \, \frac{\partial}{\partial h^i} \, \log\left[Z(\boldsymbol{\beta},\boldsymbol{h})\right]
\end{equation}
We named $m_i$ such mean values, since they are similar to the magnetizations in ferromagnetism, whose appearance breaks whatever symmetry the microscopic event manifold $\Omega$, might have had. Indeed the important  general result is that in the limit $h_i \to 0$ the magnetizations $m_i$ do not vanish everywhere, rather they become well defined functions $m_i^0(\boldsymbol{\beta})$ of the standard  temperatures $\boldsymbol{\beta}$:
\begin{eqnarray}\label{gorlagorla}
  0 \neq m_i^0(\boldsymbol{\beta}) & \equiv &  - \, \lim_{h_i\to 0} \,\frac{\partial}{\partial h_i} \, \log\left[Z(\boldsymbol{\beta},\boldsymbol{h})\right]\nonumber\\
  &=&< \sqrt{\boldsymbol{\mathfrak{C}}[\mathbb{H}_i]}>_0 \, \equiv \,\frac{1}{Z(\boldsymbol{\beta})} \, \int_{\Omega}  \boldsymbol{\mathrm{d}\mu}(p)\, \sqrt{\boldsymbol{\mathfrak{C}}[\mathbb{H}_i](p)} \, \exp\left[-\, \boldsymbol{\beta}\cdot \boldsymbol{\mathcal{P}}_{\boldsymbol{\mathcal{C}}}(p)\right] 
\end{eqnarray}
\par
In other words, even if one had not introduced the extra generalized temperatures, \textit{i.e} the magnetic fields $h_i$ one might have discovered that, in certain ranges of the $\boldsymbol{\beta}$, the mean values of square root Casimir functions $\sqrt{\boldsymbol{\mathfrak{C}}[\mathbb{H}_i](p)}$ do not vanish, rather they take the values $m_i^0(\boldsymbol{\beta})$ that are fully entitled to be named the \textbf{spontaneous magnetizations}. Hence the introduction of the magnetic fields $h_i$ is not optional, rather it is mandatory, since, anyhow, one has spontaneous magnetizations.
\par 
This result is completely new and opens very interesting perspectives for further investigations and study that have an unexpected and potentially deep impact on the overall structure of AI procedures. In order to appreciate this point let us recall some facts and observations. 
\begin{enumerate}
  \item On one side, in his recent very interesting popularizing book \cite{bennetto}, Max Bennett  remarked the following: \textit{While most AI advancements that occurred in the early 2000s involved applications of supervised-learning models, many of the recent advancements have been applications of generative models. Deepfakes, AI-generated art, and language models like GPT-3 are all examples of generative models at work. Helmholtz suggested that much of human perception is a process of inference-a process of using a generative model to match an inner simulation of the world to the sensory evidence. \dots It turns out that there is, in fact, an abundance of evidence that the neocortical microcircuit is implementing such a generative model.} 
  \item In their very interesting papers \cite{groundwork,neurocoding} on \textit{Categorical perception} in Biological and Artificial Neural Networks, Bonnasse-Gahot and Nadal consider the internal geometrical representation of the exterior world, in relation with the distribution of given exterior entities into categories. Such geometrical internal representation is presumably realized, to use the terminology of Bennett, in the \textit{microcircuitry of the neocortical columns} in mammal brains, while it is provided by appropriate unsupervised generative algorithms in artificial neural networks. In their study of such an issue, Bonnasse-Gahot and Nadal utilize
      Information Geometry and focus on the analysis of the local geometry around the boundaries between different categories.       
  \item The boundaries between categories, when one pursues a
  \textit{classification task} by means of a \textit{supervised Cartan  Neural Network}, are provided by 
  the theory of \textbf{separators}, recently introduced by two of us in in collaboration with other authors, in \cite{tassellandum}. \textbf{Separators are codimension one, homogeneous but not symmetric submanifolds} of the microscopic K\"ahler manifolds constituting each layer of the network. 
  \item According to  Bennett's vision, \textit{pattern recognition}, alias \textit{categorical perception}, and 
      \textit{the generative construction of geometrical representations}, namely \textit{imagination}, are two different operational modalities of the same neocortical microcircuitry in biological brains or of the same artificial neural network, the \textit{awake modus operandi} the former, the \textit{sleeping modus operandi} the latter, namely \textit{supervised} or \textit{unsupervised} in the artificial case. Hence the relation between the categorical partition of the same manifold induced by separators and its foliation induced by spontaneous or forced (at non vanishing generalized magnetic fields $\mathbf{h}$) magnetizations, emerges to strategic prominence. 
  \item The extra observables whose mean values are the magnetizations are  associated with  Casimir functions of the Paint-Group and of its subgroups. It follows that the already envisaged similarity grouping of data according to their distribution into Paint fibres might be realized by spontaneous magnetization mechanism. Like in physical ferromagnets where, below the critical temperature one observes the microscopic \textit{spin up} and \textit{spin down islands}, in the same way, the Souriau like Gibbs distributions might trace out \textit{islands} in the manifold where certain features are dominant and other are depressed and viceversa.  
\end{enumerate}
\subsection{Further investigation directions}
In view of the above remarks we think that the results we obtained  have opened up an entirely new vision on Information Geometry and a new paradigm for generalized Soriau-like thermodynamics that requires a vast spectrum of new investigations and studies. The immediate development lines that we intend to pursue are the following ones:
\begin{description}
\item[a)] Careful investigation of the macroscopic K\"ahler geometry whose K\"ahler potential is the Legendre transform, according with eq.(\ref{legendroK}), of the real symplectic potential, identified with the 
    stochastic hamiltonian $G(\boldsymbol{\lambda}) \, = \, -\log[Z(\boldsymbol{\lambda})]$. In relation with this we
    note the very inspiring fact that such K\"ahler potential has a structure very similar to that of special geometries since it is obtained from a symplectic potential with the following structure:  
    \begin{eqnarray}\label{gramellino}
      G(\boldsymbol{\lambda})& = & \beta_0 + \log[\mathbb{P}^{1}(\boldsymbol{\lambda})]+
      \log[\mathbb{P}^{2\nu+2}(\boldsymbol{\lambda})] \nonumber\\
       \mathbb{P}^{2\nu+2}(\lambda)&=& \text{homogeneous polynomial of degree $2\nu+2$ in $\boldsymbol{\lambda}$}\nonumber\\
       \mathbb{P}^{1}(\lambda)&=& \text{linear polynomial of degree $1$ in $\boldsymbol{\lambda}$}
       \end{eqnarray}
       Yet as we explain in the forthcoming paper \cite{terzatemperatura} the differences are quite significant and several surprises follow from the specific form of the mentioned polynomials.
\item[b)] Careful investigation of the symmetry breaking patterns related with the \textit{magnetic fields} $\mathbf{h}$ associated with Casimir functions of sequence subalgebras.
\item[c)] Investigation of the relation between $\mathcal{SK}_{3+q}$ Special K\"ahler symmetric spaces 
    and the Special K\"ahler Homogeneous but not symmetric spaces $L(-1,q)$ (see \cite{deWit:1995tf,toineugenio,SKGaggio3,SKGaggio2,SKGaggio1,specHomgeoA2,specHomgeoA1})
    that appear to be provided by a deformed metric with a smaller group of isometries on the same solvable Lie group 
    metrically equivalent to the CV manifolds (work in progress \cite{toinepietromario}). The pattern of symmetry breaking that leads to such homogeneous non-symmetric metric might be related with the symmetry breaking introduced by the "\textbf{sponatenous magnetization}"
    of Casimir functions.
\item[d)] Explicit algorithmic construction of Cartan Neural Networks based on CV-manifolds and investigation of the statistical distributions on their layers by means of the new here developed thermodynamics \cite{r2paperone}
\end{description}
\subsection{A final comment} Our results point into the direction of showing that the whole package of mathematical structures and conceptions associated with the \textit{Special Geometries} developed in the context of \textit{Supergravity Theory} has a potentially very much strategic role to play in a sound and systematic geometric reformulation of Machine Learning algorithms. 
\newpage
\appendix
\section{The partition function for the case $\mathrm{SO(2,3)/SO(2)\times SO(3)}$ by explicit integration in solvable coordinates}
\label{integso23}
In this appendix we provide all the detailed information and formulae necessary for the explicit calculation of the partition function for the $q=1$ Calabi-Vesentini manifold (alias Siegel plane $\mathbb{SH}_{2}$) discussed in section \ref{theTSmanifoldo}, whose partition function was already addressed in \cite{geotermico} arriving only at a partially analytic, partially numerical result. Here we show how, by interchanging the order of integrations, we can complete the integration task and produce the closed analytic  form (\ref{partfunzina}) of the partition function that, as already stated in the main text, fully agrees with the general result obtained in section \ref{sashasecta}. These results combined with the 
differential Ward Identity discussion of section \ref{guardarobaide} provide a very important check on the consistency of the various approaches and on the robustness of
the general form for the partition function eq.(\ref{zolotayaformula}).
\par
\subsection{Integration the old way}
\label{vecchiaintegrala}
We begin by recalling the incomplete result obtained in section 7.2 of \cite{geotermico}. After reduction to the two independent compact temperatures $\beta_9,\beta_{10}$ and using  eq.(\ref{carambola}) for the explicit form of the moment maps, the argument of the exponential in the partition function integrand is the following:
\begin{eqnarray}\label{planetario}
 \mathfrak{A}\, \equiv \, \boldsymbol{\beta}_c\cdot \boldsymbol{\mathfrak{P}}(\mathbf{\Upsilon}) & = & \beta_{9} \, \boldsymbol{\mathfrak{P}}_9(\mathbf{\Upsilon}) \, + \,  \beta_{10} 
  \boldsymbol{\mathfrak{P}}_{10}(\mathbf{\Upsilon}) \nonumber\\
  &=&\frac{1}{64} e^{-\Upsilon_1-\Upsilon_2} \left(\beta_{10} 
   \left(-4 e^{2 \Upsilon_2} \left(\Upsilon_6^2+4\right)-2
   e^{2 \Upsilon_1} \left(\left(\Upsilon_6^2+4\right)
   \Upsilon_3^2+2 \sqrt{2} \Upsilon_5 \Upsilon_6 \Upsilon_3+2
   \left(\Upsilon_5^2+4\right)\right)\right.\right. \nonumber\\
   &&\left.\left.-e^{2
   \left(\Upsilon_1+\Upsilon_2\right)} \left(8 \Upsilon_4^2-4
   \sqrt{2} \Upsilon_5 \Upsilon_6 \Upsilon_4+\left(\Upsilon_5^2+4\right)
   \left(\Upsilon_6^2+4\right)\right)-16\right)\right.\nonumber\\
   &&\left. +\beta_{9} 
   \left(-4 e^{2 \Upsilon_2} \left(\Upsilon_6^2+4\right)-2
   e^{2 \Upsilon_1} \left(\left(\Upsilon_6^2+4\right)
   \Upsilon_3^2+2 \sqrt{2} \Upsilon_5 \Upsilon_6 \Upsilon_3+2
   \left(\Upsilon_5^2+4\right)\right)\right.\right.\nonumber\\
   &&\left.\left.  +e^{2
   \left(\Upsilon_1+\Upsilon_2\right)} \left(8 \Upsilon_4^2-4
   \sqrt{2} \Upsilon_5 \Upsilon_6 \Upsilon_4+\left(\Upsilon_5^2+4\right)
   \left(\Upsilon_6^2+4\right)\right)+16\right)\right)
\end{eqnarray}
For calculation convenience it is  useful to redefine $\Upsilon_{1,2} \, = \, \log[\rho_{1,2}]$. In this way we get:
\begin{eqnarray}
\label{acco}
  \mathfrak{A} &=&\frac{N_A}{D_A} \nonumber\\
  N_A&=& \rho _1^2 \left(-\left(\rho _2^2 \left(8
   \Upsilon_4^2-4 \sqrt{2} \Upsilon_5 \Upsilon_6
   \Upsilon_4+\left(\Upsilon_5^2+4\right)
   \left(\Upsilon_6^2+4\right)\right) (\beta_{10} -\beta_{9}
   )\right.\right. \nonumber\\
  &&\left.\left.  +2 \left(\left(\Upsilon_6^2+4\right) \Upsilon_3^2+2
   \sqrt{2} \Upsilon_5 \Upsilon_6 \Upsilon_3+2
   \left(\Upsilon_5^2+4\right)\right) (\beta_{10} +\beta_{9}
   )\right)\right)-4 \left(4 (\beta_{10} -\beta_{9}
   )+\rho _2^2 \left(\Upsilon_6^2+4\right) (\beta_{10}
   +\beta_{9} )\right)  \nonumber\\
   D_A &= & 64 \rho _1 \rho _2 \nonumber\\
\end{eqnarray}
One has to calculate the 6-integrals of $\exp[\mathfrak{A}]$ on the four nilpotent coordinates $\Upsilon_3,\Upsilon_4,\Upsilon_5,\Upsilon_6$ and on the exponentials of the non-compact Cartan fields $\rho_1,\rho_2$. The strategy adopted in \cite{geotermico} was that of calculating first the integrals from $-\infty$ to $\infty$ of the four nilpotent coordinates leaving the integrals from $0$ to $\infty$ of $\rho_1,\rho_2$ as last integrations. Hence we defined:
\begin{equation}\label{defiogotof}
  \mathfrak{F}\left(\rho_1,\rho_2,\beta_{10},\beta_{9}\right)
  \, \equiv\, \int_{-\infty}^{\infty} \mathrm{d}\Upsilon_6 \, \int_{-\infty}^{\infty} \mathrm{d}\Upsilon_5 \, \int_{-\infty}^{\infty} \mathrm{d}\Upsilon_4
  \,
  \int_{-\infty}^{\infty} \mathrm{d}\Upsilon_3 \exp\left[-\mathfrak{A} \right]
\end{equation}
and we obtained:
\begin{alignat}{2}\label{cartamura}
 &\mathfrak{F}(\rho_1,\rho_2,\beta_{10},\beta_{9}) \nonumber\\
 &= \frac{64 \pi ^{3/2} \sqrt{\beta_{10} -\beta_{9} } \exp \left(-\frac{\beta_{10} ^2+(\beta_{10} -\beta_{9} )
   \left(\rho _1^2 \left(\rho _2^2 (\beta_{10} -\beta_{9} )+\beta_{10} +\beta_{9} \right)+\rho _2^2
   (\beta_{10} +\beta_{9} )\right)-6 \beta_{10}  \beta_{9} +\beta_{9} ^2}{8 \rho _1 \rho _2 (\beta_{10} -\beta_{9}
   )}\right) }{\sqrt{\rho _2 (\beta_{10} +\beta_{9} )} \left(\frac{\rho _1 (\beta_{10} -\beta_{9} )}{\rho
   _2^2 (\beta_{10} -\beta_{9} )+\beta_{10} +\beta_{9} }\right){}^{3/2} \left(\rho _2^2 (\beta_{10} -\beta_{9}
   )+\beta_{10} +\beta_{9} \right){}^{3/2}} \times\nonumber\\
   &         \times K_0\left(\frac{\left((\beta_{10} -\beta_{9} ) \rho _1^2+\beta_{10} +\beta_{9} \right)
   \left((\beta_{10} -\beta_{9} ) \rho _2^2+\beta_{10} +\beta_{9} \right)}{8 (\beta_{10} -\beta_{9} ) \rho _1 \rho
   _2}\right)
\end{alignat}
where $K_0(x)$ is the Bessel function of type $K$ and index $0$.
The last integral:
\begin{equation}\label{comancho}
  \mathfrak{Z}(\beta_{10},\beta_9) \, = \, \int_{0}^{\infty}\mathrm{d}\rho_1 
  \, \int_{0}^{\infty}\mathrm{d}\rho_2 \, \, \mathfrak{F}\left(\rho_1,\rho_2,\beta_{10},\beta_{9}\right)
\end{equation}
proved instead too much complicated in order to be evaluated analytically, although we could demonstrate that it is convergent provided the two temperatures $\beta_{9,10}$ satisfy the same constraints that where necessary for the nilpotent integrals mentioned in eq.(\ref{defiogotof}) to converge, namely:
\begin{equation}\label{miserabile}
  \beta_{10}>\beta_9 \geq 0
\end{equation}
\subsection{Integration the new way}
\label{nuovaintegrala}
We consider next the new strategy that leads to an explicit analytical integration.
Starting from  eq.(\ref{planetario}) in the exponent $\mathfrak{A}$ we collect  the coefficients of the sum $\beta_{10} + \beta_{9}$ and of the difference $\beta_{10} - \beta_{9}$:
\begin{equation}\label{carduccio}
  \mathfrak{A} \, = \, \left(\beta_{10} + \beta_{9}\right) \, \mathfrak{P}_+ \, + \, 
  \left(\beta_{10} - \beta_{9}\right) \, \mathfrak{P}_- \,
\end{equation}
obtaining:
\begin{eqnarray}
\label{cardogobbo}
  \mathfrak{P}_+ &=& -\frac{1}{16} e^{\Upsilon _2-\Upsilon _1} \left(\Upsilon
   _6^2+4\right)-\frac{1}{32} e^{\Upsilon _1-\Upsilon _2}
   \left(\left(\Upsilon _6^2+4\right) \Upsilon _3^2+2 \sqrt{2}
   \Upsilon _5 \Upsilon _6 \Upsilon _3+2 \left(\Upsilon
   _5^2+4\right)\right) \\
  \mathfrak{P}_- &=& -\frac{1}{64} e^{-\Upsilon _1-\Upsilon _2} \left(e^{2 \left(\Upsilon
   _1+\Upsilon _2\right)} \left(8 \Upsilon _4^2-4 \sqrt{2} \Upsilon
   _5 \Upsilon _6 \Upsilon _4+\left(\Upsilon _5^2+4\right)
   \left(\Upsilon _6^2+4\right)\right)+16\right) 
\end{eqnarray}
Introducing the change of variables:
\begin{equation}\label{cangen}
  \Upsilon_{1}\, = \, \ft 12 \left(\log[r_1]+\log[r_2]\right) \quad ; \quad \Upsilon_{2}\, = \, \ft 12 \left(\log[r_1]-\log[r_2]\right) 
\end{equation}
and the convenient intermediate renaming:
\begin{equation}\label{fragiacomo}
  \beta_{10} + \beta_9 \, =\,  X \quad ; \quad \beta_{10} - \beta_9 \, = \, Y
\end{equation}
one finds that the  integral takes the following form:
\begin{alignat}{3}\label{cammello}
  &\exp\left[-A\right] \prod_{i=1}^{6}\mathrm{d}\Upsilon_i & = &  \int_{-\infty}^{\infty} 
  \, d\Upsilon_6 \, \int_{-\infty}^{\infty} 
  \, d\Upsilon_5 \, \,\,\left( I_1 \times I_2\right) \nonumber\\
  &I_1 & = & \int_{0}^{\infty} \mathrm{d}r_1 \, \int_{-\infty}^{\infty} \mathrm{d}\Upsilon_4 \, \left(-\frac{\exp \left(-\frac{Y \left(r_1^2 \left(8 \Upsilon _4^2-4
   \sqrt{2} \Upsilon _5 \Upsilon _6 \Upsilon _4+\left(\Upsilon
   _5^2+4\right) \left(\Upsilon _6^2+4\right)\right)+16\right)}{64
   r_1}\right)}{r_1}\right)\nonumber\\
  &I_2 & = & \int_{0}^{\infty} \mathrm{d}r_2 \, \int_{-\infty}^{\infty} \mathrm{d}\Upsilon_3 \left( -\frac{\exp \left(-\frac{X \left(r_2^2 \left(\left(\Upsilon
   _6^2+4\right) \Upsilon _3^2+2 \sqrt{2} \Upsilon _5 \Upsilon _6
   \Upsilon _3+2 \left(\Upsilon _5^2+4\right)\right)+2 \left(\Upsilon
   _6^2+4\right)\right)}{32 r_2}\right)}{r_2}\,\right) 
\end{alignat}
Explicitly one finds:
\begin{eqnarray}
\label{staifresco}
  I_1 &=& -\frac{4 \sqrt{2} \pi  e^{-\frac{1}{4} \sqrt{\Upsilon _5^2+\Upsilon
   _6^2+4} Y}}{Y} \\
  I_2 &=& -\frac{16 \sqrt{2} \pi  e^{-\frac{1}{4} \sqrt{\Upsilon _5^2+\Upsilon
   _6^2+4} X}}{\Upsilon _6^2 X+4 X} 
\end{eqnarray}
so that the final integral to be computed is:
\begin{equation}\label{fringe}
  Z(X,Y) \, = \,\int_{-\infty}^{\infty} 
  \, d\Upsilon_6 \, \int_{-\infty}^{\infty} 
  \, d\Upsilon_5 \, \,\,\frac{128 \pi ^2 e^{-\frac{1}{4} \sqrt{\Upsilon _5^2+\Upsilon _6^2+4}
   (X+Y)}}{\left(\Upsilon _6^2+4\right) X Y}
\end{equation}
Transforming the $\mathbb{R}^2$ coordinates $\{\Upsilon_5,\Upsilon_6\}$ to polar coordinates:
\begin{equation}\label{polarefreddo}
  \Upsilon_5 \, = \, r \, \cos[\theta] \quad ; \quad \Upsilon_6 \, = \, r \, \sin[\theta]
\end{equation}
we obtain:
\begin{eqnarray}
\label{cavernicolo}
  Z(X,Y) &=& \int_{0}^{\infty}\mathrm{d}r\, \int_{0}^{2\pi} \mathrm{d}\theta \,\,\left(-\frac{256 \pi ^2 r e^{-\frac{1}{4} \sqrt{r^2+4} (X+Y)}}{r^2 X Y \cos
   (2 \theta )+r^2 (-X) Y-8 X Y}\right)\nonumber\\
   &=& \int_{0}^{\infty}\mathrm{d}r\,  \left(\frac{128 \pi ^3 r e^{-\frac{1}{4} \sqrt{r^2+4} (X+Y)}}{\sqrt{r^2+4}
   X Y} \right)  \nonumber\\
   & = & \frac{512 \pi ^3 e^{-\frac{X}{2}-\frac{Y}{2}}}{X Y (X+Y)}
\end{eqnarray}
and recalling the temporary position (\ref{fragiacomo}) we get:
\begin{equation}\label{tallone}
 Z(X,Y)\, = \, Z(\beta_{10},\beta_9) \, = \,  \frac{256 e^{-\beta _{10}} \pi ^3}{\beta _{10}{}^3-\beta _{10} \beta _9{}^2}
\end{equation}
which is the result anticipated in eq. (\ref{partfunzina})
\section{Details on the case $\mathrm{SO(2,4)/SO(2)\times SO(4)}$}
\label{so2com4}
 Within the Tits Satake Universality Class $M_{CV}^{[2,q]}$ defined in eq.(\ref{quaresima}) we choose the explicit case $q=2$ in order to illustrate the properties of the whole class.
 Indeed the Paint Group symmetry is evident already at $q=2$ and it suffices to extend the sum over the Paint index from $q=2$ to any value of $q$ and one retrieves the general formulae.
 The  symmetric space $\mathcal{M}^{[2,2]}_{CV}$ of our example has dimension 8, which, therefore, is also the dimension of the corresponding solvable group $\mathcal{S}_8$ and of the corresponding solvable Lie algebra $Solv_{[2,2]}$:
 \begin{eqnarray}\label{laborroto}
 \mathcal{M}^{[2,2]}_{CV}&=&\frac{\mathrm{SO(2,4)}}{\mathrm{SO(2) \times SO(4)}} \nonumber\\
  \text{dim} \left[\mathcal{M}^{[2,2]}\right]& = & 8 \, = \, \text{dim} [\mathcal{S}_{[2,2]}] \, = \, \text{dim} [Solv_{[2,2]}]
 \end{eqnarray}
 The chosen basis of generators of the solvable Lie algebra are displayed in table \ref{baldovino} (for more details on their construction and normalization see \cite{pgtstheory}).
\begin{table}[htb]
  \centering
  \begin{alignat*}{7}
   T_1 & =  & \left(
\begin{array}{cccccc}
 1 & 0 & 0 & 0 & 0 & 0 \\
 0 & 0 & 0 & 0 & 0 & 0 \\
 0 & 0 & 0 & 0 & 0 & 0 \\
 0 & 0 & 0 & 0 & 0 & 0 \\
 0 & 0 & 0 & 0 & 0 & 0 \\
 0 & 0 & 0 & 0 & 0 & -1 \\
\end{array}
\right) &\quad ;  \quad & T_2& =  & \left(
\begin{array}{cccccc}
 0 & 0 & 0 & 0 & 0 & 0 \\
 0 & 1 & 0 & 0 & 0 & 0 \\
 0 & 0 & 0 & 0 & 0 & 0 \\
 0 & 0 & 0 & 0 & 0 & 0 \\
 0 & 0 & 0 & 0 & -1 & 0 \\
 0 & 0 & 0 & 0 & 0 & 0 \\
\end{array}
\right) \nonumber\\
   T_3 & =  & \left(
\begin{array}{cccccc}
 0 & \frac{1}{\sqrt{2}} & 0 & 0 & 0 & 0 \\
 0 & 0 & 0 & 0 & 0 & 0 \\
 0 & 0 & 0 & 0 & 0 & 0 \\
 0 & 0 & 0 & 0 & 0 & 0 \\
 0 & 0 & 0 & 0 & 0 & -\frac{1}{\sqrt{2}} \\
 0 & 0 & 0 & 0 & 0 & 0 \\
\end{array}
\right) &\quad ;  \quad   & T_4& =  & \left(
\begin{array}{cccccc}
 0 & 0 & 0 & 0 & \frac{1}{\sqrt{2}} & 0 \\
 0 & 0 & 0 & 0 & 0 & -\frac{1}{\sqrt{2}} \\
 0 & 0 & 0 & 0 & 0 & 0 \\
 0 & 0 & 0 & 0 & 0 & 0 \\
 0 & 0 & 0 & 0 & 0 & 0 \\
 0 & 0 & 0 & 0 & 0 & 0 \\
\end{array}
\right) \nonumber\\
  T_5 & =  & \left(
\begin{array}{cccccc}
 0 & 0 & \frac{1}{\sqrt{2}} & 0 & 0 & 0 \\
 0 & 0 & 0 & 0 & 0 & 0 \\
 0 & 0 & 0 & 0 & 0 & -\frac{1}{\sqrt{2}} \\
 0 & 0 & 0 & 0 & 0 & 0 \\
 0 & 0 & 0 & 0 & 0 & 0 \\
 0 & 0 & 0 & 0 & 0 & 0 \\
\end{array}
\right) &\quad ;  \quad  & T_6& =  & \left(
\begin{array}{cccccc}
 0 & 0 & 0 & \frac{1}{\sqrt{2}} & 0 & 0 \\
 0 & 0 & 0 & 0 & 0 & 0 \\
 0 & 0 & 0 & 0 & 0 & 0 \\
 0 & 0 & 0 & 0 & 0 & -\frac{1}{\sqrt{2}} \\
 0 & 0 & 0 & 0 & 0 & 0 \\
 0 & 0 & 0 & 0 & 0 & 0 \\
\end{array}
\right) \nonumber\\
   T_7 & =  & \left(
\begin{array}{cccccc}
 0 & 0 & 0 & 0 & 0 & 0 \\
 0 & 0 & \frac{1}{\sqrt{2}} & 0 & 0 & 0 \\
 0 & 0 & 0 & 0 & -\frac{1}{\sqrt{2}} & 0 \\
 0 & 0 & 0 & 0 & 0 & 0 \\
 0 & 0 & 0 & 0 & 0 & 0 \\
 0 & 0 & 0 & 0 & 0 & 0 \\
\end{array}
\right)&\quad ;  \quad  & T_8& =  & \left(
\begin{array}{cccccc}
 0 & 0 & 0 & 0 & 0 & 0 \\
 0 & 0 & 0 & \frac{1}{\sqrt{2}} & 0 & 0 \\
 0 & 0 & 0 & 0 & 0 & 0 \\
 0 & 0 & 0 & 0 & -\frac{1}{\sqrt{2}} & 0 \\
 0 & 0 & 0 & 0 & 0 & 0 \\
 0 & 0 & 0 & 0 & 0 & 0 \\
\end{array}
\right) \nonumber\\
 \end{alignat*}
  \caption{The generators of the solvable Lie algebra $Solv_8$}\label{baldovino}
\end{table}
Following the conventions and the theory exposed in \cite{pgtstheory,TSnaviga} a generic element of the solvable Lie algebra is parameterized as follows:
\begin{equation}\label{genelement}
  Solv_8 \, \ni \, \mathbf{X}(\boldsymbol{\Upsilon}) \, = \, \Upsilon^1 \,T_1\, + \, \Upsilon^2 \,T_2\, + \, \Upsilon^3 \,T_3\, + \, 
  \Upsilon^4\,T_4\, + \, \Upsilon^{5,1}\,T_5\, + \, \Upsilon^{5,2}\,T_6\, + \,\Upsilon^{6,1}\,T_7\, + \, \Upsilon^{6,2}\,T_8
\end{equation}
 The reason for the special naming of the solvable coordinates $\boldsymbol{\Upsilon}$ is the distinction between the long roots (generators $T_{3,4}$ associated with roots $\alpha_{3,4}$) that have no multiplicity and the short ones that have multiplicity and transform in the fundamental representation of the Paint Group $\mathrm{G_{Paint}}$ (see \cite{pgtstheory} for details). For the chosen example the Paint Group is just $\mathrm{SO(2)}$ and the solvable generators $T_{5,6}$ form the doublet of painted roots $\alpha_5$, while the solvable generators $T_{7,8}$ form the  doublet  of painted roots $\alpha_6$ in the root system of $\so(2,3)\simeq \sym (4,\mathbb{R})$.
 \par
 Following the conventions and notations of  \cite{pgtstheory,TSnaviga}, the $\Sigma$ exponential map from the solvable Lie algebra to the solvable group yields the generic element of the solvable group manifold in  the  form presented in eq. (3.56) of \cite{pgtstheory}, that we repeat here for reader's convenience:
{\scriptsize
\begin{eqnarray}\label{exempli2}
  &\mathbb{L}(\boldsymbol\Upsilon)^{[2,1]} \, = \,& \nonumber\\
  & \left(
\begin{array}{cccccc}
 e^{\Upsilon _1} & \frac{e^{\Upsilon _1} \Upsilon _3}{\sqrt{2}} & \frac{1}{2}
   e^{\Upsilon _1} \left(\sqrt{2} U_1+\Upsilon _3 V_1\right) & \frac{1}{2}
   e^{\Upsilon _1} \left(\sqrt{2} U_2+\Upsilon _3 V_2\right) & -\frac{1}{8}
   e^{\Upsilon _1} \left(4 \mathbf{U}\cdot \mathbf{V}+\sqrt{2} \left(\Upsilon _3 \mathbf{V}^2-4 \Upsilon
   _4\right)\right) & -\frac{1}{4} e^{\Upsilon _1} \left(\mathbf{U}^2+2 \Upsilon _3
   \Upsilon _4\right) \\
 0 & e^{\Upsilon _2} & \frac{e^{\Upsilon _2} V_1}{\sqrt{2}} & \frac{e^{\Upsilon
   _2} V_2}{\sqrt{2}} & -\frac{1}{4} e^{\Upsilon _2} \mathbf{V}^2 & -\frac{e^{\Upsilon _2}
   \Upsilon _4}{\sqrt{2}} \\
 0 & 0 & 1 & 0 & -\frac{V_1}{\sqrt{2}} & -\frac{U_1}{\sqrt{2}} \\
 0 & 0 & 0 & 1 & -\frac{V_2}{\sqrt{2}} & -\frac{U_2}{\sqrt{2}} \\
 0 & 0 & 0 & 0 & e^{-\Upsilon _2} & -\frac{e^{-\Upsilon _2} \Upsilon _3}{\sqrt{2}}
   \\
 0 & 0 & 0 & 0 & 0 & e^{-\Upsilon _1} \\
\end{array}
\right)&\nonumber\\
\end{eqnarray}
}
\par
In eq.(\ref{exempli2}) we have used the notation $U^i = \Upsilon^{5,i}$ and  $V^i = \Upsilon^{6,i}$  (in our case $i=1,2$) which puts into evidence the existence of two Paint vectors $\mathbf{U},\mathbf{V}$ in the case $r=2$ and the Paint Group covariant structure of the solvable group element $\mathbb{L}(\boldsymbol\Upsilon)^{[2,q]}$. Indeed from eq.(\ref{exempli2}) it is immediate to deduce the general form of the matrix for any value of $q$. 
\par
Starting from eq.(\ref{exempli2}) we easily calculate all the further required items necessary for our argumentation. To begin with we calculate the left-invariant $1$-form:
\begin{equation}\label{tettaforma}
  \Theta \equiv \mathbb{L}^{-1} \mathrm{d}\mathbb{L}
\end{equation}
and then we project it onto the coset generators 
in the orthogonal decomposition of the full $\mathbb{U}$ Lie algebra:
\begin{equation}\label{carnicchio}
  \so(2,4) \, = \, \underbrace{\so(2)\oplus\so(4)}_{\mathbb{H}} \oplus \mathbb{K}
\end{equation}
The list of $K$ generators is given in table \ref{clodoveo}.
\begin{table}[htb]
  \centering
  \begin{alignat*}{7}
   K^1 & =  &\left(
\begin{array}{cccccc}
 1 & 0 & 0 & 0 & 0 & 0 \\
 0 & 0 & 0 & 0 & 0 & 0 \\
 0 & 0 & 0 & 0 & 0 & 0 \\
 0 & 0 & 0 & 0 & 0 & 0 \\
 0 & 0 & 0 & 0 & 0 & 0 \\
 0 & 0 & 0 & 0 & 0 & -1 \\
\end{array}
\right) &\quad ;  \quad & K^2& =  &\left(
\begin{array}{cccccc}
 0 & 0 & 0 & 0 & 0 & 0 \\
 0 & 1 & 0 & 0 & 0 & 0 \\
 0 & 0 & 0 & 0 & 0 & 0 \\
 0 & 0 & 0 & 0 & 0 & 0 \\
 0 & 0 & 0 & 0 & -1 & 0 \\
 0 & 0 & 0 & 0 & 0 & 0 \\
\end{array}
\right) \nonumber\\
   K^3 & =  & \left(
\begin{array}{cccccc}
 0 & \frac{1}{\sqrt{2}} & 0 & 0 & 0 & 0 \\
 \frac{1}{\sqrt{2}} & 0 & 0 & 0 & 0 & 0 \\
 0 & 0 & 0 & 0 & 0 & 0 \\
 0 & 0 & 0 & 0 & 0 & 0 \\
 0 & 0 & 0 & 0 & 0 & -\frac{1}{\sqrt{2}} \\
 0 & 0 & 0 & 0 & -\frac{1}{\sqrt{2}} & 0 \\
\end{array}
\right) &\quad ;  \quad   & K^4& =  & \left(
\begin{array}{cccccc}
 0 & 0 & 0 & 0 & \frac{1}{\sqrt{2}} & 0 \\
 0 & 0 & 0 & 0 & 0 & -\frac{1}{\sqrt{2}} \\
 0 & 0 & 0 & 0 & 0 & 0 \\
 0 & 0 & 0 & 0 & 0 & 0 \\
 \frac{1}{\sqrt{2}} & 0 & 0 & 0 & 0 & 0 \\
 0 & -\frac{1}{\sqrt{2}} & 0 & 0 & 0 & 0 \\
\end{array}
\right) \nonumber\\
  K^5 & =  & \left(
\begin{array}{cccccc}
 0 & 0 & \frac{1}{\sqrt{2}} & 0 & 0 & 0 \\
 0 & 0 & 0 & 0 & 0 & 0 \\
 \frac{1}{\sqrt{2}} & 0 & 0 & 0 & 0 & -\frac{1}{\sqrt{2}} \\
 0 & 0 & 0 & 0 & 0 & 0 \\
 0 & 0 & 0 & 0 & 0 & 0 \\
 0 & 0 & -\frac{1}{\sqrt{2}} & 0 & 0 & 0 \\
\end{array}
\right)&\quad ;  \quad  & K^6& =  & \left(
\begin{array}{cccccc}
 0 & 0 & 0 & \frac{1}{\sqrt{2}} & 0 & 0 \\
 0 & 0 & 0 & 0 & 0 & 0 \\
 0 & 0 & 0 & 0 & 0 & 0 \\
 \frac{1}{\sqrt{2}} & 0 & 0 & 0 & 0 & -\frac{1}{\sqrt{2}} \\
 0 & 0 & 0 & 0 & 0 & 0 \\
 0 & 0 & 0 & -\frac{1}{\sqrt{2}} & 0 & 0 \\
\end{array}
\right)\nonumber\\
   K^7 & =  & \left(
\begin{array}{cccccc}
 0 & 0 & 0 & 0 & 0 & 0 \\
 0 & 0 & \frac{1}{\sqrt{2}} & 0 & 0 & 0 \\
 0 & \frac{1}{\sqrt{2}} & 0 & 0 & -\frac{1}{\sqrt{2}} & 0 \\
 0 & 0 & 0 & 0 & 0 & 0 \\
 0 & 0 & -\frac{1}{\sqrt{2}} & 0 & 0 & 0 \\
 0 & 0 & 0 & 0 & 0 & 0 \\
\end{array}
\right)&\quad ;  \quad  & K^8& =  & \left(
\begin{array}{cccccc}
 0 & 0 & 0 & 0 & 0 & 0 \\
 0 & 0 & 0 & \frac{1}{\sqrt{2}} & 0 & 0 \\
 0 & 0 & 0 & 0 & 0 & 0 \\
 0 & \frac{1}{\sqrt{2}} & 0 & 0 & -\frac{1}{\sqrt{2}} & 0 \\
 0 & 0 & 0 & -\frac{1}{\sqrt{2}} & 0 & 0 \\
 0 & 0 & 0 & 0 & 0 & 0 \\
\end{array}
\right) \nonumber\\
 \end{alignat*}
  \caption{The coset generators of the of $\mathrm{SO(2,4)}/\mathrm{SO(2) }\times \mathrm{SO(4)}$}\label{clodoveo}
\end{table}
Since the $K^i$ are normalized in such a way that $\mathrm{Tr}(K^i\cdot K^j)\, = \, \delta^{ij}$ we can immediately calculate the vielbein as:
\begin{equation}\label{cardigan}
  V^i \, = \, \text{Tr} \left(K^i \Theta \right) \quad ; \quad i\, = \, 1,\dots , 8 
\end{equation}
The vielbein are, as they should, linear combinations, with constant coefficient of the left-invariant $1$-forms $e^i$, defined by the decomposition:
\begin{equation}\label{defiforme}
  \Theta \, = \, \sum_{i=1}^8 \, e^i \, T_i
\end{equation}
The latter have the following explicit appearance 
\begin{eqnarray}
\label{mc1formeA}
  e^1 &=& \mathrm{d}\Upsilon_1  \nonumber\\
  e^2  &=& \mathrm{d}\Upsilon_2 \nonumber\\
  e^3 &=& \mathrm{d}\Upsilon_3+\Upsilon_3 (\mathrm{d}\Upsilon_1-\mathrm{d}\Upsilon_2) \nonumber\\
  e^4 &=& \frac{1}{4} \left(\Upsilon_{6,1}^2 (-\mathrm{d}\Upsilon_3)+\Upsilon_3 \Upsilon_{6,1}^2
   \mathrm{d}\Upsilon_2-2 \sqrt{2} \Upsilon_{6,1} \mathrm{d}\Upsilon_{5,1}-\Upsilon_{6,2}^2
   \mathrm{d}\Upsilon_3+\Upsilon_3 \Upsilon_{6,2}^2 \mathrm{d}\Upsilon_2-2 \sqrt{2} \Upsilon_
   {6,2} \mathrm{d}\Upsilon_{5,2}\right.\nonumber\\
   &&\left.+\left(-\Upsilon_3 \Upsilon_{6,1}^2-2 \sqrt{2} \Upsilon_{5,1} \Upsilon_{6,1}-\Upsilon_3 \Upsilon_{6,2}^2-2 \sqrt{2} \Upsilon_{5,2} \Upsilon_{6,2}+4 \Upsilon_4\right) \mathrm{d}\Upsilon_1+4 \mathrm{d}\Upsilon_4+4 \Upsilon_4
   \mathrm{d}\Upsilon_2\right)\nonumber\\
\end{eqnarray}
\begin{eqnarray}
\label{mc1formeB}
  e^{5,1} &=& \mathrm{d}\Upsilon_{5,1}+\frac{\Upsilon_{6,1} (\mathrm{d}\Upsilon_3-\Upsilon_3 \mathrm{d}\Upsilon_2
   )}{\sqrt{2}}+\left(\Upsilon_{5,1}+\frac{\Upsilon_3 \Upsilon_{6,1}}{\sqrt{2}}\right) \mathrm{d}\Upsilon_1 \nonumber\\
  e^{5,2} &=& \mathrm{d}\Upsilon_{5,2}+\frac{\Upsilon_{6,2} (\mathrm{d}\Upsilon_3-\Upsilon_3 \mathrm{d}\Upsilon_2)}{\sqrt{2}}+\left(\Upsilon_{5,2}
  +\frac{\Upsilon_3 \Upsilon_{6,2}}{\sqrt{2}}\right) \mathrm{d}\Upsilon_1 \nonumber\\
  e^{6,1} &=& \mathrm{d}\Upsilon_{6,1}+\Upsilon_{6,1} \mathrm{d}\Upsilon_2 \nonumber\\
  e^{6,2} &=& \mathrm{d}\Upsilon_{6,2}+\Upsilon_{6,2} \mathrm{d}\Upsilon_2 
\end{eqnarray}
One finds the following relation between the vielbein and the left invariant $1$-forms:
\begin{equation}\label{carmelitano}
  V^i \, = \, \nu^i_A \, e^A\quad ; \quad \nu \, = \, \left(
\begin{array}{cccccccc}
 1 & 0 & 0 & 0 & 0 & 0 & 0 & 0 \\
 0 & 1 & 0 & 0 & 0 & 0 & 0 & 0 \\
 0 & 0 & \frac{1}{2} & 0 & 0 & 0 & 0 & 0 \\
 0 & 0 & 0 & \frac{1}{2} & 0 & 0 & 0 & 0 \\
 0 & 0 & 0 & 0 & \frac{1}{2} & 0 & 0 & 0 \\
 0 & 0 & 0 & 0 & 0 & \frac{1}{2} & 0 & 0 \\
 0 & 0 & 0 & 0 & 0 & 0 & \frac{1}{2} & 0 \\
 0 & 0 & 0 & 0 & 0 & 0 & 0 & \frac{1}{2} \\
\end{array}
\right)
\end{equation}
This result determines the form of the $\kappa$ matrix in this case and consequently
the expression of the $\mathrm{SO(2,4)}$  invariant metric on the symmetric space (\ref{laborroto}) in terms of the left-invariant $1$-forms. Indeed we have:
\begin{eqnarray}\label{barnabone}
  ds^2 & =& \kappa_{AB}\, \, e^A \times e^B \nonumber\\
  \kappa & \equiv & \nu^T\cdot\nu \, = \, \left(
\begin{array}{cccccccc}
 1 & 0 & 0 & 0 & 0 & 0 & 0 & 0 \\
 0 & 1 & 0 & 0 & 0 & 0 & 0 & 0 \\
 0 & 0 & \frac{1}{4} & 0 & 0 & 0 & 0 & 0 \\
 0 & 0 & 0 & \frac{1}{4} & 0 & 0 & 0 & 0 \\
 0 & 0 & 0 & 0 & \frac{1}{4} & 0 & 0 & 0 \\
 0 & 0 & 0 & 0 & 0 & \frac{1}{4} & 0 & 0 \\
 0 & 0 & 0 & 0 & 0 & 0 & \frac{1}{4} & 0 \\
 0 & 0 & 0 & 0 & 0 & 0 & 0 & \frac{1}{4} \\
\end{array}
\right)
\end{eqnarray}
As in all the other cases the left-invariant $1$-forms satisfy a set of  Maurer Cartan equations: specifically those displayed in eq. (\ref{maurocartus}) (with $q=2$) and already discussed in  section \ref{equivosolvo} .
\subsection{The K\"ahler $2$-form}
As we stressed in \cite{geotermico}, the reason why the symmetric spaces (\ref{quaresima}) are all K\"ahler manifolds is the presence in the isotropy compact subgroup
$\mathrm{H_c }\, = \, \mathrm{SO(2) \times H^\prime}$ of the factor $\mathrm{SO(2)} \simeq \mathrm{U(1)}$ and the arrangement of the coset generator vector space $\mathbb{K}$ into a representation $(2\mid \mathbf{v})$ where $2$ is the doublet of $\mathrm{SO(2)}$ and  $\mathbf{v}$ is some irreducible representation of the other factor $ \mathrm{H^\prime}$. All coset manifolds where such situation is realized are K\"ahler manifolds, since the generator of $\mathrm{SO(2)}$ in the $\mathbb{K}$ representation of $\mathrm{H}_c$ can be identified with the complex structure and leads to the explicit expression of the closed K\"ahler $2$-form. In our specific case (\ref{laborroto}) the $\mathrm{SO(2)}$-generator in the fundamental representation of $\mathrm{SO(2,4)}$ is the following one:
\begin{equation}\label{Xci}
  X_c \, = \, \left(
\begin{array}{cccccc}
 0 & \frac{1}{2} & 0 & 0 & -\frac{1}{2} & 0 \\
 -\frac{1}{2} & 0 & 0 & 0 & 0 & \frac{1}{2} \\
 0 & 0 & 0 & 0 & 0 & 0 \\
 0 & 0 & 0 & 0 & 0 & 0 \\
 \frac{1}{2} & 0 & 0 & 0 & 0 & -\frac{1}{2} \\
 0 & -\frac{1}{2} & 0 & 0 & \frac{1}{2} & 0 \\
\end{array}
\right)
\end{equation}
and its representation on the space of $K^i$ generators and hence on the vielbein $V^i$ is the following one:
\begin{equation}\label{XcOnK}
 J_c \, =\, \left(
\begin{array}{cccccccc}
 0 & 0 & -\frac{1}{\sqrt{2}} & \frac{1}{\sqrt{2}} & 0 & 0 & 0 & 0 \\
 0 & 0 & \frac{1}{\sqrt{2}} & \frac{1}{\sqrt{2}} & 0 & 0 & 0 & 0 \\
 \frac{1}{\sqrt{2}} & -\frac{1}{\sqrt{2}} & 0 & 0 & 0 & 0 & 0 & 0 \\
 -\frac{1}{\sqrt{2}} & -\frac{1}{\sqrt{2}} & 0 & 0 & 0 & 0 & 0 & 0 \\
 0 & 0 & 0 & 0 & 0 & 0 & -1 & 0 \\
 0 & 0 & 0 & 0 & 0 & 0 & 0 & -1 \\
 0 & 0 & 0 & 0 & 1 & 0 & 0 & 0 \\
 0 & 0 & 0 & 0 & 0 & 1 & 0 & 0 \\
\end{array}
\right)
\end{equation}
The matrix $J_c$ 
is obtained from the adjoint action of $X_c$ on the $K^i$ coset generators, namely from:
\begin{equation}\label{aggiungoazio}
  J_c^{\phantom{c}ij} \, = \, \ft 12 \text{Tr} \left( \left[X_c \, , \, K^i\right] \cdot K^j\right)
\end{equation}
and it squares to minus the identity:
\begin{equation}\label{carolonagalbani}
  J_c\cdot J_c \, = \, - \, \mathbf{1}_{8\times 8}
\end{equation}
hence it acts as a complex structure on the cotangent bundle (implying the same for the tangent bundle). Correspondingly the K\"ahler $2$-form can be written as:
\begin{equation}\label{formaggiomagro}
  \boldsymbol{\mathcal{K}} \, = \, \sum_{a=1}^8 \sum_{b=1}^8 \, J^c_{ab} V^a\wedge V^b \, = \, -\frac{e^{1}\wedge e^{3}}{\sqrt{2}}+\frac{e^{1}\wedge e^{4}}{\sqrt{2}}+\frac{e^{2}\wedge
   e^{3}}{\sqrt{2}}+\frac{e^{2}\wedge e^{4}}{\sqrt{2}}-\frac{1}{2} \,\sum_{i=1}^{q} e^{5,i}\wedge
   e^{6,i}
\end{equation}
where, once again we have utilized the example under investigation to put the Paint invariance structure into evidence. In the present case the index $i$ takes only two values $i=1,2$ but the formula (\ref{formaggiomagro})
applies to all values of $q$, namely to the entire Tits Satake universality class. In particular for $q=1$, eq.(\ref{formaggiomagro}) coincides, to the expression of the K\"ahler $2$-form  in the case of the genus $g=2$ the Siegel half-plane. One  easily verifies that $\boldsymbol{\mathcal{K}}$ is closed and of maximal rank:
\begin{equation}\label{domenicano}
  \mathrm{d} \boldsymbol{\mathcal{K}}  \, = \, 0 \quad ; \quad \boldsymbol{\mathcal{K}}\wedge\boldsymbol{\mathcal{K}}\wedge \boldsymbol{\mathcal{K}}\wedge\boldsymbol{\mathcal{K}} \, = \, \text{const} \, \times \,
  e^1 \wedge e^2 \wedge \dots \wedge e^{6,2} 
\end{equation}
just as a consequence of the Maurer Cartan equations (\ref{maurocartus}).
\par
The manifold (\ref{laborroto}) equipped with the K\"ahler $2$-form becomes a symplectic manifold $(\mathcal{M}^{[2,2]},\boldsymbol{\mathcal{K}})$ and because of metric equivalence we can say that the solvable Lie group manifold $\mathcal{S}_{[2,2]}$ is a symplectic manifold
$(\mathcal{S}_{[2,2]},\boldsymbol{\mathcal{K}})$.
\newpage
\begin{landscape}
\section{The alternative solvable coordinate basis evidencing the maximal abelian ideal}
\label{auxiliaryappo}
In this subsection we present the explicit form of the solvable Lie group coset representative for the Calabi-Vesentini manifold $\mathcal{M}^[{2,2}]$ and some other auxiliary items necessary for the discussion of section \ref{baracco}
{\tiny{
\begin{eqnarray}\label{newsolvab}
&&\hat{\mathbb{L}}(\mathbf{W}) \,= \, \nonumber\\
&&  \left(
\begin{array}{cccccc}
 e^{w_1} & \frac{e^{w_2} w_3}{\sqrt{2}} & \frac{1}{2} \left(\sqrt{2}
   w_{5,1}+e^{w_2} w_3 w_{6,1}\right) & \frac{1}{2} \left(\sqrt{2}
   w_{5,2}+e^{w_2} w_3 w_{6,2}\right) & \frac{1}{8} \left(\sqrt{2}
   e^{-w_2} \left(4 w_4-e^{2 w_2} w_3
   \left(w_{6,1}^2+w_{6,2}^2\right)\right)-4 \left(w_{5,1}
   w_{6,1}+w_{5,2} w_{6,2}\right)\right) & -\frac{1}{4} e^{-w_1}
   \left(w_{5,1}^2+w_{5,2}^2+2 w_3 w_4\right) \\
 0 & e^{w_2} & \frac{e^{w_2} w_{6,1}}{\sqrt{2}} & \frac{e^{w_2}
   w_{6,2}}{\sqrt{2}} & -\frac{1}{4} e^{w_2}
   \left(w_{6,1}^2+w_{6,2}^2\right) & -\frac{e^{-w_1} w_4}{\sqrt{2}}
   \\
 0 & 0 & 1 & 0 & -\frac{w_{6,1}}{\sqrt{2}} & -\frac{e^{-w_1}
   w_{5,1}}{\sqrt{2}} \\
 0 & 0 & 0 & 1 & -\frac{w_{6,2}}{\sqrt{2}} & -\frac{e^{-w_1}
   w_{5,2}}{\sqrt{2}} \\
 0 & 0 & 0 & 0 & e^{-w_2} & -\frac{e^{-w_1} w_3}{\sqrt{2}} \\
 0 & 0 & 0 & 0 & 0 & e^{-w_1} \\
\end{array}
\right)\nonumber\\
\end{eqnarray}}}
\end{landscape}

\end{document}